\documentclass[12pt,twoside]{article}
\usepackage{epsfig}
\usepackage{amssymb,amsbsy,amsmath}
\newcommand{\AMDRef}{OHM92,OHM93,OnH96}
\newcommand{\BUURef}{BKG84,KJS85,BBC86,GRS87,BeG88,BTW89}
\newcommand{\FMDRef}{Fel90,FBS95,ScF96}
\newcommand{\PauliPotRef}{Wil77,BoG88,BoG88a,DDR87,DoR88,DoR89}
\newcommand{\QMDRef}{AiS86,Aic91,KOM92,PKS92,BHS95,Kon96,MNI96}
\newcommand{\figref}[1]{fig.~\protect\ref{#1}}
\newcommand{\xref}[1]{\protect\ref{#1}}

\newcommand{\fmref}[1]{(\protect\ref{#1})}
\newcommand{\mycaption}[2]{%
\begin{center}\begin{minipage}{145mm}%
\caption[]{#1}\label{#2}%
\end{minipage}\end{center}%
}%
\newcommand{\element}[2]{$^{#1}$#2}
\def\leap{\raisebox{-.6ex}{$\stackrel {<}{\sim}$}} 
\def\V0{\stackrel{\circ}{V}}
\def\v0{\stackrel{\circ}{v}}
\def\half{{\frac{1}{2}}\;}

\newcommand{\CM}{{\mbox{\scriptsize cm}}}

\newcommand{\erf}{\mbox{erf}}
\newcommand{\fm}{\mbox{fm}}
\newcommand{\MeV}{\mbox{MeV}}
\newcommand{\dint}{\mbox{d}}

\renewcommand{\Re}[1]{\mbox{Re}\left(#1\right)\;}
\renewcommand{\Im}[1]{\mbox{Im}\left(#1\right)\;}

\newcommand{\sign}{\mbox{sgn}}

\newcommand{\Tr}{\mbox{Tr}}

\newcommand{\Operator}[1]{\smash{\raisebox{-1.1ex}{
$\!\!\stackrel{\displaystyle #1}{\sim}$}}}
\newcommand{\LittleOperator}[1]{\smash{\raisebox{-1.1ex}{
$\!\!\stackrel{\scriptstyle #1}{\sim}$}}}
\newcommand{\EinsOp}
           {\;\smash{\raisebox{-1.1ex}{$\!\!\stackrel{\!\mbox{1}
            \hspace{-0.4ex}\rule[0.0ex]{0.06ex}{1.60ex}}{\sim}$}}}
\newcommand{\EinsMa}
           {\;\smash{\raisebox{-0.5ex}{$\!\!\stackrel{\!\mbox{1}
            \hspace{-0.4ex}\rule[0.0ex]{0.06ex}{1.60ex}}{}$}}}
\newcommand{\OpHHO}
           {\Operator{H}_{{\vphantom{A}}^{\mbox{\scriptsize HO}}}}
\newcommand{\OphHO}
           {\Operator{h}_{{\vphantom{A}}^{\mbox{\scriptsize HO}}}}

\newcommand{\Rerg}{\Operator{R}_{\mbox{erg}}}
\newcommand{\HHO}{\Operator{H}_{{\vphantom{A}}^{\mbox{\scriptsize HO}}}}
\newcommand{\VI}{\Operator{V}_{{\vphantom{A}}^{\mbox{\scriptsize I}}}}
\newcommand{\EnsembleMean}[1]{\langle\langle \; {#1}\; 
            \rangle\rangle_{\big|\;\scriptstyle{T}} \;}
\newcommand{\ErgodicMean}[1]{\overline{\langle \; {#1}^{\vphantom{A}}
            \;\rangle}_{\big|
            \scriptstyle{\erw{\LittleOperator{H}}}} \;\;}

\def\bra#1{\langle \, {#1} \, | \;}
\def\ket#1{\; | \, {#1} \, \rangle}
\newcommand{\braket}[2]{\langle \, {#1} \, | \, {#2} \, \rangle}

\def\erw#1{\;\langle \; {#1} \; \rangle\;}

\def\summn{\sum_{m,n}^A\;}

\def\sumklmn{\sum_{k,l,m,n}^A\;}
\def\OO{\Big( {\mathcal O}_{mk}{\mathcal O}_{nl} - {\mathcal O}_{ml}{\mathcal O}_{nk} \Big)}
\def\brak{\langle q_k | \;}

\def\ketl{\; | q_l \rangle}
\def\bram{\langle q_m | \;}

\def\ketn{\; | q_n \rangle}
\def\brakl{\langle q_k q_l | \;}
\def\ketkl{\; | q_k q_l \rangle}
\def\ketlk{\; | q_l q_k \rangle}

\def\ketmn{\; | q_m q_n \rangle}

\def\prodklkl{\langle q_k q_l|q_k q_l\rangle}
\def\prodkllk{\langle q_k q_l|q_l q_k\rangle}
\def\prodkmkm{\langle q_k q_m|q_k q_m\rangle}
\def\prodkmmk{\langle q_k q_m|q_m q_k\rangle}

\def\prodlmlm{\langle q_l q_m|q_l q_m\rangle}
\def\prodlmml{\langle q_l q_m|q_m q_l\rangle}

\def\prodkl{\langle q_k|q_l\rangle}

\def\ppqmy#1{\frac{\partial \, {#1}}{\partial q_{\mu}}\;}

\def\ppqny#1{\frac{\partial \, {#1}}{\partial q_{\nu}}\;}

\newcommand{\ddt}{\frac{d}{dt}\;}

\newcommand{\rvec}{\vec{r}}
\newcommand{\xvec}{\vec{x}}
\newcommand{\pvec}{\vec{p}}

\newcommand{\kvec}{\vec{k}}

\newcommand{\vecrk}{\vec{r}_{k}\,}
\newcommand{\vecrl}{\vec{r}_{l}\,}

\newcommand{\Bsim}{\raisebox{-1.1ex}{
            $\!\!\stackrel{\displaystyle B}{\sim}$}}
\newcommand{\rhosim}{\raisebox{-1.1ex}{
            $\!\!\stackrel{\displaystyle \rho}{\sim}$}}

\newcommand{\Hsim}{\raisebox{-1.1ex}{
            $\!\!\stackrel{\displaystyle H}{\sim}$}}
\newcommand{\Psim}{\raisebox{-1.1ex}{
              $\!\!\stackrel{\displaystyle P}{\sim}$}}

\newcommand{\vsim}{\raisebox{-1.1ex}{
              $\!\!\stackrel{\displaystyle v}{\sim}$}}
\newcommand{\Vsim}{\raisebox{-1.1ex}{
              $\!\!\stackrel{\displaystyle V}{\sim}$}}

\newcommand{\Tsim}{\raisebox{-1.1ex}{
              $\!\!\stackrel{\displaystyle T}{\sim}$}}
\newcommand{\Usim}{\raisebox{-1.1ex}{
              $\!\!\stackrel{\displaystyle U}{\sim}$}}
\newcommand{\Gsim}{\raisebox{-1.1ex}{
               $\!\!\stackrel{\displaystyle G}{\sim}$}}
\newcommand{\kvecsim}{\raisebox{-1.1ex}{
               $\!\!\vec{\stackrel{\displaystyle k}{\sim}}$}}

\newcommand{\Jvecsim}{\raisebox{-1.1ex}{
               $\!\!\vec{\stackrel{\displaystyle J}{\sim}}$}}

\newcommand{\Lvecsim}{\raisebox{-1.1ex}{
               $\!\!\vec{\stackrel{\displaystyle L}{\sim}}$}}

\newcommand{\Svecsim}{\raisebox{-1.1ex}{
               $\!\!\vec{\stackrel{\displaystyle S}{\sim}}$}}
\newcommand{\xvecsim}{\raisebox{-1.1ex}{
                $\!\!\vec{\stackrel{\displaystyle x}{\sim}}$}}
\newcommand{\Xvecsim}{\raisebox{-1.1ex}{
                $\!\!\vec{\stackrel{\displaystyle X}{\sim}}$}}
\newcommand{\sigvecsim}{\raisebox{-1.1ex}{
                $\!\!\vec{\stackrel{\displaystyle \sigma}{\sim}}$}}

\begin{document}
%
\typeout{   --- >>>   GSI-Preprint-97-16   <<<   ---   }
\typeout{   --- >>>   GSI-Preprint-97-16   <<<   ---   }
\typeout{   --- >>>   GSI-Preprint-97-16   <<<   ---   }
\title{Fermionic Molecular Dynamics}
 
\author{H. Feldmeier\thanks{email: h.feldmeier\char'100gsi.de,
              WWW:~http://www.gsi.de/$\sim$feldm}\;
and J. Schnack\thanks{email: j.schnack\char'100gsi.de,
              WWW:~http://www.gsi.de/$\sim$schnack}\\[2mm]
Gesellschaft f\"ur Schwerionenforschung mbH, \\ 
Postfach 110 552, D-64220 Darmstadt}

\date{}

\maketitle

\begin{abstract}
\noindent
A quantum molecular model for fermions is investigated which
works with antisymmetrized many--body states composed of
localized single--particle wave packets. The application to the
description of atomic nuclei and collisions between them shows
that the model is capable to address a rich variety of observed
phenomena. Among them are shell effects, cluster structure and
intrinsic deformation in ground states of nuclei as well as
fusion, incomplete fusion, dissipative binary collisions and
multifragmentation in reactions depending on impact parameter
and beam energy.  Thermodynamic properties studied with long
time simulations proof that the model obeys Fermi--Dirac
statistics and time averaging is equivalent to ensemble
averaging. A first order liquid--gas phase transition is observed
at a boiling temperature of $T\approx5~\MeV$ for finite nuclei of
mass $16\dots40$.\\[20mm]


\end{abstract}

\newpage
\tableofcontents
\parskip1.5ex
\newpage
\section{Introduction and summary}

Heavy--ion reactions show typical dissipative phenomena at low
beam energies (a few MeV per nucleon above the Coulomb barrier)
\cite{ScH84,GoN80}. With increasing impact parameter the
complete fusion reactions go over into deeply inelastic
collisions where the scattered nuclei have lost large fractions
of the initial energy and converted that into intrinsic
excitation energy.  The exchange of nucleons causes the mass
and charge numbers of the outgoing nuclei to fluctuate around a
mean value. At the same time energy loss and scattering angle
show fluctuations like a Brownian movement. For the description
of these phenomena the so called particle exchange picture
\cite{BBN78, Ran79,Fel87} which is based on the
assumption of a Fermi gas with long mean free paths turned out
to be rather successful.
 
Around the same time the time--dependent Hartree--Fock model (TDHF)
\cite{DDK85,MaC76,KDM77,DFF78} was conceptually
and numerically developed to a stage where heavy--ion collisions
could be calculated. Although TDHF is a microscopic quantal
model, which is supposed to describe systems with long mean free
path and slow collective motion of the mean field, slow compared
to the Fermi velocity such that the particles and their mean
field can always be in equilibrium, it turned out that only the
dissipation of the collective energy could be described but not
the fluctuations \cite{KDM77} which inevitably go along with any
dissipation. TDHF results are very close to a classical
trajectory picture with friction and very small fluctuations.

The surprising failure of the more quantal and more microscopic
TDHF model compared to the more phenomenological particle
exchange picture, both being based mainly on independent
particle motion, originates in the inability of the TDHF state
to react to small fluctuations. Consider a symmetric system: if
there is a fluctuation in one direction, let us say a nucleon
passes from the left to the right hand nucleus, there is of
course in a many--body quantum state, always with the same
probability the fluctuation in the opposite direction. Thus the
mean field which averages over all configurations, will not
adjust to the new situation, namely that there is now one more
particle in the right hand nucleus, because with exactly the
same probability a particle went out to the other nucleus and
thus the mean particle number stays the same. The same holds
true for recoil effects which would give a fluctuation to the
collective relative momentum. The time evolution with the
Hartree--Fock Hamiltonian conserves global symmetries of the
initial state. The mean field in TDHF is the same for all
macro--channels (mass numbers, charge numbers, relative momenta
etc.) which causes a "spurious cross channel coupling"
\cite{GLD80} suppressing almost all fluctuations.

In the particle exchange picture a random transfer of a nucleon
will cause a jump to a new mean field situation where the
acceptor nucleus is now enlarged in volume such that the new
radius is that of a nucleus with one more particle. This way
the decision is made, which of the many possibilities in the
many--body quantum state is realized and this new situation is
then evolved further in time. Of course with the same
probability the particle could have jumped to the other side,
then this would be the new situation with a larger mean field on
the other side which would be evolved further in time.  This
picture is much more realistic for dissipative phenomena because
the exact solution of the Schr\"odinger equation develops random
phases between the macro--channels such that they cannot
interfere on a macroscopic scale.

When the beam velocity becomes comparable to the Fermi velocity
the mean field picture of binary collisions should break
down. In addition the mean free path of the nucleons becomes
shorter than the diameter of the system due to the larger
intrinsic excitation energy. In BUU--type models
\cite{\BUURef} a random Boltzmann collision term has been
introduced to account for the shorter mean free path, but the
long range part of the two--body interaction is still treated in
a mean field fashion.  The next step has been to replace the
mean--field part by classical molecular dynamics with soft
phenomenological two--body potentials \cite{AiS86,Aic91}.  These
Quantum Molecular Dynamics (QMD) approaches have two main
conceptional difficulties. First, the mean field (long range
part of the interaction) and the collision term (short range
repulsion) should be treated self--consistently
\cite{JAO92,KOM92}. Second, the Pauli principle is reduced to
the numerically enforced prohibition of over--occupation of
phase--space cells. Other approaches try to replace the Pauli
principle by adding a two--body "Pauli potential" in order to
avoid too large one--body phase--space densities
\cite{\PauliPotRef}. The difficulty is that the Pauli principle
requires the many--body state to be antisymmetric with respect
to particle exchange, and this is outside the classical notions.

In the following sections we discuss a model, named Fermionic
Molecular Dynamics (FMD) \cite{\FMDRef}, which combines the
microscopic quantal features including the Pauli principle with
the properties of classical molecular dynamics. Wave packets
replace the classical points in phase space and thus introduce
quantum properties into classical molecular dynamics.
Especially the antisymmetrization of the many--body state has
many consequences genuine in quantum and absent in classical
mechanics.

A classical system of particles (molecules) which interact via a
two--body interaction of van der Waals type --- repulsive at
short and attractive at large distance --- often behaves in a
deterministic chaotic way. Small deviations in the initial
conditions lead to exponentially diverging trajectories in the
many--body phase space. Situations with global symmetries, for
instance the left--right symmetry in the TDHF solution discussed
above, are of measure zero in the classical case. Each particle
follows a single trajectory which cannot split into two or more
with certain amplitudes like in the quantum case.  This property
is also common to FMD since there is only one wave packet per
nucleon for which the equations of motion decide in a
deterministic way where to move, for example whether it will
join the right or left hand nucleus. This "quantization" of the
particle number density is the main difference to TDHF if a
single Slater determinant is used. In a forthcoming publication
we are introducing antisymmetric many--body states which are not
single Slater determinants anymore, but take care of the short
range correlations caused by the repulsive core in the
interaction.

After discussing the concept of the model and the resulting
properties we calculate ground states of nuclei. In
section~\xref{Sec-3} an important insight is that the
antisymmetrization not only takes care of the Pauli principle,
but also delocalizes the wave packets and introduces shell
effects. The ground states turn out to be rather similar to
deformed Hartree--Fock states.

The dynamical properties of the FMD equations of motion are
investigated in section~\xref{Sec-4}. Calculations of heavy--ion
collisions at $6~A\MeV$ show that the typical fluctuations of
dissipative collisions, which are absent in TDHF, are seen in
FMD. Increasing the beam energy to $32~A\MeV$ which corresponds
to a relative velocity of about the Fermi velocity leads to a
break down of the mean field. The very same model which
describes fusion and binary dissipative collisions now predicts
multifragmentation. 

A microscopic transport model developed for non-equilibrium
situations should of course also possess the correct equilibrium
properties. In the last section we proof that the thermo{\it static}
properties calculated with the FMD trial state are those of
Fermi--Dirac statistics. 
Time averaging over dynamical calculations defining the 
thermo{\it dynamic} properties not only shows that FMD equilibrates
towards the Fermi distribution, but even allows to investigate
the liquid--gas phase transition of small nuclei.

\section{The concept of Fermionic Molecular Dynamics}
\label{concept}

The concept of Fermionic Molecular Dynamics is based on the
molecular dynamics picture in which subgroups of the
investigated system are described by their centre of mass
coordinates.  These subgroups are composite bound objects like
molecules or in our case nucleons. The molecules are interacting
by two--body potentials which approximate the complicated
interactions between the constituents of the molecules. The
two--body potentials, which include to a certain degree also the
polarization induced in one molecule by the presence of the
other one, depend in general on the relative distance, the
relative velocity and the relative orientation of the molecules.

Consider for example the interaction $V_{ww}$ between water
molecules which have a dipole moment. First, it is a function of
the distance but also of the relative orientation (see
\figref{P-2.0-1}). Second, depending on the relative velocity
the electron clouds will be able to adjust more or less completely
to the adiabatic situation. This will result in a velocity
dependence.
\begin{figure}[tttt]
\unitlength1mm
\begin{picture}(120,70)
\put(30,0){\epsfig{file=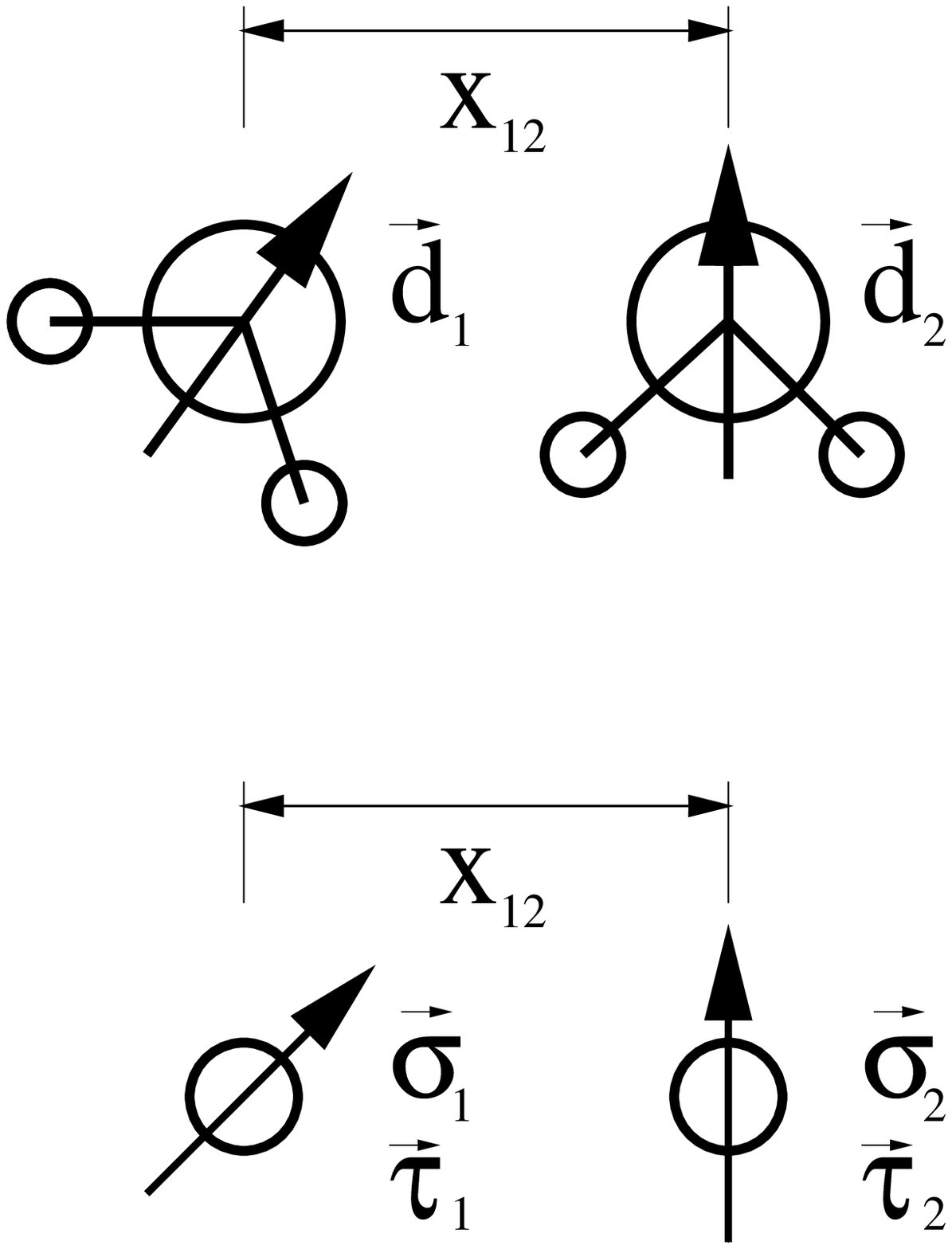,height=70mm}}
\put(100,60){$\mbox{H}_2\mbox{O}$ molecules}
\put(100,50){$V_{ww}(\vec{x}_{12},\vec{v}_{12},\vec{d}_{1},\vec{d}_{2})$}
\put(100,20){protons and neutrons}
\put(100,10){$V_{nn}(\vec{x}_{12},\vec{v}_{12},\vec{\sigma}_{1},
                    \vec{\sigma}_{2},\vec{\tau}_{1},\vec{\tau}_{2})$}
\end{picture}
\begin{center}\begin{minipage}{145mm}
\mycaption{Interactions between water molecules and nucleons 
(protons or neutrons) are rather complex.}{P-2.0-1}
\end{minipage}\end{center}
\end{figure} 

In our case the complex molecule is the nucleon which consists
of three quarks surrounded by a cloud of virtual mesons. Like
the water molecule the nucleon is not spherically symmetric but
has an orientation, the spin. In addition there are two kinds of
molecules, protons and neutrons, which introduces a further
degree of freedom, the isospin. Altogether the picture of a
nucleon--nucleon potential in itself is only an approximation
and there is a certain degree of freedom in its choice.

In classical molecular dynamics the molecules are treated as classical
distinguishable particles which means that one has to solve Newton's
equations of motion for the classical centre of mass coordinates. The
many--body state is given by the trajectory in many--body phase space.

Here Fermionic Molecular Dynamics differs substantially because
its many--body state is described by an antisymmetrized
many--body wave function for the centre of mass coordinates and
the spin degrees of freedom of all nucleons. As already
explained in the introduction, in nuclei the nucleons cannot be
localized in phase space well enough for a non--quantal
treatment. The same holds true for the spin variables. Unlike
the water molecule, where the dipole moment has only small
quantum fluctuations around its mean value, the Hilbert space
for the spin is only two--dimensional and hence one cannot
construct a localized wave--packet for the spin orientation. A
spin $\frac{1}{2}$ degree of freedom is always quantal.

All this together does not allow to treat nucleons in nuclei as
classical particles. Therefore FMD is using localized
single--particle wave packets for the centre of mass coordinates
instead of points in phase space. The indistinguishability of
the particles, which has such prominent consequences like
Fermi motion, Fermi pressure, or Pauli blocking, is introduced
by using a many--body state which is antisymmetric with respect
to particle exchange.
Therefore, all effects which
arise from the fact that the particles are indistinguishable
fermions are taken into account and by construction the
Pauli principle cannot be violated.  The use of wave packets
guarantees that the uncertainty relation cannot be violated
either. It actually turns out that due to antisymmetrization the
localization of the particles is suspended when the wave packets
are overlapping. Thus, shell model states as well as plane waves
are included in $\ket{Q}$ even if it is built from Gaussian
shaped single--particle states (see section \ref{shellstructure})

This means that the trial state contains the essential quantum
properties of a Fermi gas. In addition if the system is dilute
and the distance between the particles is  much larger
than the widths of their wave packets one obtains the classical limit in
the sense of Ehrenfest. Hence the FMD trial state has enough
freedom to cover large areas of quantum physics and connects
seamlessly to classical physics.  

So the first part of the FMD concept is to include all necessary
physics in the many--body trial 
state $\ket{Q}$ but keeping this many--body state still numerically
treatable.

The second part of the concept is to deduce the equations of motion from
the time--dependent quantum variational principle
\begin{eqnarray}
\delta \int_{t_1}^{t_2} \! \! \dint t \;
\bra{Q(t)}\; i \frac{d}{dt} - \Operator{H}\; \ket{Q(t)}
=
0\ .
\end{eqnarray}
This variational principle \cite{KrS81}, which will be explained in more detail
below, ensures automatically the conservation laws, provided the trial state
has enough freedom as shown in section \ref{conslaw}. 

Altogether, the concept is well defined and the success of the
model is granted if one has the proper Hamiltonian $\Operator{H}$ and
a rich enough trial state. The art consists in finding a trial
state which contains the essential degrees of freedom without
becoming numerically intractable. As one does not know a priori
what the important physical phenomena will be, one needs
experimental guidance and physical intuition to set up the trial
state. Even for the same Hamiltonian $\Operator{H}$ there can be
different optimal sets of trial states, depending on the
excitation energy or the part of the phase space the system is
occupying.

If the model fails, the concept is to reexamine the trial state and to include
more or other degrees of freedom. But  also the Hamiltonian is not
sacred as it is in many cases only an approximation to a much more
complex underlying microscopic picture.

There is also an interplay between the trial state and the effective
Hamiltonian. In nuclear physics the nucleon--nucleon interaction is
strongly repulsive at short distances. This important feature contributes
to nuclear saturation and hence is part of the essential physics which
should also be present in the trial state. The antisymmetrized product
state (single Slater determinant) which we use in this paper cannot
account for the depletion of the many--body wave function at small relative
distances between the particles. Therefore, we are using an effective
interaction with a moderate repulsion at short distances in the spirit
of a Brueckner G--Matrix treatment.

\subsection{Parameterized many--body trial state}
\label{trialstate}
\hyphenation{mo-le-cule}

Instead of classical points in phase space Fermionic Molecular
Dynamics deals with wave packets which are localized in phase
space. Each nucleon (molecule) is represented by a Gaussian wave
packet parameterized by the set 
\\
$q(t)=\{\rvec(t),\pvec(t),a(t),\chi(t),\phi(t),\xi\}$:
\begin{equation} \hspace*{-8mm} 
\braket{\xvec}{q(t)} =
\exp\left\{ \, -\; \frac{(\,\xvec-\rvec(t)\,)^2}{2\,a(t)}
 + i\pvec(t)\cdot\xvec \right\} 
\otimes\!\ket{\chi(t),\phi(t)}\! \otimes\! \ket{\xi} \ ,
\label{gaussian} \end{equation}
which in quantum mechanics is the closest analogue to a classical
particle described by a single point $(\rvec(t),\pvec(t))$ in phase--space.

In FMD the match to classical positions and momenta are
the parameters $\rvec(t)$ and $\pvec(t)$ which determine the
mean values of the position and momentum operator of the
single particle state:
\begin{equation}\label{meanrp}  
\rvec(t)=\frac{\bra{q(t)} \xvecsim \ket{q(t)}}{\braket{q(t)}{q(t)}}
\ \ , \ \
\pvec(t)=\frac{\bra{q(t)} \kvecsim \ket{q(t)}}{\braket{q(t)}{q(t)}}\ .
\end{equation}
Due to the quantum mechanical uncertainty relation the wave
packet can be either narrow in coordinate space and wide in
momentum space or vice versa. This non--classical degree of
freedom is taken care of by the complex width parameter
$a(t)=a_R(t)+ia_I(t)$. It determines via its real part
$a_R(t)$ the variance of the
momentum distribution $\sigma_K^2(t)$ by the relation
\begin{equation} 
\frac{3}{2a_R(t)} =
\frac{\bra{q(t)}(\kvecsim-\pvec(t))^2\ket{q(t)}}
                         {\braket{q(t)}{q(t)}}
                  = 3 \sigma_K^2(t) \ .
\label{varp} \end{equation}
Since the wave packet is spherical the widths are equal in all
three Cartesian directions. The imaginary part $a_I(t)$ appears in the
expression for the spatial width $\sigma_X^2(t)$ as
\begin{equation} 
\frac{3}{2} \frac{a_R^2(t)+a_I^2(t)}{a_R(t)} =
\frac{\bra{q(t)}(\xvecsim-\rvec(t))^2\ket{q(t)}}{\braket{q(t)}{q(t)}}
                 = 3 \sigma_X^2(t)
\label{varr} \end{equation}
and determines in how far the wave packet is of minimal uncertainty.
The product of the variances
\begin{equation} 
\sigma_X^2 \ \sigma_K^2 =
\frac{1}{4}\left(1+\frac{a_I^2}{a_R^2}\right)
\label{uncert}
\end{equation}
shows that for $a_I=0 $ one has the minimum--uncertainty packet where
$\sigma_X \sigma_K =\half$,
while for $a_I \neq 0$
the uncertainty can become arbitrarily large.
This means that the particle occupies more than $(\hbar/2\pi)^3$
of phase space volume but at a lower phase space density, such that
other fermions can find place at the same area in phase space. 
 
Besides the parameters for the spatial part of the wave packet there
are two parameters $\chi(t)$ and $\phi(t)$
for the spin degree of freedom. If one
parameterizes the trial spin--state by
\begin{equation} 
\braket{m_s}{\chi(t),\phi(t)} =
\left\{ \begin{array}{l@{\quad:\quad}l}
 \cos\frac{\chi(t)}{2}                    &  m_s=\half \\
 \sin\frac{\chi(t)}{2}\,\mbox{e}^{i\phi(t)} & m_s=-\half
       \end{array} \right.\ ,
\label{spinor} \end{equation}
the relation between the parameters and the corresponding spin
operators is
\begin{equation} 
\vec{\sigma}(t)=\frac{\bra{q(t)}\sigvecsim
\ket{q(t)}}{\braket{q(t)}{q(t)}} \ ,
\label{spin} \end{equation}
where $\vec{\sigma}(t) = (\sin\chi(t)\cos\phi(t),\sin\chi(t)\sin\phi(t),
\cos\chi(t))$
is a vector in the 3--dimensional real space and the
quantization axis is the 3--axis.  Thus $\vec{\sigma}(t)$ can be
regarded as the ``classical'' spin direction just like
$\rvec(t)$ or $\pvec(t)$, although there is no classical spin
degree of freedom which can vary only its direction but not its
magnitude as it is the case for $\vec{\sigma}(t)$.
 
In principle the same parameterization can be chosen for the
isospin part $\ket{\xi}$.  A time--dependent isospin would mean
that for example due to the exchange of charged pions neutrons
can dynamically transform into protons and vice versa.  Isospin
symmetry is, however, only an approximate symmetry, the Coulomb
interaction and the difference in proton and neutron mass break
it. A linear superposition of protons and neutrons in one
wave packet would impose undesired artificial symmetries. For
example only the proton component would feel the Coulomb
repulsion and its centre of the wave packet would be accelerated
away from the centre for the neutron component. Therefore,
either one gives each component its own wave packet or one does
not allow proton neutron mixing in the trial state. In this
paper we shall not consider rotations in isospin space but
assume $\ket{\xi}$ to be independent of time and either
$\ket{\xi}=\ket{proton}$ or $\ket{\xi}=\ket{neutron}$.
 
To describe a system with $A$ fermions we construct a Slater
determinant $\ket{Q(t)}$ with these parameterized single--particle
states $\ket{q_k(t)}$
\begin{equation} 
\ket{q_1(t),q_2(t),\cdots,q_A(t)} \equiv
    \ket{Q(t)} =
    \frac{1}{\braket{\widehat{Q(t)}}{\widehat{Q(t)}}^{\half}}
  \ \ket{\widehat{Q(t)}} \ ,
\label{many} \end{equation}
where the antisymmetrized but not normalized state
$\ket{\widehat{Q(t)}}$ is given by
\begin{equation} 
    \ket{\widehat{Q(t)}} =
 \frac{1}{A!}\sum_{all\ \pi} \mbox{sgn}(\pi)
 \ \ket{q_{\pi(1)}(t)}\otimes\ket{q_{\pi(2)}(t)}
                        \otimes\cdots\otimes\ket{q_{\pi(A)}(t)}\ .
\label{permut} \end{equation}
The sum runs over all permutations $\pi$ and sgn($\pi$) is the sign of
the permutation.
It should be noted that 
$q_k(t)=\{\rvec_k(t),\pvec_k(t),a_k(t),\chi_k(t),\phi_k(t),\xi_k\}$
denotes the set of parameters specifying the single--particle
state with number $k$. The parameter set for the many--body state
thus reads
\begin{eqnarray}
Q(t)&=&\{\rvec_1(t),\pvec_1(t),a_1(t),\chi_1(t),\phi_1(t),\xi_1;
\rvec_2(t),\cdots;\rvec_A(t),\cdots,\xi_A\}\nonumber\\
\nonumber\\ 
    &=& \{\ q_{\nu}(t)\ |\ \nu=1,\cdots,\mbox{N}A\}\ ,
\label{set}\end{eqnarray}
where N is the number of real parameters per particle, in our case N=10.
Whenever $q$ carries a Greek index it is an individual parameter,
whereas a Latin index implies that $q_k$ is the whole set for the 
state $\ket{q_k}$.

Due to antisymmetrization FMD is constrained to the
antisymmetric subspace of the Hilbert space and hence the Pauli
principle is a priori incorporated. Furthermore, the projection
(\ref{permut}) from a product state onto the antisymmetric
subspace destroys for overlapping Gaussians the localization of
the particles and introduces shell model states.  This will be
discussed and explicitly shown in section \ref{shellstructure}.
If the single--particle states $\ket{q_k}$ are not overlapping,
the antisymmetrization has no effect anymore and the particles
are localized in the individual wave packets. In this limit we
return to classical Newtonian mechanics for $\rvec_k(t)$ and
$\pvec_k(t)$, which however can still be coupled to the
non--classical variables $\vec{\sigma}_k(t)$ for the spin
directions and the widths $a_k(t)$.  Of course also in this
limit the particles are indistinguishable and it is not possible
to decide which particle occupies which Gaussian packet.

Even though eqs. (\ref{meanrp}) - (\ref{varr}) and (\ref{spin})
provide unique relations between the parameters and the expectation
values of the corresponding operators, this is no longer true for the
antisymmetrized many--body state. To illustrate this important aspect
let us regard a two--body state of two packets which for simplicity
have the same real width parameter $a_1=a_2=a_R$ and the same spin, but
different $\rvec_1,\rvec_2$ and $\pvec_1, \pvec_2$
\begin{equation} 
\ket{q_1, q_2}_a=\frac{1}{\sqrt{2N}}
\big\{ \ket{q_1}\otimes \ket{q_2} -\ket{q_2}\otimes \ket{q_1} \big\}
\ .
\label{2state}\end{equation}
with the normalization
\begin{align} 
N
&=
\braket{q_1}{q_1}\braket{q_2}{q_2}-|\braket{q_1}{q_2}|^2\\
&=
\braket{q_1}{q_1}\braket{q_2}{q_2}
\left(
1-\exp\left\{ -\xi_{12}^2\right\}
\right) \ , \qquad
\xi_{12}^2=(\rvec_1-\rvec_2)^2/a_R-(\pvec_1-\pvec_2)^2 a_R \ ,\nonumber
\label{2norm}
\end{align}
where $\xi_{12}$ measures the distance in phase space.

The operator for the position of particle 1, 
$\xvecsim(1)=\xvecsim\otimes\EinsOp$,
(which by the way is not an observable because it is not symmetric
under particle permutation) has the following expectation value
\begin{align}
&
_a\bra{q_1, q_2} \xvecsim(1)\ket{q_1, q_2}_a
\\
&
=\frac{1}{2N}\left\{\bra{q_1}\xvecsim\ket{q_1}+
                    \bra{q_2}\xvecsim\ket{q_2}-
     2 {\mbox{Re} (\bra{q_1}\xvecsim\ket{q_2}\braket{q_2}{q_1})} \right\}
\nonumber \\
&
= \half(\rvec_1+\rvec_2)
\ . \nonumber
\label{x1}\end{align}
The result is not $\rvec_1$ as presumed from Ehrenfest type
arguments, however, 
\\
$\left< \xvecsim(1)+\xvecsim(2)\right> =\rvec_1+\rvec_2$.

Another example is the relative distance between two identical
fermions, $\xvecsim(1)-\xvecsim(2)$, (again not an observable, but
used for example in its classical meaning in the collision term of
QMD \cite{Aic91}) with the expectation value
\begin{equation}
_a\bra{q_1, q_2} \xvecsim(1)-\xvecsim(2) \ket{q_1, q_2}_a =0.
\label{x1-x2}\end{equation}
If the spin of the two fermions is the same, the two--body state has to
be antisymmetric in coordinate space and hence this expectation value is
always zero, independent on the actual value of $\rvec_1$ and
$\rvec_2$. The same holds true if the spins are anti-parallel. If the
spins are not parallel the "distance" depends
on the spin orientations, which only reflects that
$\rvec_1-\rvec_2$ is not the mean distance between the two fermions.

An observable is for example the rms--distance
$|\xvecsim(1)-\xvecsim(2)|$ which, 
however, also depends on the spin directions and the relative
momentum. The reason is that for the component of the trial state in which
the two fermions have total spin $S=0$ the radial part of the wave
function is symmetric, while for the three $S=1$ components it is
antisymmetric. Hence the rms--distance is different for the singlet
and the triplet component. In addition the exchange term contributes
less if the relative momentum becomes large. 
The Pauli principle, which demands that two fermions
cannot be in the same phase space cell, is of course the origin of
these classically not existing correlations in the rms--distance.

Another example for an observable which is even used to motivate a so
called "Pauli potential" is the kinetic energy \cite{BoG88a,DDR87}. For
equal spins and equal widths $a_1=a_2=a_R$ we get 
\begin{align}
\left< \ \Tsim \ \right> 
=& \frac{1}{2m} \
_a\bra{q_1, q_2} \kvecsim^2(1)+\kvecsim^2(2) \ket{q_1, q_2}_a 
\\
=&
\frac{1}{2m}\frac{1}{N}\left\{\bra{q_1}\kvecsim^2\ket{q_1}+
                    \bra{q_2}\kvecsim^2\ket{q_2}-
     2 {\mbox{Re} (\bra{q_1}\kvecsim^2\ket{q_2}\braket{q_2}{q_1}) } \right\} 
\nonumber \\
=&
\frac{1}{2m} \frac{\half (\pvec_1-\pvec_2)^2+(\rvec_1-\rvec_2)^2/a_R
\exp\left\{-\xi_{12}^2\right\} }
{1-\exp\left\{-\xi_{12}^2\right\}}
\nonumber \\
&
+\frac{1}{4m}(\pvec_1+\pvec_2)^2  + \frac{3}{2ma_R}
\nonumber .
\label{Kin2} 
\end{align}
One sees that the kinetic energy is not simply
$(\pvec_1^{\,2}+\pvec_2^{\,2})/(2m)$ but depends also on the
"relative distance" $\rvec_1-\rvec_2$ and on the distance in
phase space $\xi_{12}$. But one has to be aware that for wave
packets which are not of minimal uncertainty (i.e. the imaginary
part $a_I$ of the width parameter $a$ is not zero) the whole
expression is completely different and $|\braket{q_2}{q_1}|^2$
can be zero even when the packets fully overlap spatially. A
non-vanishing $a_I$ is nothing exceptional, it describes for
example the well--known spreading of the packets in free
space. Therefore, the constraint $a_I=0$ is not appropriate for
the dynamical case and leads actually to unphysical scatterings
and strong hindrance of evaporation.

In section \ref{shellstructure} we shall show for an $A$--body
system that even if all parameters for the momenta vanish,
i.e. $\pvec_1=\pvec_2= \cdots =\pvec_A=0$, the kinetic energy is
neither zero nor the sum of the zero-point energies $3/(2m
a_R)$, but the momentum distribution is that of a Fermi gas with
a sharp edge at the Fermi momentum.

To summarize this discussion, one should always carefully distinguish
between parameters and physical observables. The parameters are in
general not the expectation values of the corresponding operators.
\subsection{Time--dependent variational principle}
\label{TDVP}

After having set up the many--body trial state $\ket{Q(t)}$ one
has to construct equations of motion for the set of parameters
$Q(t)=\{\; q_{\nu}(t)\ |\ \nu=1,2,3,\cdots \; \}$ which are the
generalized coordinates of the system. As shown in the previous
section the parameters may loose their original physical meaning
due to antisymmetrization. Therefore one must not simply use the
classical equations of motion for the position and momenta. In
addition, even without antisymmetrization, classical physics
does not tell how the complex width parameter $a_k$ should be
evolved in time. Also the equations for the angles $(\chi_k,
\phi_k)$ of the spin direction $\vec{\sigma}$, which turn out to
be of the Bargmann--Michel--Telegdi type \cite{BMT59}, are not
self--evident.
 
Therefore, the equations of motion for the Fermionic Molecular
Dynamics model are derived from the following time--dependent
variational principle
\begin{equation}
\delta \int_{t_1}^{t_2} \! \! dt \;
\bra{Q(t)}\; i \frac{d}{dt} - \Hsim\; \ket{Q(t)} \ =\ 0
\label{var}
\end{equation}
in which the trial state $\ket{Q(t)}$ is to be varied. This has
the advantage that the geometry of the manifold of trial states
is automatically taken care of, irrespective of how the
parameters of the states are defined. Furthermore, conservation
laws follow in a transparent way from invariance properties of
the trial state and the Hamiltonian, see section \ref{conslaw}.

The variation has to be performed with respect to each parameter
$q_{\nu}(t)$ with the end points kept fixed, i.e.
$\delta q_{\nu}(t_1) = \delta q_{\nu}(t_2) = 0 $.
The operator $\Hsim$ is the total Hamiltonian of the system.
(Throughout the paper operators in Hilbert space will be underlined
with a twiddle to distinguish them from parameters or expectation
values.) 
 
The Euler--Lagrange equations
\begin{eqnarray} 
\frac{d}{dt}\frac{\partial{\mathcal L}}{\partial \dot{q}_{\nu}}
 - \frac{\partial{\mathcal L}}{\partial q_{\nu}} = 0
\qquad , \qquad \nu=1,2,\cdots,N
\label{Lagr1}
\end{eqnarray}
which result from the variation (\ref{var}) are written in terms of the
Lagrange function
\begin{eqnarray} 
{\mathcal L} \left( Q(t), \dot{Q}(t) \right) &:=&
\bra{Q(t)}\; i \frac{d}{dt} - \Hsim \;\ket{Q(t)} \nonumber \\
&=& {\mathcal L}_0\left(Q(t),\dot{Q}(t)\right)
   \   - \ {\mathcal H}(Q(t)) \ ,
\label{Lagr}
\end{eqnarray}
with
\begin{eqnarray} 
{\mathcal L}_0 \left( Q(t), \dot{Q}(t) \right) :=
\bra{Q(t)} i \frac{d}{dt} \ket{Q(t)} = 
\sum_\nu \ \bra{Q(t)} 
i\frac{\partial}{\partial q_\nu}   \ket{Q(t)} \
\dot{q}_{\nu}\ ,
\label{L0}\end{eqnarray}
in which $\dot{Q}(t)=\{\,\dot{q}_{\nu}(t)\equiv dq_{\nu}/dt\ |\ 
\nu=1,2,\cdots\,\}$ is the set of generalized velocities
and ${\mathcal H}(Q(t))$ is the Hamilton function defined as the expectation
value of the Hamiltonian $\Hsim$:
\begin{equation}
{\mathcal H}(Q(t))=\bra{Q(t)} \Hsim \ket{Q(t)} \ .
\end{equation}
Different from classical mechanics the Lagrange function (\ref{Lagr})
is linear in the velocities $\dot{q}_{\nu}$ but at the same time
the set $Q(t)=\{q_{\nu}(t)\}$ contains
both, coordinates and momenta.
 
Using the general structure (\ref{Lagr}) of the Lagrange
function the Euler--Lagrange equations (\ref{Lagr1}) in their
most general form can be written as
\begin{equation} 
\sum_\nu {\mathcal A}_{\mu \nu}(Q)\ \dot{q}_{\nu}
= - \ppqmy{{\mathcal H}(Q)}
\label{Bewgl}
\end{equation}
or, if ${\mathcal A}_{\mu\nu}$ is not singular \cite{CPY91}
the equations of motion are
\begin{equation} 
\dot{q}_{\mu} = -\sum_\nu {\mathcal A}_{\mu \nu}^{-1}(Q)
                                       \  \ppqny{{\mathcal H}(Q)}\ ,
\label{Bewglq}
\end{equation}
where
\begin{equation} 
{\mathcal A}_{\mu\nu}(Q) = -{\mathcal A}_{\nu\mu}(Q) =
\frac{\partial^{2}{\mathcal L}_{0}}{\partial \dot{q}_{\mu}
\partial q_{\nu}} -
\frac{\partial^{2}{\mathcal L}_{0}}{\partial \dot{q}_{\nu} \partial q_{\mu}}
 \ ,
\end{equation}
is a skew symmetric matrix, which depends in general on all
variables $Q(t)=\{q_{\nu}(t)\}$. For details see ref. \cite{FBS95}.
 
With help of the matrix ${\mathcal A}_{\mu\nu}$ one can define generalized 
Poisson brackets \cite{KrS81} as
\begin{equation} 
 \{ {\mathcal H},{\mathcal B} \} :=
\sum_{\mu,\nu} \ \frac{\partial {\mathcal H}}{\partial q_{\mu}} \
{\mathcal A}^{-1}_{\mu\nu} \
                 \frac{\partial {\mathcal B}}{\partial q_{\nu}} \ ,
\label{Poisson}\end{equation}
such that the time derivative of an expectation value
\begin{equation} 
{\mathcal B}(t) = \bra{Q(t)} \Bsim \ket{Q(t)}
\label{ExpB}
\end{equation}
of a time--independent operator $\Bsim$ calculated with the
trial state $\ket{Q(t)}$ is given by
\begin{eqnarray} 
\ddt {\mathcal B}(t)
  &=& \ddt \bra{Q(t)} \Bsim \ket{Q(t)}
= \sum_{\nu} \dot{q}_{\nu} \ \ppqny{\mathcal B} \nonumber \\
&=&
\sum_{\mu,\nu} \ \frac{\partial {\mathcal H}}{\partial q_{\mu}} \
{\mathcal A}^{-1}_{\mu\nu} \
                 \frac{\partial {\mathcal B}}{\partial q_{\nu}}
         = \{ {\mathcal H},{\mathcal B}  \}
\ . 
\label{Opbgl}\end{eqnarray}
Equation (\ref{Opbgl}) has the symplectic manifold structure of
Hamiltonian dynamics \cite{Arn89}, but in the general case the
parameters $q_{\nu}$ cannot be grouped into pairs of canonical
variables.  However, according to Darboux's theorem \cite{Arn89}
canonical variables exist locally. They are non--linear
functions of the parameters $q_{\nu}$ and have to be constructed
such that ${\mathcal A}_{\mu\nu}^{-1}$ assumes the canonical
form
\begin{equation} 
{\mathcal A}_{\mu\nu}^{-1}=
\left(
\begin{array}{cc}    0  & -\EinsMa    \\ \EinsMa &   0 
\end{array} 
\right)\ ,
\label{Astandard}\end{equation}
where $\EinsMa\ $ is the unit matrix. Their choice, however, is
not unique. 

\subsection{Conservation laws}
\label{conslaw}
 
After having solved the equations of motion (\ref{Bewglq}) for
the parameters $Q(t)=\{q_{\nu}(t)\ |\ \nu=1,\cdots,\mbox{N}\}$
the trial state $\ket{Q(t)}$ is known at all times. Thus one can
calculate the expectation value ${\mathcal G}(t) = \bra{Q(t)}
\Gsim \ket{Q(t)}$ of an arbitrary time--independent operator
$\Gsim$.  With the definition (\ref{Poisson}) for the Poisson
brackets the time derivative of this expectation value can be
written as
\begin{equation}
\ddt {\mathcal G} = \{ {\mathcal H},{\mathcal G} \} \  .
\end{equation}
The expectation value is conserved in time if \cite{BLL89}
\begin{equation}
\{ {\mathcal H},{\mathcal G} \}  =
\sum_{\mu,\nu} \ \frac{\partial {\mathcal H}}{\partial q_{\mu}} \
{\mathcal A}^{-1}_{\mu\nu} \
                 \frac{\partial {\mathcal G}}{\partial q_{\nu}} = 0
\end{equation}
Since
${\mathcal A}^{-1}_{\mu\nu}$
is skew symmetric the energy $\mathcal H$ itself is always
conserved by the equations of motion, provided they are derived
from the variational principle (\ref{var}). This
is completely independent on the choice of the trial state .
 
For other constants of motion
we consider a unitary transformation with $\Gsim$ as the
hermitean generator
\begin{equation}
\Usim = \exp\left(\,i\,\epsilon \; \Gsim \, \right) \quad , \quad
\epsilon \quad \mbox{real} \ .
\end{equation}
If $\Usim$ maps the set of
trial states onto itself
\begin{equation}
\Usim \ket{Q}=\ket{Q'} \quad \in \quad \left\{ \ket{Q} \right\}\ ,
\end{equation}
then, as a result of the equations of motion, it can be shown
\cite{FBS95} that
\begin{equation} 
 \{ {\mathcal H},{\mathcal G} \} =
            \bra{Q(t)} i\left[\Hsim,\Gsim \right]\ket{Q(t)} \ .
\label{gener}\end{equation}
This means that for this class of generators the generalized
Poisson bracket is just the expectation value of the
commutator with $i\Hsim$.

Relation (\ref{gener}) is very useful for two reasons.
First, if $\Gsim$ commutes with the Hamiltonian $\Hsim$ and
$\exp(i\epsilon \Gsim) \ket{Q}=\ket{Q'}$ then
$\bra{Q(t)} \Gsim \ket{Q(t)}$
is automatically a constant of motion.
 
Second, this relation is an important guidance for the choice of
the trial state $\ket{Q}$.  If one wants the model to obey
certain conservation laws then the set of trial states should be
invariant under the unitary transformations generated by the
constants of motion. For example, total momentum conservation
implies that a translated trial state is again a valid trial
state.  This is fulfilled for the trial states specified in
section \ref{trialstate}.  The Gaussians defined in
eq. (\ref{gaussian}) can be translated or Galilei boosted, the
latter taking care of the conservation of the centre of gravity.
 
Conservation of total spin $\Jvecsim =\Lvecsim+\Svecsim$ 
is guaranteed if rotation of the trial state in
coordinate and spin space  results again
in a trial state. This implies that the Gaussian (\ref{gaussian})
has to have either a spherical shape or the width parameter
has to be replaced by a complete tensor with 12 real parameters.
It also means that all spin directions in $\ket{\phi,\chi}$
have to be allowed, otherwise the rotation would in general lead
out of the manifold of trial states. 
 
If $\Gsim$ does not commute with $\Hsim$, relation \fmref{gener}
sheds some light on the quality of the variational principle
\fmref{var}.  It says that under the premise that
$\exp(i\epsilon \Gsim)$ does not map out of the set of trial
states the time derivative of the expectation value of $\Gsim$
calculated with a trial state is exact.

\begin{equation}
\ddt{\mathcal G}(t)=\ddt {\mathcal G}_{exact}(t)\ ,
\label{deviation}\end{equation}
where the exact solution with the initial state $\ket{Q(t)}$ is
\begin{equation}
{\mathcal G}_{exact}(t+\tau)=\bra{Q(t)}
\mbox{e}^{\displaystyle{i\tau} \Hsim}\ \Gsim\
\mbox{e}^{\displaystyle{-i\tau} \Hsim} 
\ket{Q(t)} \ .
\end{equation}
Due to the fact that $\ket{Q(t+\tau)}$ is only the approximate time
evolution of $\ket{Q(t)}$ for some time $\tau$ later
${\mathcal G}(t+\tau)$ will begin to deviate from ${\mathcal
G}_{exact}(t+\tau)$ for larger $\tau$.

The kinetic energy $\Tsim$ is such a generator. Since our trial
state \fmref{many} is the exact solution of the Schr\"odinger
equation without interactions it fulfills
\begin{equation}
\exp(-i\tau \Tsim) \ket{Q(t)} = \ket{Q(t+\tau)}\ .
\end{equation}
With the two--body interaction included, $\ket{Q(t)}$ is not an exact 
solution anymore, but the expectation value of the total kinetic energy,
which then is not a conserved quantity any longer, is well
approximated in the sense of equation \fmref{deviation}.

\subsection{Two--body Hamiltonian}
\subsubsection{Effective nucleon--nucleon interaction}

Up to now the interaction $\Vsim$ contained in the Hamiltonian
was not specified. Since in this paper we shall investigate
small and medium--heavy nuclei, we choose an effective two--body
potential suited for mass numbers up to about 50.  This
interaction is repulsive at small and attractive at larger
distances.  The repulsive core is, however, rather weak and one
should regard the potential as a phenomenological ansatz for a
G--matrix rather than the free nucleon--nucleon interaction which
has a very strong repulsion for distances smaller than 0.5 fm.
In this paper we are using a potential of the form
\begin{eqnarray}
\label{V-Def}
\Operator{V}(i,j)
&=&
\Operator{V}_{a}
\left(
w_a + m_a \Operator{P}^{{r}} 
+ b_a \Operator{P}^{{\sigma}} + h_a \Operator{P}^{{\tau}}
\right)
\\
&+&
\Operator{V}_{b}
\left(
w_b + m_b \Operator{P}^{{r}} 
+ b_b \Operator{P}^{{\sigma}} + h_b \Operator{P}^{{\tau}}
\right)
\nonumber\\
&+&
\Operator{V}_{c}
\left(
w_c + m_c \Operator{P}^{{r}} 
+ b_c \Operator{P}^{{\sigma}} + h_c \Operator{P}^{{\tau}}
\right)
\nonumber
\end{eqnarray}
where
$\Operator{P}^{{r}},\Operator{P}^{{\sigma}},\Operator{P}^{{\tau}}$
denote the various exchange operators for coordinate, spin and
isospin. The radial dependences are of Gaussian type given by
\begin{eqnarray}
\bra{\xvec_i,\xvec_j}\Vsim_{a,b,c}(i,j)\ket{\xvec_k,\xvec_l}
&=&
V_{a,b,c}\;\mbox{exp}\left\{-\frac{(\xvec_i-\xvec_j)^2}{r_{a,b,c}^2}\right\}
\\
&& \times
\delta^3(\xvec_i-\xvec_k) \delta^3(\xvec_j-\xvec_l)
\nonumber \,
\end{eqnarray}
which allows to calculate all matrix elements and their
derivatives analytically.
The parameters
\begin{align}
&
V_{a} = -7.10\MeV\ ; \quad r_{a} = 1.16\fm
\\
&
V_{b} =-31.90\MeV\ ; \quad r_{b} = 2.22\fm
\nonumber \\
&
V_{c} =+81.65\MeV\ ; \quad r_{c} = 0.735\fm
\nonumber \\
&
w_a=8.700,\quad m_a=5.610,\quad b_a=7.860,\quad h_a= -21.170
\nonumber \\
&
w_b=0.133,\quad m_b=0.514,\quad b_b=0.085,\quad h_b= 0.268
\nonumber\\
&
w_c=1.000,\quad m_c=0.000,\quad b_c=0.000,\quad h_c= 0.000
\nonumber
\end{align}
were determined in order to reproduce the binding energies of a
wide range of medium--heavy isotopes.

\subsubsection{Coulomb interaction}

The Coulomb interaction is included in the Hamilton operator,
it is given by
\begin{eqnarray}
\label{VCoulomb}
\bra{\xvec_i,\xvec_j}\Vsim_{c}(i,j)\ket{\xvec_k,\xvec_l}
&=&
\frac{1.44\MeV\fm}{|\xvec_i-\xvec_j|}\;\Psim^p \otimes \Psim^p
\delta^3(\xvec_i-\xvec_k) \delta^3(\xvec_j-\xvec_l)
\ ,
\end{eqnarray}
where $\Psim^p$ denotes the projection operator on the protons
\begin{eqnarray}
\Operator{P}^p \ket{q_m}
&=&
\half (1 + \xi_m) \ket{q_m}
\ .
\end{eqnarray}
The isospin variable $\xi_m$ takes values $\xi_m=1$ for protons
and $\xi_m=-1$ for neutrons, respectively.

In order to speed up computing time the expectation values of
the two--body potentials are approximated as given in the
appendix and spin degrees have been kept fixed in the following
calculations.

\section{Ground states in FMD}
\label{Sec-3}

The ground state of a nucleus is the many--body state
$\ket{Q_{GS}}$ in which the energy 
${\mathcal H}=\bra{Q_{GS}}\Hsim\ket{Q_{GS}}$ is an absolute minimum with
respect to variation of all parameters $q_\nu$, therefore
\begin{equation}
\frac{\partial}{\partial q_{\nu}}\, {\mathcal H}=0 \ .
\label{stat.}\end{equation}
This implies that the FMD ground state is completely
time--independent (up to an overall phase) and the time derivatives
of all parameters vanish because, by definition, all
generalized forces $\partial{\mathcal H}/{\partial q_\mu}$ are zero and hence
\begin{equation}
\dot{q}_{\nu} = - \sum_{\mu}\ {\mathcal A}_{\nu\mu}^{-1}\
 \frac{\partial{\mathcal H}}{\partial q_{\mu}}=0 \ .
\end{equation}
Requirement (\ref{stat.})
not only determines the positions $\rvec_k$ and momenta $\pvec_k$ but also
the complex widths $a_k$ and the spin directions $(\chi_k,\phi_k)$.
 
Since we do not correct for the centre of mass motion in the
dynamical calculation we use the expectation value of the
Hamilton operator as the ground state energy and do not subtract
the centre of mass energy. The centre of mass energy is of the
order of $10\MeV$ for all isotopes, so that its contribution to
the energy per particle vanishes for larger mass numbers. An
ansatz for the many--body state where centre of mass and relative
motion separate is introduced in ref. \cite{KiD96}.

Besides the ground state energy we also calculate the
root--mean--square radius of the charge distribution. 
\begin{eqnarray}
E_{GS}&=&\bra{Q_{GS}}\Hsim\ket{Q_{GS}}
\\
R_{rms}^2 &=& \frac{1}{Z}\, 
\sum_{i=1}^A\;\bra{Q_{GS}}(\xvecsim(i)-\Xvecsim_{CM})^2
                       \Psim^p(i)\ket{Q_{GS}} + R^2_{proton}\  \ ,
\end{eqnarray}
where the centre of mass position operator is
\begin{eqnarray}
\Xvecsim_{CM} =  \frac{1}{A}\sum_{i=1}^A\; \xvecsim(i)\ .
\end{eqnarray}
The operator $\Psim^p(i)$ projects on protons and
$R_{proton}=0.876$ fm takes the finite charge radius of the
proton into account.

\subsection{Ground states of nuclei}
\label{groundstates}

In table~\xref{T-3.1-1} we summarize ground state binding
energies and charge radii and compare them with experimental
results. The interaction \fmref{V-Def} describes a wide range of
isotopes with a satisfying accuracy.
\begin{table}[hhhh]
\begin{center}
\begin{tabular}{|c||c|c|c|c||c|c|}
\hline
      & \multicolumn{2}{c|}{experiment}
      & \multicolumn{2}{c||}{FMD}
      & experiment & FMD\\
Isotope& $E_{GS}$ & $E_{GS}/A$ & $\erw{\Operator{H}}$ 
      & $\erw{\Operator{H}}/A$ 
      & $R_{rms}$ & $R_{rms}$ \\
      & ( MeV ) &  ( MeV )  & ( MeV ) & ( MeV ) & ( fm ) & ( fm ) \\
\hline
\hline
\element{4}{He}  & -28.296 & -7.07 &-28.32 &-7.08& 1.63 & 1.63\\
\hline
\element{12}{C}  & -92.163 & -7.68 &-91.36 &-7.61& 2.42 & 2.79\\
\hline
\element{16}{O}  & -127.62 & -7.98 &-127.7 &-7.98& 2.73 & 2.84\\
\hline
\element{19}{F}  & -147.80 & -7.78 &-143.3 &-7.54&      & 2.87\\
\hline
\element{24}{Mg} & -198.256 & -8.26 &-196.7 &-8.19& 2.95 & 3.17\\
\hline
\element{27}{Al} & -224.952 & -8.33 &-216.9 &-8.03& 2.95 & 3.13\\
\hline
\element{28}{Si} & -236.537 & -8.44 &-238.3 &-8.51& 3.04 & 3.24\\
\hline
\element{32}{S}  & -271.783 & -8.49 &-273.4 &-8.54& 3.20 & 3.35\\
\hline
\element{40}{Ca} & -342.056 & -8.55 &-349.0 &-8.72& 3.50 & 3.44\\
\hline
\end{tabular}\end{center}
\mycaption{Ground state energies and charge radii in FMD. 
The experimental binding energies are taken from
ref. \cite{YKS85,MNM95} and the charge radii from
ref. \cite{LaB67,Ang91}.  
All experimental errors are at most in the last digit.}{T-3.1-1} 
\end{table}

In \figref{P-3.1-1} and \figref{P-3.1-2} we display the density
of different ground states in coordinate and momentum space,
respectively.  The densities are defined as
\begin{eqnarray}
\rho_x(\xvec) = \bra{\xvec}\rhosim^{(1)}\ket{\xvec}
\ \ \ \mbox{and} \ \ \
\rho_k(\kvec) = \bra{\kvec}\rhosim^{(1)}\ket{\kvec}\ ,
\end{eqnarray}
where $\rhosim^{(1)}$ is the one--body density operator.  The
crosses indicate the centres $\rvec_l$ and $\pvec_l$ of the wave
packets.  One should however keep in mind that a Slater
determinant is invariant under transformations among the
occupied single--particle states. In the coordinate
representation the density is integrated over the
$z$--direction, the momentum representation is shown as a cut at
$k_z=0$.
\begin{figure}[hhhh]
\unitlength1mm
\begin{picture}(120,80)
\put(30,-2){\epsfig{file=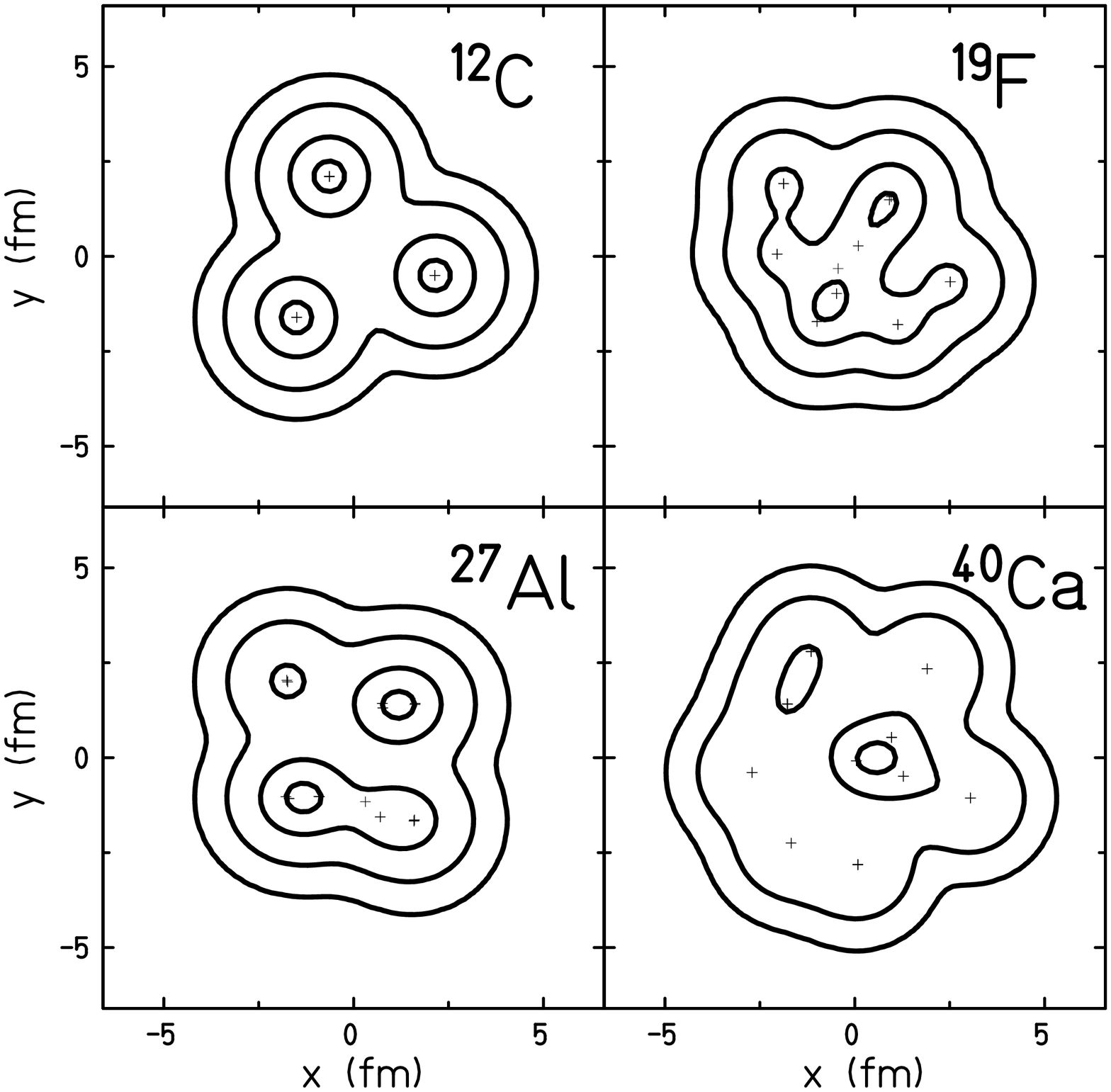,height=80mm}}
\end{picture}
\mycaption{
Contour plot of spatial densities integrated over $z$--direction
for different ground states. Crosses indicate centroids of wave
packets. Contour lines are at 0.9, 0.5, 0.1 and 0.01 of the
maximum density.
}{P-3.1-1}
\end{figure}
\begin{figure}[hhhh]
\unitlength1mm
\begin{picture}(120,80)
\put(30,-2){\epsfig{file=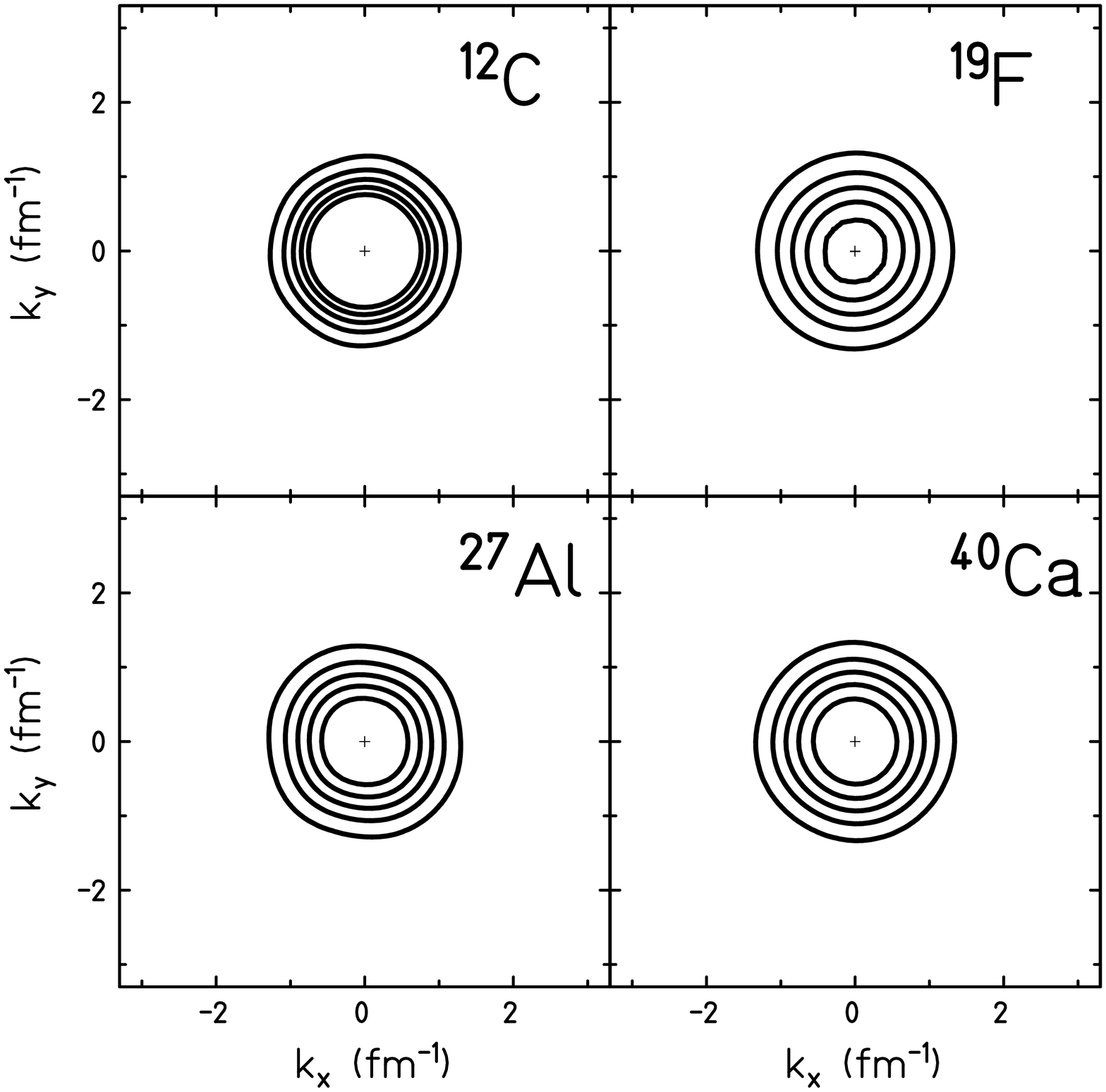,height=80mm}}
\end{picture}
\mycaption{
Contour plot of momentum distributions cut at $k_z=0$
for different ground states. Crosses indicate centroids of wave
packets. Contour lines are at 0.03, 0.06, 0.09, 0.12 and 0.15 $\fm^3$.
}{P-3.1-2}
\end{figure}

In \figref{P-3.1-1} one sees that the ground states are
intrinsically deformed.  The \element{12}{C} nucleus for
instance arranges as three $\alpha$--clusters, whereas this
$\alpha$--symmetry is broken in
\element{19}{F} and \element{27}{Al}.
The true ground state, which is an
eigenstate of the total spin, is a superposition of all
orientations of the intrinsically deformed ground state
\cite{FBS95}.

The momentum distributions shown in \figref{P-3.1-2} reflect the
Fermi motion of the ground states. Although all single--particle
wave--packets are stationary (and even centred at $\vec{p}=0$)
the system possesses Fermi motion.  In FMD the Fermi motion is a
quantum mechanical zero--point motion and not a random motion of
the packet centroids.  In the ground state $\rvec_k$ and
$\pvec_k$ and all other parameters are time--independent
otherwise it would be not the ground state of the system.
Using a product state for the many--body system \cite{\QMDRef} it
might be possible to use a momentum dependent Pauli potential in
order to mock up the momentum distribution. Nevertheless such a
system has the thermodynamic properties of distinguishable
particles, which are different from those of a Fermi system
especially at low excitations.

\subsection{Shell structure in FMD}
\label{shellstructure}

It is not immediately obvious that FMD includes shell--model
features like the nodal structure of single--particle orbits
since the states are localized in coordinate and momentum space.
But due to the invariance of a Slater determinant under linearly
independent transformations among the occupied single--particle
states, after antisymmetrization, any set of single--particle
states which is complete in the occupied phase space is as good
as any other. This applies also to non--orthogonal states.  To
illustrate this we take four one--dimensional real Gaussians
with the same real width parameter $a$ and zero mean momentum
and displace them by $d=0.75\sqrt{a}$ (see l.h.s. of
\figref{P-3.2-1}).
\begin{figure}[hhht]
\unitlength1mm
\begin{picture}(120,50)
\put(5,-10){\epsfig{file=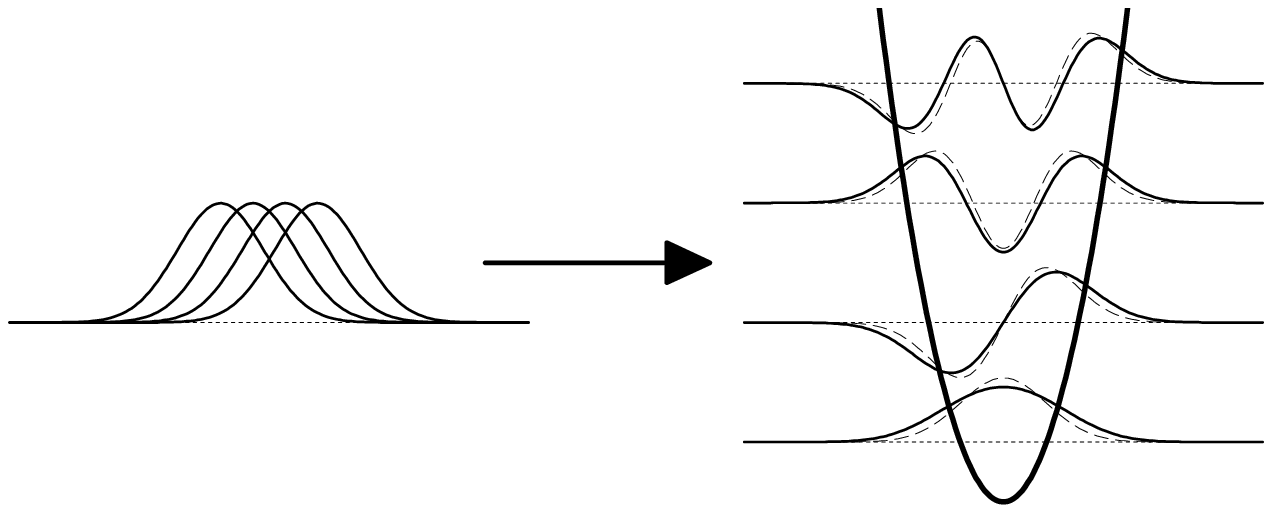,height=60mm}}
\end{picture}
\mycaption{
Antisymmetrization of the four Gaussians on the left hand side
leads to harmonic oscillator states. Dashed lines are the exact 
eigenstates of the oscillator.
}{P-3.2-1}
\end{figure} 
The one--body density can be written in terms of orthonormal
states $\ket{\psi_m}$ as
\begin{equation}
\label{Onebds}
\rhosim^{(1)}
 = \sum^A_{k,l=1} \ket{q_k}\ {\mathcal O}_{kl}\  \bra{q_l} 
= \sum^A_{m=1} \ket{\psi_m} \bra{\psi_m}  \ ,
\end{equation}
where the orthonormal eigenstates of $\rhosim^{(1)}$ are given by
\begin{equation}
\ket{\psi_m}
 = \sum^A_{k=1} \ket{q_k}\ ({\mathcal O}^{\frac{1}{2}})_{km} 
\end{equation}
and ${\mathcal O}_{kl}$ is the inverse of the overlap matrix
$\prodkl$.  They are displayed on the right hand side of
\figref{P-3.2-1} and compared to harmonic oscillator eigenstates
(dashed lines).  One sees that the occupied single--particle
states $\ket{\psi_m}$ consist of an s-, p- ,d- and an f-state,
all very close to harmonic oscillator states.  The difference
between both sets can be made arbitrarily small by letting
$d/\sqrt{a}$ approach zero.

A second example is illustrated in \figref{P-3.2-2}, where we
consider 100 equally spaced Gaussians in one dimension
\cite{FeS93}.  Again all mean momenta are zero and the width $a$
is real.  In the upper part of \figref{P-3.2-2} the width
$\sqrt{a}$ is 0.2 of the mean distance $d$ so that the wave
packets are well separated. Therefore the spatial density
$\rho_x$ and the momentum density $\rho_k$ are not changed by
antisymmetrization.  In the lower part the width has been
increased to $\sqrt{a}=d$. Without antisymmetrization (dash
dotted line) the spatial density is uniform and the momentum
distribution is that of a single packet. After
antisymmetrization (full lines) one obtains the typical shell
model oscillations in coordinate space and a Fermi distribution
in momentum space.  It is amazing to see how in
eq.~(\xref{Onebds}) the superposition of Gaussians by means of
the inverse overlap matrices can create a fully occupied
momentum state, see for example in \figref{P-3.2-2} the lower right
momentum distribution at $k=0.8k_F$, where the individual
Gaussians give practically zero probability to measure this
momentum.  We also calculated the eigenstates of the kinetic
energy in the occupied space and got perfect sinusoidal waves.
\begin{figure}[hhht]
\unitlength1mm
\begin{picture}(120,95)
\put(0,0){\epsfig{file=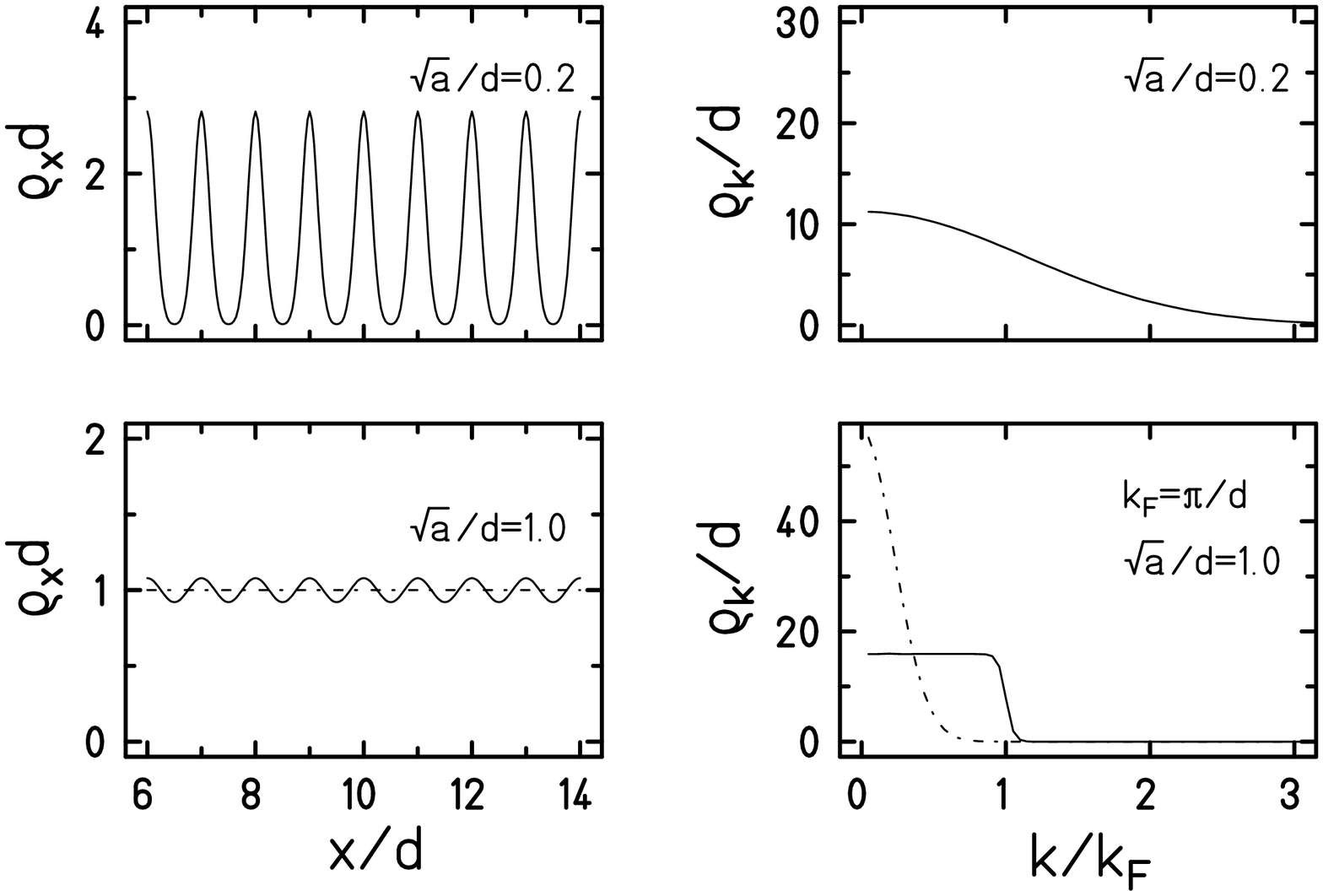,height=95mm}}
\end{picture}
\mycaption{
Upper part: section of spatial density of hundred Gaussians
(not overlapping in coordinate space) and 
corresponding momentum distribution (same for all). 
Antisymmetrization does not change the distribution.
Lower part: same as above but for overlapping Gaussians.
Full line with antisymmetrization, dash--dotted line without.
For details see text.}{P-3.2-2}
\end{figure}

These two examples illustrate nicely that even localized 
single--particle states with zero mean momentum 
build up FMD many--body trial states which describe
the harmonic oscillator shell model or even the Fermi motion
of a gas of fermions in which plane waves are occupied up to 
the Fermi momentum.

If one wants to simulate this effect by a "Pauli potential",
disregarding the momentum distribution in each wave packet,
the resulting ground state momentum distribution is 
unsatisfactory \cite{DDR87}. 

\section{Heavy ion reactions}
\label{Sec-4}

It is the aim of Fermionic Molecular Dynamics to describe the
phenomena seen in heavy--ion collisions at laboratory energies of
$E\leap200~A\MeV$, as there are fusion, incomplete fusion,
dissipative reactions, evaporation of nucleons and
fragmentation. In order to gain a predictive power that goes
beyond global phenomena dominated by conservation laws three
basic conditions are indispensable.

The first condition concerns the ground state properties. Many
observables like for instance fragment multiplicities depend
strongly on ground state energies and radii. Thermodynamic
properties like the specific heat are related to the exchange
symmetry of the many--body state. Equilibration is strongly
influenced by the mean free path of the constituents, which is
usually much larger for Fermi systems than for classical
systems.

The second condition is that the trial state has to have the
necessary degrees of freedom for the phenomena one wants to
account for. Mean positions and momenta of the single--particle
wave--packets are obvious degrees of freedom.  For reasons of
simplification the width is often chosen as real and time
independent, but it could be shown that this degree of freedom
is not only useful to reproduce the free motion exactly, but is
needed to describe evaporation of nucleons and fragmentation of
nuclei \cite{FBS95,Sch93}.

The third important condition, which is related to the second,
refers to symmetries. The deterministic equations of motion of
FMD preserve all symmetries in the initial state under which the
Hamilton operator is invariant. It is therefore desirable to use
a Hamilton operator that breaks as many symmetries as possible.
This Hamilton operator should not only contain a central
potential, as it is the case in the present calculations, but
also spin--orbit and tensor interactions, which are expected to
break spin symmetries of the ground states during the
dynamics. In addition short range correlations have to be
considered, which lead to more momentum transfer in the
reaction. These short range correlations are responsible for
hard collisions of nucleons, a role that is played by the
fluctuating collision term in QMD or AMD. First steps into this
direction are already taken \cite{Sch96}.

The model is chosen to be a molecular dynamics model because it
addresses the large fluctuations observed especially in
multifragmentation reactions. In the model the molecular aspect
is expressed through the localization of single--particle
wave--packets; this can be regarded as a quantization of the
particle number at any time. It means, fragments have always
integer particle number, in contrast to time--dependent
Hartree--Fock, see for instance \cite{KnS84,KnW88}.  Another
aspect is that the FMD single--particle wave--packets are not
allowed to split, therefore tunneling through a barrier cannot
be described. But this quantal process is of less importance in
fragmentation reactions because it takes much longer than the
involved time scales.

The equations of motion of FMD are deterministic, therefore the
event ensemble does not arise from random fluctuations of the
collision term during the time evolution, but from the average
over all orientations of the intrinsically deformed ground
states. Although these orientations superimpose coherently in
the true ground state their relative phases are randomized
during the collision so that they add up incoherently in the
exit channels. The same holds true for the summation over
different impact parameters
\cite{FBS95,FeS95}.

\subsection{Deeply inelastic reactions}

As a first example for heavy--ion collisions described with FMD a
dissipative reaction is presented. The reaction of
\element{19}{F} and \element{27}{Al} at a laboratory energy of
$5.9~A\MeV$ was investigated at the SMP Tandem accelerator in
Catania \cite{PAB96}.  At this low energy the reaction is
dominated by two processes. For impact parameters $b$ up to
about $5~\fm$ the system fuses completely, for larger $b$ the
reaction shows the typical dissipative phenomena of deeply
inelastic collisions.  The experiment focussed on the latter
process.
\begin{figure}[h]
\unitlength1mm
\begin{picture}(140,60)
\put(30,0){\epsfig{file=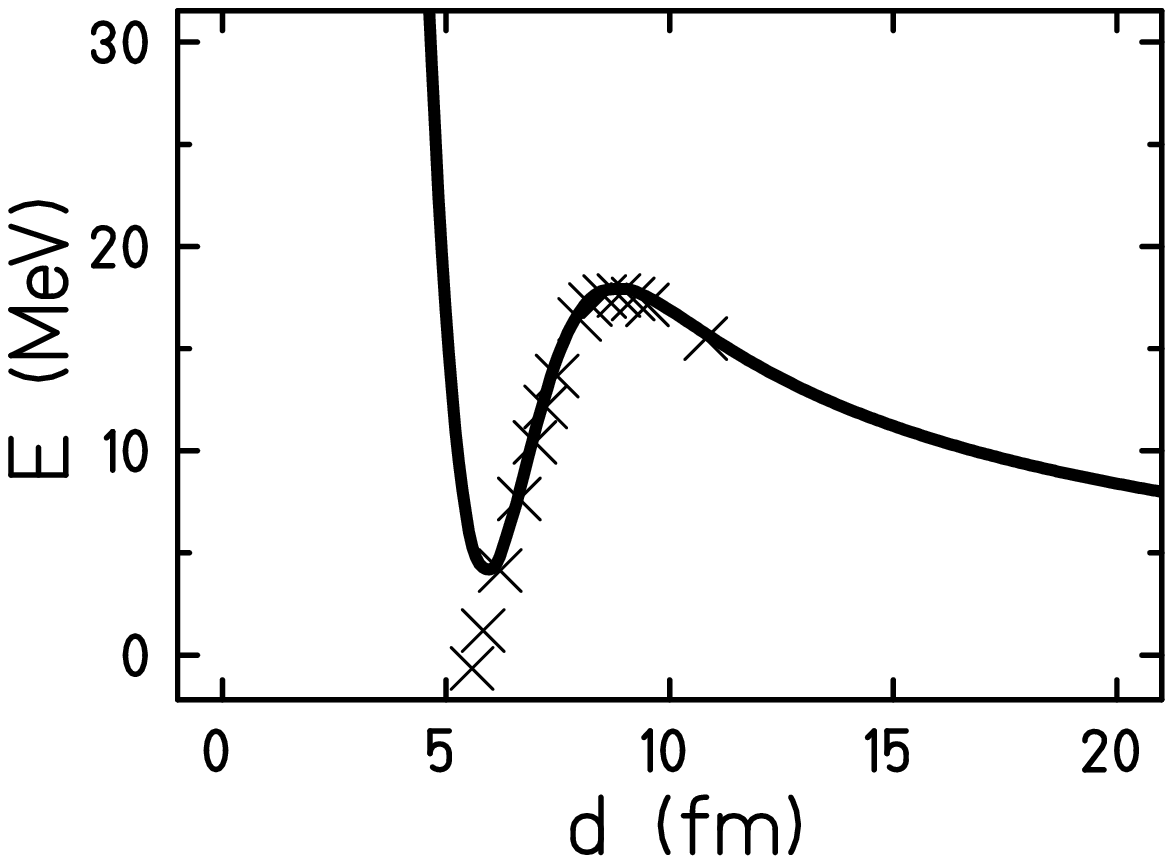,height=55mm}}
\end{picture}
\mycaption{
\element{19}{F}--\element{27}{Al} static nucleus--nucleus potential.
The solid line shows the FMD result using the nucleon--nucleon
interaction \fmref{V-Def} and frozen ground states, the crosses
indicate the parameterization by Krappe, Nix and Sierk.
}{P-4.1-0}
\end{figure} 
In order to understand the reaction a view on the
nucleus--nucleus potential might be helpful. In \figref{P-4.1-0}
the solid line displays the total energy of the
\element{19}{F}--\element{27}{Al} system (less the ground state
energies) as a function of the distance between the two centres
of mass as it arises from the two--body interaction
\fmref{V-Def} and \fmref{VCoulomb}.  
The many--body state of the two nuclei was taken
to be the respective ground states and not changed as a function
of the distance. Therefore, antisymmetrization induces a strong
repulsion below $d=5~\fm$. Above $d=5~\fm$ this static
interaction compares nicely to the parameterization of the
nucleus--nucleus interaction by Krappe, Nix and Sierk
\cite{KNS79}, which was adjusted to bulk properties of nuclei
and successfully applied in earlier studies of heavy--ion
dynamics, e.g. in ref. \cite{Fel87}.

\begin{figure}[hhh]
\unitlength1mm
\begin{picture}(140,65)
\put( 10,0){\epsfig{file=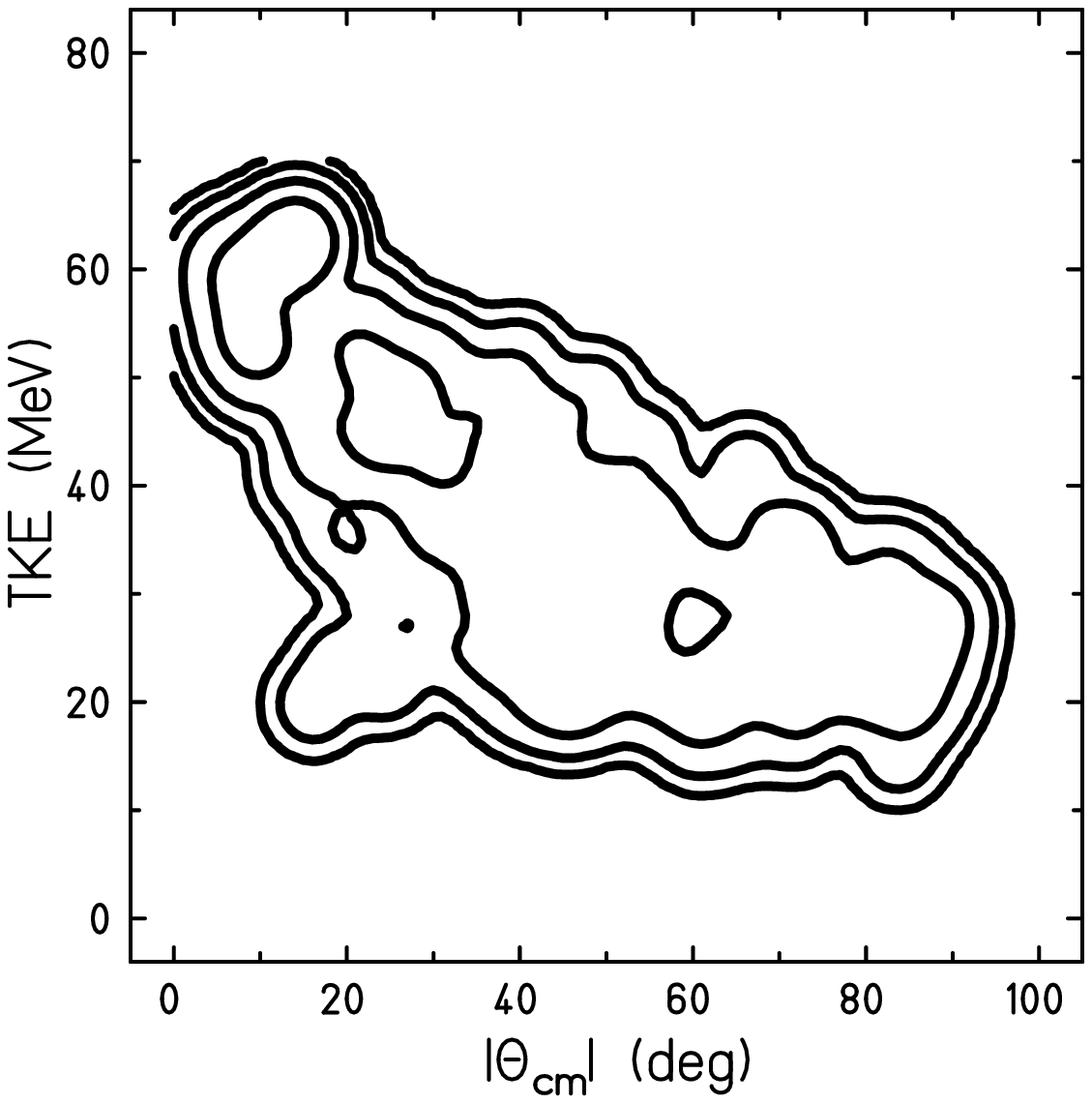,height=60mm}}
\put( 85,0){\epsfig{file=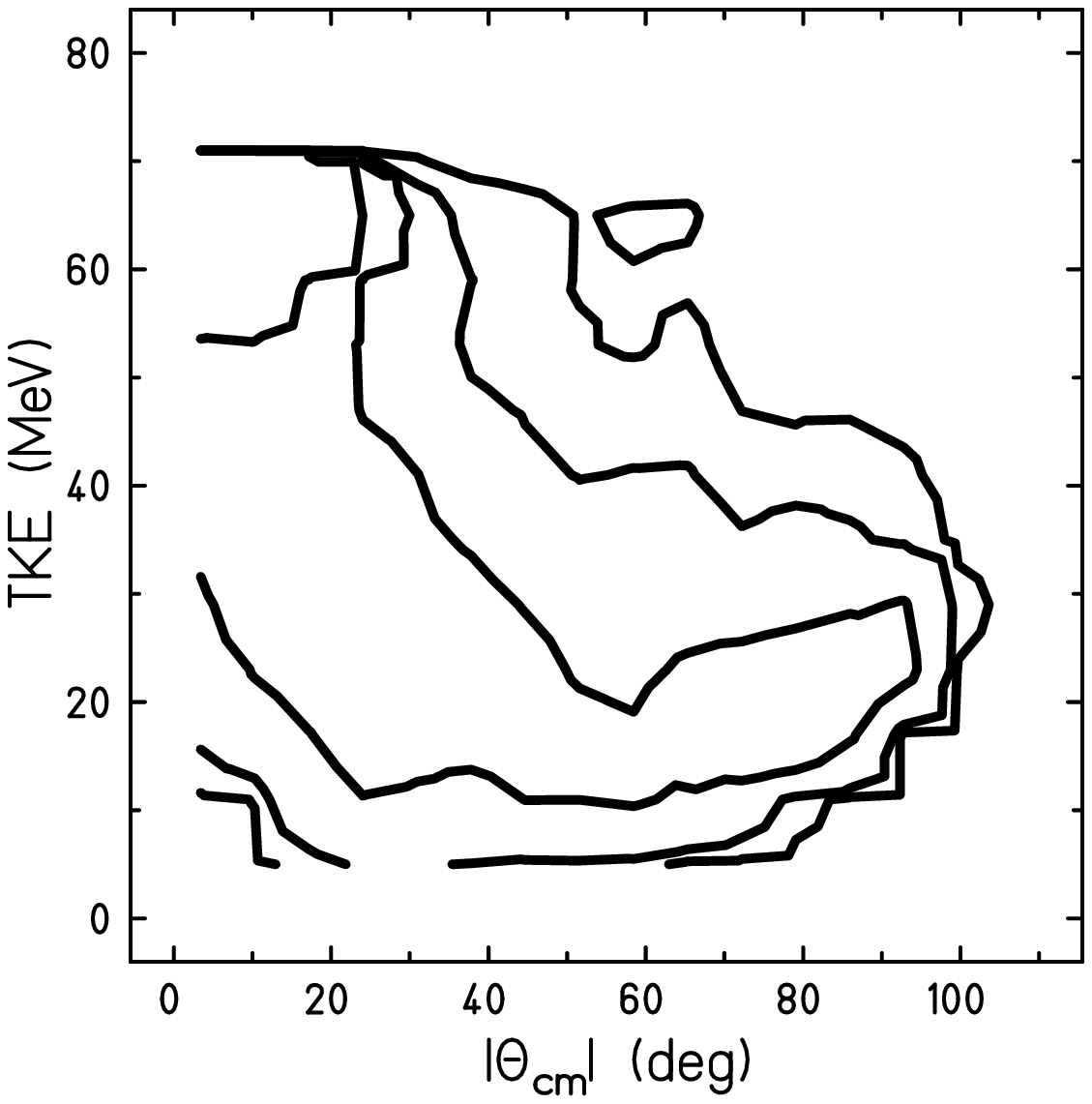,height=60mm}}
\end{picture}
\mycaption{
$d^2\sigma/(d\theta_\CM d\mbox{TKE})$ for
\element{19}{F}--\element{27}{Al} collisions at 5.9~$A$\MeV:
FMD calculations are
displayed on the l.h.s., experimental results on the r.h.s..
Subsequent contours differ by a factor of $5$.
}{P-4.1-1}
\end{figure} 
In the impact parameter range of $4.8~\fm\le b \le 10~\fm$ 509
events were generated with FMD. The results were filtered with
the experimental angular acceptance 
$3^o\le \theta_{\mbox{\scriptsize Lab}} \le 54^o$.
Below $b=4.8~\fm$ the system was fusing.

Figure \xref{P-4.1-1} shows the double differential cross
section $d^2\sigma/(d\theta_\CM d\mbox{TKE})$ as a function of
the centre of mass scattering angle $\theta_\CM$ and the total
kinetic energy TKE of the two scattered nuclei.  On the left
hand side the FMD calculations are presented as a contour
plot. Each event $(\theta_\CM,\mbox{TKE})$ contributes as a
Gaussian ($\Delta|\theta_\CM|=2^o$ and
$\Delta\mbox{TKE}=2~\MeV$) in order to smoothen the
distribution. For impact parameters smaller than the one leading
to a grazing collision ($|\theta_\CM|\approx 15^o$,
$\mbox{TKE}\approx65~\MeV$) the approaching nuclei feel the
nuclear attraction, $\theta_\CM$ gets smaller and becomes
negative. The total kinetic energy $\mbox{TKE}$ is decreasing
due to internal excitation of the nuclei.  The major part of the
cross section is observed at negative angles. For even smaller
impact parameters the nuclei stick together so long that they
appear at positive angles again. One observes a flat
distribution in $\theta_\CM$ at the Viola energy of
$\mbox{TKE}\approx 20~\MeV$.  On the right hand side the
experimental result is given.  Comparing both sides of
\figref{P-4.1-1} one sees that FMD is capable to describe the
dissipative character of the reaction dynamics.

\begin{figure}[h]
\unitlength1mm
\begin{picture}(140,65)
\put( 10,0){\epsfig{file=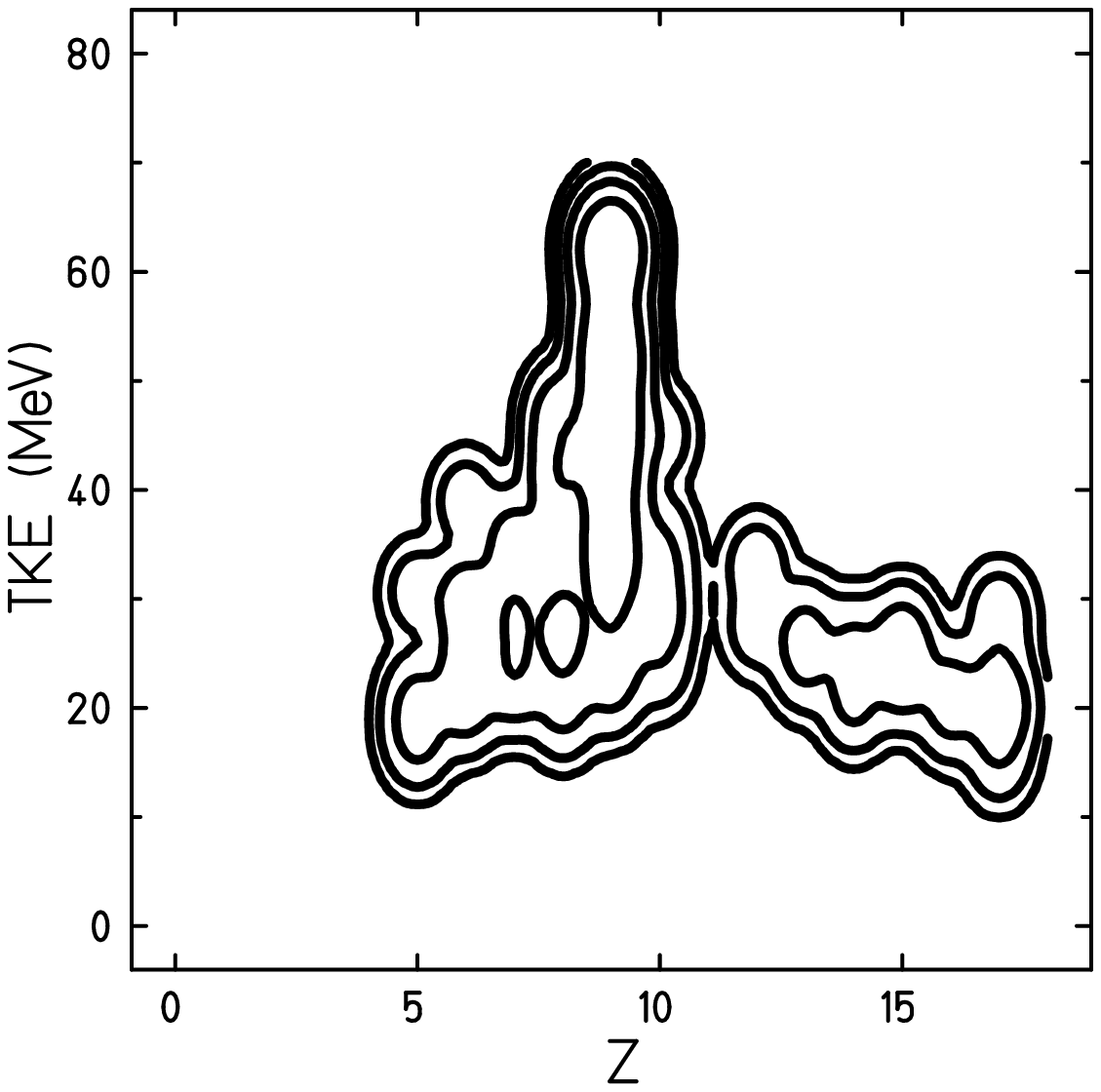,height=60mm}}
\put( 85,0){\epsfig{file=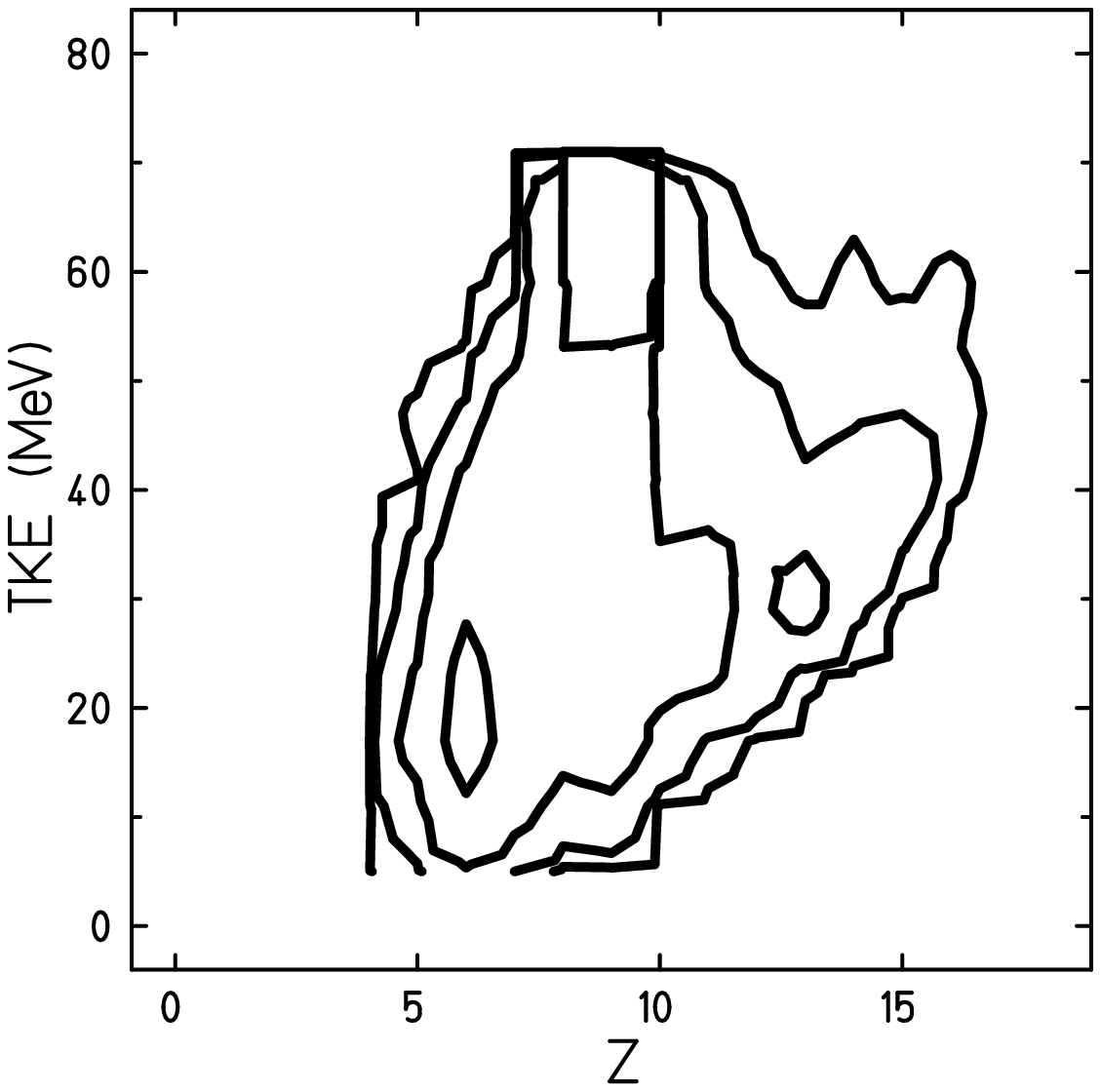,height=60mm}}
\end{picture}
\mycaption{
Diffusion plot $d^2\sigma/(dZ\, d\mbox{TKE})$ for
\element{19}{F}--\element{27}{Al} collisions at $5.9~A\MeV$:
FMD calculations are
displayed on the l.h.s., experimental results on the r.h.s..
Subsequent contours differ by a factor of $5$.
}{P-4.1-2}
\end{figure} 
The diffusion plot $d^2\sigma/(dZ\, d\mbox{TKE})$,
\figref{P-4.1-2}, samples fragments in the mentioned angular
range of $3^o\le \theta_{\mbox{\scriptsize Lab}} \le 54^o$,
which are mostly projectile like. For the
experiment, r.h.s. of \figref{P-4.1-2}, the data are limited by
an energy threshold seen at low $\mbox{TKE}$ and high $Z$,
which is not imposed on the analysis of the FMD simulations.
The FMD events $(Z,\mbox{TKE})$ are smoothened with
$\Delta~Z=0.5$ and $\Delta\mbox{TKE}=2~\MeV$.

Both contour plots, FMD (l.h.s.) and experiment (r.h.s.), show
the typical broadening of the charge distribution with
increasing dissipated energy and the lack of charge drift for
TKE above the Viola energy of $\mbox{TKE}\approx 20~\MeV$ for
the completely relaxed events. A drift to smaller $Z$ values,
which means that the \element{19}{F} nucleus is giving away
nucleons to the \element{27}{Al}, is seen in the measured and
the calculated cross sections for large energy losses where all
the initial kinetic energy is dissipated and the reaction lasts
very long.

\begin{figure}[h]
\unitlength1mm
\begin{picture}(140,65)
\put(45,0){\epsfig{file=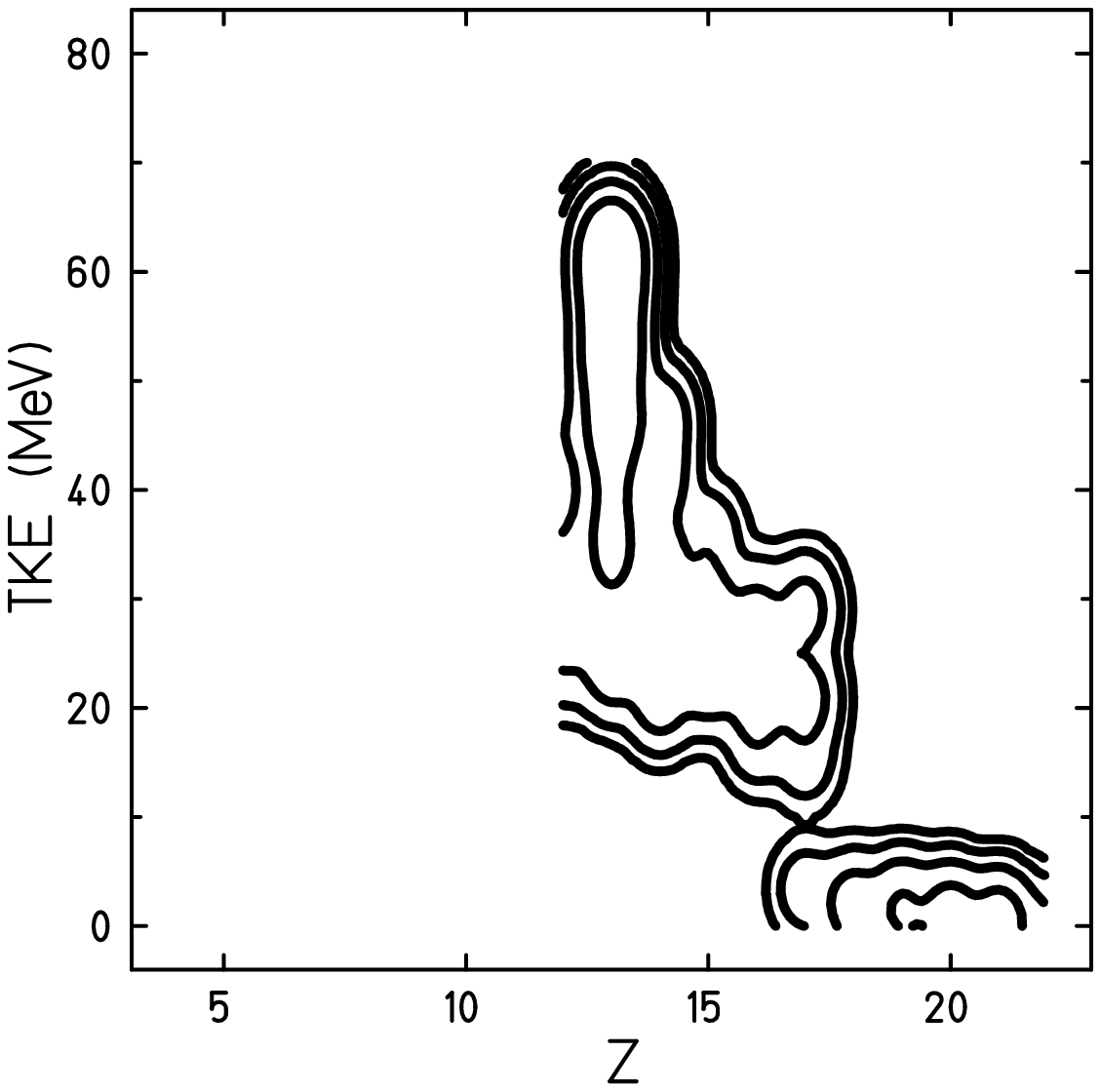,height=60mm}}
\end{picture}
\mycaption{
Diffusion plot $d^2\sigma/(dZ\, d\mbox{TKE})$ for
\element{19}{F}--\element{27}{Al} collisions at $5.9~A\MeV$
calculated with FMD for target like fragments ($Z\ge12$).
Subsequent contours differ by a factor of $5$.
}{P-4.1-3}
\end{figure} 
Figure \xref{P-4.1-3} presents the second part of the diffusion
plot for $Z\ge~12$ which has not been measured. Besides the
deeply inelastic reactions one now also sees at TKE values close
to zero the evaporation residues from fusion reactions with
charges ranging from $Z=16$ to the total charge of $Z=22$ with a
maximum around $Z=20$.  Here one should note that the FMD time
evolution was followed only up to $1200~\fm/c$ so that the loss
of charges due to evaporation was not complete yet. Nevertheless
the broad distribution of evaporation residues shows that FMD is
not unrealistic in this respect.

\subsection{Multifragmentation}

In this section the same system of \element{19}{F} and
\element{27}{Al} is investigated, but at an energy of
$32A\MeV$ where multifragmentation is expected.

How multifragmentation happens in heavy ion collisions is still
a matter of debate. Explanations reach from nucleation over self
organization, spinodal decomposition to cold break-up and
survival of initial correlations. For an overview see
ref. \cite{Hir94}. A key question is the time scale of
the reaction.  Slow processes like nucleation or self
organization are hindered if the expansion of the whole system
is too fast. Another issue is the relaxation time for thermal
and chemical equilibrium which is important when statistical
models are used to explain multifragmentation
\cite{BBI95,Gro90,Fri88}. 

If one considers the decay of excited spectator matter which has
not been compressed one expects a slow expansion so that there
might be enough time for global equilibration. On the other hand
the excitation energies are not so high, so that the mean free
path, due to the Pauli principle, is still not small compared to
the diameter of the nuclear system. Therefore it is not obvious
that global thermal equilibrium is achieved.

For the participant matter the compression is much stronger and
the excitation energy much higher. This provides a short mean
free path, but the whole system is expanding and cooling fast so
that the time available for equilibration is rather short and it
is questionable if there are enough collisions to ensure local
equilibrium until freeze out.

It is very difficult to measure temperature and flow profiles 
\cite{Poc95,NHW95,Rei96} and even harder or impossible to infer
experimentally on the time scale of the evolution of the
system. Therefore, microscopic transport models which do not
assume equilibration are needed for a better understanding. These
models should go beyond the mean field approach, which is a kind
of equilibrium assumption in itself, so that in principle they
are capable to describe many--body correlations like the
formation of fragments.  QMD, AMD and FMD are molecular dynamics
models which assert this claim. How equilibrium is achieved can
then be studied by comparing distributions, for example of mass,
charge, kinetic energy etc, with equilibrium distributions.

FMD calculations show that correlations play an important role.
Since the time evolution of FMD is deterministic, correlations
or symmetries can be broken only either in the ground state,
e.g. in \element{27}{Al} there is no $\alpha$--symmetry, but in
\element{28}{Si} there is, or during the time evolution if the
Hamilton operator breaks these symmetries dynamically. In models
like AMD or QMD a randomly fluctuating collision term, which
models the short--range repulsion as a Langevin force,
destroys existing correlations and symmetries. We do not want to
follow this line, because it is unknown to which extent
correlations of the initial state survive during the collision
and it even might be an important mechanism for cluster
formation in nature.

In the previous section we investigated beam energies of about
$6~A\MeV$ which led to dissipative reactions with two out-coming
nuclei with about the same mass number as in the entrance
channel. The energy was not high enough to break neither the
fused nor the scattered nuclei into pieces.

Now we chose an energy of $32~A\MeV$ that corresponds to a
relative velocity between the nuclei of about the Fermi
velocity. One expects that at this large collective velocity a
common mean field cannot be established any more.
\begin{figure}[tttt]
\unitlength1mm
\begin{picture}(140,150)
\put(0,0){\epsfig{file=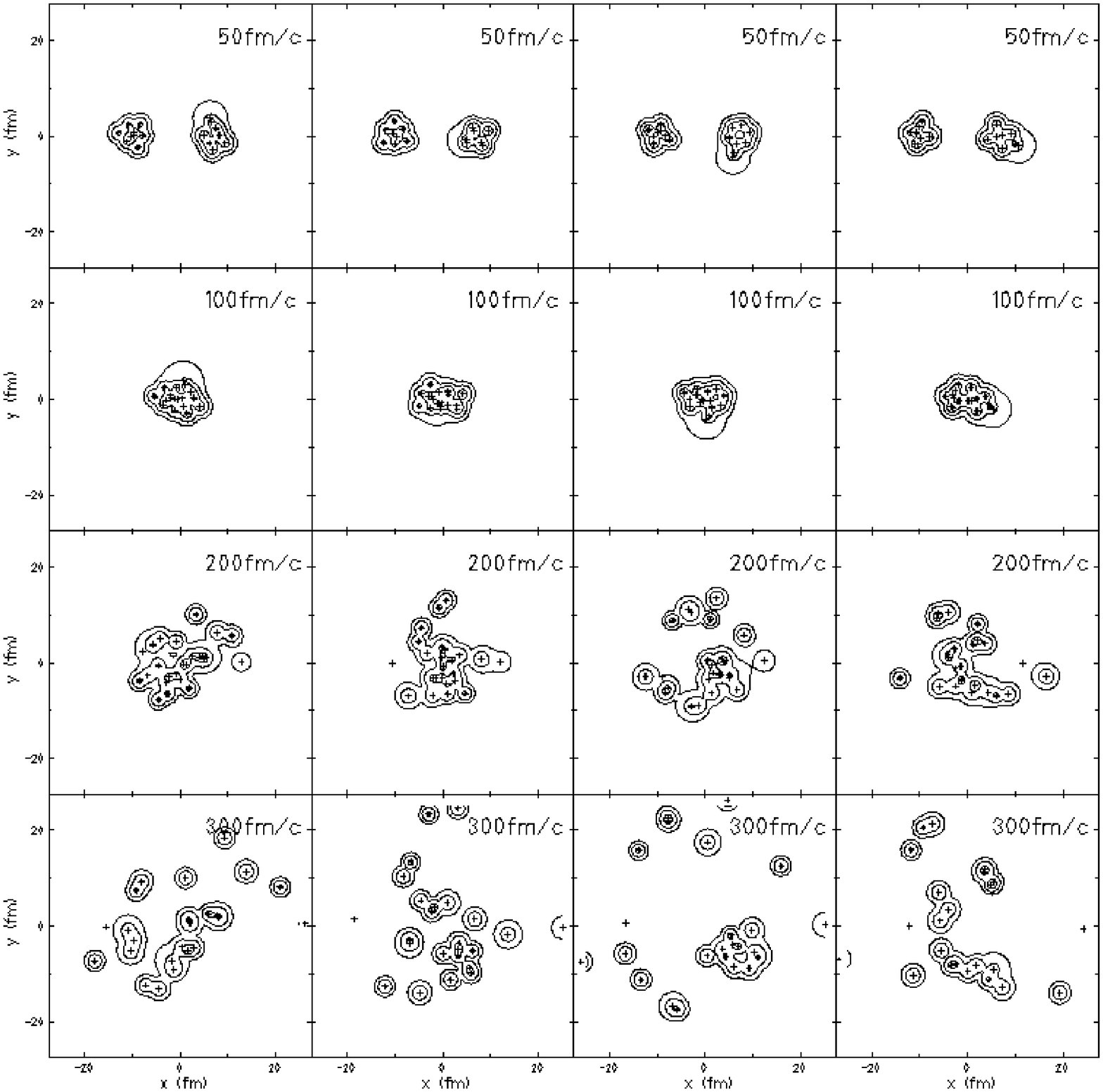,width=150mm}}
\end{picture}
\mycaption{
\element{19}{F}--\element{27}{Al} collisions at $32~A\MeV$,
$b=0.5~\fm$: the one--body density in coordinate space integrated
over $z$ is shown. The contour lines depict the density at
$0.01, 0.1, 0.5 \fm^{-2}$.
}{P-4.2-1}
\end{figure} 

The following figures show a variety of events as contour plots
of the one--body density in coordinate space. This density is
integrated over the $z$--direction. Figure \xref{P-4.2-1},
\xref{P-4.2-2} and \xref{P-4.2-3} present runs at different
impact parameters. Within a figure different columns show runs
with initial states that differ in the orientation of the
intrinsically deformed ground states.
\begin{figure}[tttt]
\unitlength1mm
\begin{picture}(140,150)
\put(0,0){\epsfig{file=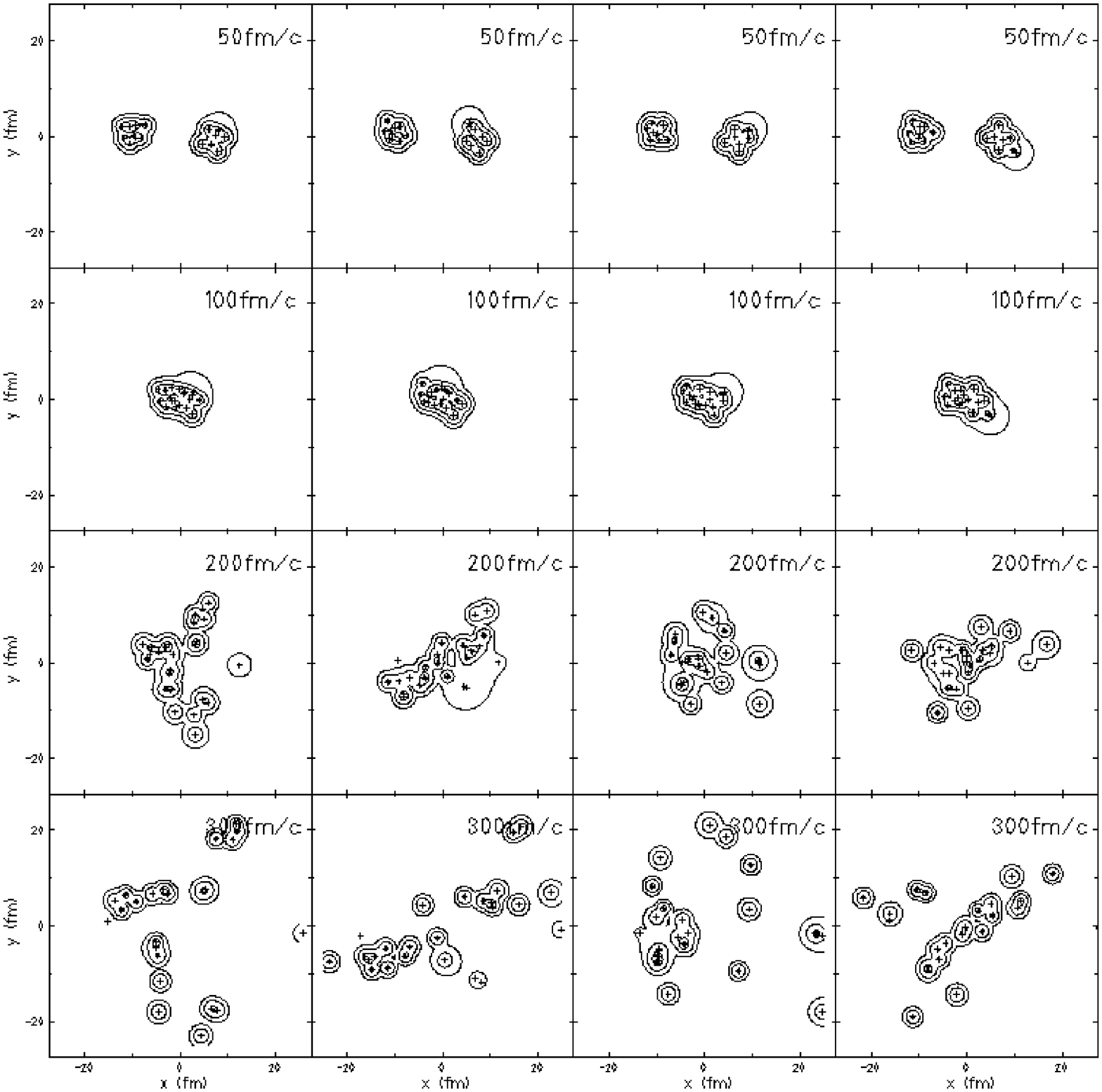,width=150mm}}
\end{picture}
\mycaption{
\element{19}{F}--\element{27}{Al} collisions at $32~A\MeV$,
$b=1.5~\fm$: for explanation see \figref{P-4.2-1}.
}{P-4.2-2}
\end{figure} 
Figure \xref{P-4.2-1} presents four time evolutions at an impact
parameter of $b=0.5~\fm$. The time is given in the upper right
corners. One sees that at this impact parameter and energy a
rather long living source is created which fragments into pieces
of all sizes. In \figref{P-4.2-2} for a higher impact parameter
of $b=1.5~\fm$ the situation is similar but the source is
stretched.
\begin{figure}[tttt]
\unitlength1mm
\begin{picture}(140,150)
\put(0,0){\epsfig{file=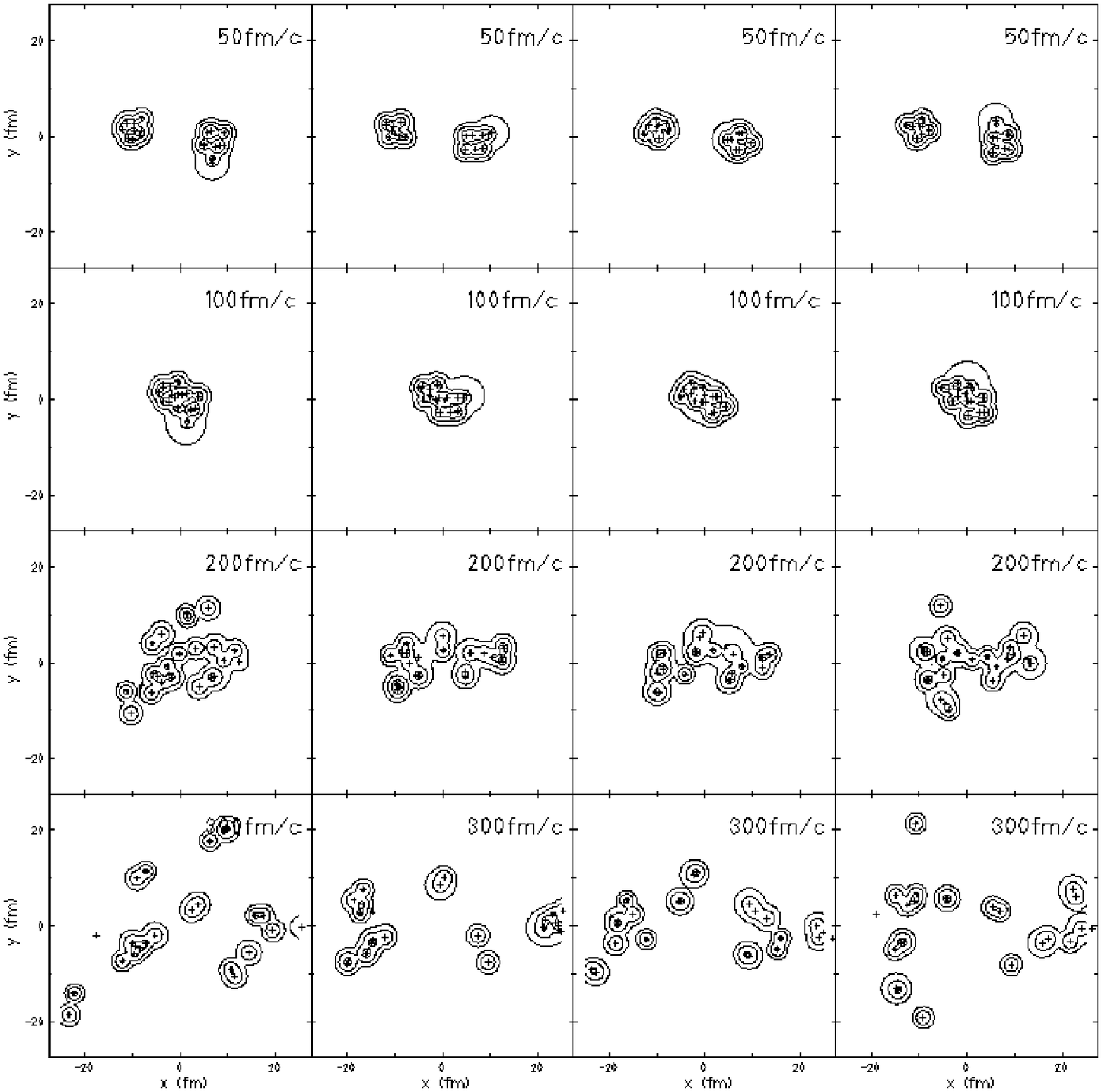,width=150mm}}
\end{picture}
\mycaption{
\element{19}{F}--\element{27}{Al} collisions at $32~A\MeV$,
$b=2.5~\fm$: for explanation see \figref{P-4.2-1}.
}{P-4.2-3}
\end{figure} 
Figure \xref{P-4.2-3} finally shows runs with an impact parameter of
$b=2.5~\fm$ where the angular distribution of the fragments is
more forward--backward peaked than for the central events where
it is isotropic.

Analyzing the time evolution of a cluster one sees that the
correlations between the wave packets which finally compose the
fragment can be followed back for rather long time.

In the ground state the wave packets arrange in phase space such
that the total energy is minimized. This FMD ground state is an
intrinsic state in which the relative positions and orientations
of the wave packets reflect many--body correlations.  If one
destroys these correlations by randomly moving all position
parameters $\vec{r}_k$ by only $0.2~\fm$, which leaves the
one--body density almost unchanged, a \element{56}{Fe} nucleus
for example achieves already $90~\MeV$ excitation energy.  A
typical $Q$--value to break up lighter fragments (up to about
\element{16}{O}) into smaller clusters is $10~\MeV$. Therefore,
for not too high excitation energies where the mean free path
between hard nucleon--nucleon collisions is not yet small
compared to the diameter of the fragment, one expects the
many--body correlations which characterize a bound cluster to
survive to a large extend.  Since the nucleons are
indistinguishable one can of course not say which nucleon from
the projectile or target ends up in a given final fragment. Even
the wave packets sometimes exchange rapidly $\vec{r}$ and
$\vec{p}$, reminding of a Landau--Zener crossing.

It is also interesting to note that in most of the runs
several $\alpha$--particles are created (small spheres in the
plots), which is in accord with experimental findings at a
similar energy of $E=35~A\MeV$ in
\element{40}{Ca}--\element{40}{Ca} collisions \cite{HGW94}.

At this point a few remarks concerning the time--dependent width
might be in place, since this degree of freedom is used
exclusively in FMD, but not in AMD or QMD and led to a
discussion about its role \cite{KiD96,CCG96}.

It was shown not only in our own investigations
\cite{Sch93,FeS93,FBS95}, but also in AMD calculations \cite{\AMDRef},
that a time evolution with fixed width parameters does not
produce evaporation or fragmentation.  The reason is that
a fixed width implies a zero--point energy in each packet of
about $\varepsilon_{kin}=3/(4m_Na_R)\approx10~\MeV$.  Inside the
nucleus this energy is part of the kinetic energy in the Fermi
motion. Outside the nucleus the zero--point energy of an emitted
particle is determined by the amount of its localization at the
end of the emission process.  If the width is kept fixed the
nucleon carries besides its mean energy always these additional
$10~\MeV$, which is much more than the experimentally observed
$2\cdots3~\MeV$ and thus a wave packet with fixed width has
little probability to escape.

Compared to that a wave packet with a dynamical width escapes
from an excited nucleus by spreading in coordinate space
\cite{FeS93}, which has two effects: first, the large spatial
extend reduces the overlap with the other nucleons and the
negative potential energy tends to zero, second, the positive
zero--point energy also becomes small so that the sum of both
need not change too much during the emission.  Classically
spoken, a particle leaving the nucleus has to climb up the
potential wall at the surface loosing almost all of its kinetic
energy before escaping.  FMD gives values of about $2~\MeV$ for
the kinetic energy of evaporated particles in accord with
experimental findings.

For fragmentation the dynamical variation of the width
parameters is also needed. Here a change in the width parameters
allows the density to notch so that it may break.

In AMD this lack of evaporation and fragmentation is removed by a
subtraction of spurious zero--point energy \cite{OHM93}
which also takes care of the localization energy in the
center of mass motion of the fragments.

\section{Statistical properties of FMD}

Fragmentation reactions show large fluctuations, for example in
the mass distribution, which are beyond an ensemble averaged
mean--field treatment.  In this context it is important to
understand the statistical properties of molecular dynamic
models especially at low temperatures \cite{ScF96}.

There are two aspects. One concerns the thermo{\it static}
properties of a molecular dynamic model where the attribute
thermostatic refers to the properties of the static canonical
statistical operator, which are contained in the partition
function $Z(T)=\Tr(\exp\{-\Operator{H}/T\})$.  Once the
partition function is calculated within a given model, its
thermostatic properties can be deduced by standard methods like
partial derivatives of $\ln Z(T)$ with respect to temperature
$T$ or other parameters contained in the Hamilton operator
$\Operator{H}$.

In the case of Fermionic Molecular Dynamics the trace in the
partition function can be evaluated exactly because the model is
based on antisymmetric many--body states which form an
over-complete set covering the whole Hilbert space.  Also the
states of Antisymmetrized Molecular Dynamics (AMD)
\cite{ScF96,OHM92}
provide a representation for the unit operator.  As the
calculation of the trace does not depend on the representation
all thermostatic properties like Fermi--Dirac distribution,
specific heat, mean energy as a function of temperature
etc. ought to be correct and fully quantal using FMD or AMD
trial states.

The issue of this section is more the other and even more
important aspect, namely the dynamical behaviour of a molecular
dynamics model.  A dissipative system which is initially far
from equilibrium is expected to equilibrate towards the
canonical ensemble.  The simulation of such a system within the
model provides a crucial test of its thermo{\it dynamic}
behaviour.

The time--evolved FMD state is in general not the exact solution
of the Schr\"odinger equation,
so the correct thermo{\it static} properties do not a priori
guarantee correct thermo{\it dynamic} properties.
In other words the question is:
does the FMD state as a function of time explore the 
Hilbert space according to the canonical weight?

Since the parameters of the single--particle wave packets follow
generalized Hamilton equations of motion, one is tempted to
infer that the dynamical statistical properties might be
classical \cite{OhR93,OhR95}.  This conjecture, that classical
equations of motion always imply classical statistics, is
disproved by the following examples, in which we compare
time--averaged expectation values of wave--packet molecular
dynamics with the equivalent ones of the canonical ensemble at
the same excitation energy.

Within Fermionic Molecular Dynamics we study the equilibration
of four identical fermions enclosed in a one dimensional
harmonic oscillator.  The particles interact by a weak repulsive
two--body potential which is necessary to convert the
integrable harmonic oscillations into chaotic motion.  The
important result is that the initial many--body state, which is
far from equilibrium, approaches the canonical ensemble with
Fermi--Dirac statistics in an ergodic sense.  The time--averaged
occupation numbers of the harmonic oscillator eigenstates are
practically identical with the Fermi--Dirac distribution of the
canonical ensemble, provided the canonical ensemble is taken at
the time--independent mean excitation energy of the many--body
state.

When distinguishable particles, which are described by a product
state of wave packets, are considered, the molecular dynamic
equations for the parameters of the wave packets lead to a
Boltzmann distribution for the occupation numbers of the
single--particle eigenstates.

A further important result is, that the use of time averages
provides us with a tool for establishing relations between
well--defined quantities of a molecular dynamic model such as
excitation energy and statistical quantities like temperature.
This will be done in subsection~\xref{SubSec-5.3} where we
investigate excited nuclei and the nuclear liquid--gas phase
transition which is of great experimental
\cite{Poc95,NHW95,MoP96} and theoretical interest
\cite{Gro90,JMZ84,GKM84,BLV84,BLV85,LeB85,PaN95,ScF97a}.

\subsection{Thermostatic properties}

In order to describe the thermostatic properties of a given
system by means of model states exactly, it is necessary that
these model states span the whole Hilbert space. The question,
whether the thermostatic properties of FMD are correct, therefore
reduces to the question whether its model states are complete.
Starting from coherent states the following short explanation
proofs that this is the case.

\subsubsection{Completeness relation with coherent states}

Coherent states $\ket{\vec{z}}$ which are defined as the
eigenstates of the harmonic oscillator destruction operator
$\Operator{\vec{a}}$,
\begin{eqnarray}
\Operator{\vec{a}}\; \ket{\vec{z}}
&=&
\vec{z} \; \ket{\vec{z}}
\ ,\qquad
\OpHHO
=
\omega\; \left( \Operator{\vec{a}}^+ \, \Operator{\vec{a}} 
+ \frac{3}{2} \right)
\end{eqnarray}
form an over-complete set of states. Their completeness relation
reads 
\begin{eqnarray}
\label{KohVoll}
\EinsOp^{(1)}
&=& \int \frac{\dint^3\Re{z}\dint^3\Im{z}}{\pi^3}\;
\frac{\ket{\vec{z}}\bra{\vec{z}}}{\braket{\vec{z}}{\vec{z}}}
\\
&=& \int \frac{\mbox{d}^3r\;\mbox{d}^3p}{(2\pi)^3}\;
\frac{\ket{\vec{r},\;\vec{p}}\bra{\vec{r},\;\vec{p}}}
{\braket{\vec{r},\;\vec{p}}{\vec{r},\;\vec{p}}}
\nonumber\\
&=& \int \frac{\mbox{d}^3r\;\mbox{d}^3p}{(2\pi)^3}\;
\frac{\ket{q}\bra{q}}
{\braket{q}{q}}\ ,
\nonumber
\end{eqnarray}
where all three lines are equivalent notations; $\ket{\vec{z}}$
labeling coherent states by their eigenvalue with respect to
$\Operator{\vec{a}}$, the phase space notation
$\ket{\vec{r},\;\vec{p}}$ labeling the states by their
expectations values of the coordinate and momentum operators and
$\ket{q}$ being the FMD notation.  Coherent states are
extensively discussed in ref.~\cite{Kla85}.

Since we are dealing with fermions the spin degree of freedom
has to be considered and consequently the resolution of unity
changes to
\begin{eqnarray}
\EinsOp^{(1)}
&=& \int \frac{\mbox{d}^3r\;\mbox{d}^3p}{(2\pi)^3}\;\sum_{m_s}\;
\frac{\ket{q}\bra{q}}
{\braket{q}{q}}\ ,
\end{eqnarray}
where the sum runs over the different magnetic quantum numbers
$m_s$ which are included in the set of parameters denoted by $q$
(see eq.~\fmref{spinor}).

Proceeding one step further the unity operator in the
antisymmetric part of the two--particle Hilbert space is the
antisymmetric product of two single--particle unity operators 
\begin{eqnarray}
\EinsOp^{(2)}
&=& 
\Operator{A}^{(2)}\; 
\left(\EinsOp^{(1)}\otimes\EinsOp^{(1)}\right)\;
\Operator{A}^{(2)}
\\
&=& 
\half\Big(1 - \Operator{P}_{12}\Big)\; 
\left(\EinsOp^{(1)}\otimes\EinsOp^{(1)}\right)\;
\half\Big(1 - \Operator{P}_{12}\Big)\ ,
\nonumber
\end{eqnarray}
which may be expressed with antisymmetric two--body states
$\ket{q_1,\; q_2}_a$ as
\begin{eqnarray}
\EinsOp^{(2)}
&=& 
\int \frac{\mbox{d}^3r_1\;\mbox{d}^3p_1}{(2\pi)^3}\;\sum_{m_s(1)}\;
\int \frac{\mbox{d}^3r_2\;\mbox{d}^3p_2}{(2\pi)^3}\;\sum_{m_s(2)}\;
\frac{\ket{q_1,\; q_2}_a \; {}_a\bra{q_1,\; q_2}}
     {\braket{q_1}{q_1}\braket{q_2}{q_2}}
\\[5mm]
\mbox{where}
&&
\ket{q_1,\; q_2}_a:= 
\half\left(
\ket{q_1}\otimes\ket{q_2} -
\ket{q_2}\otimes\ket{q_1}\right)
\ .
\nonumber
\end{eqnarray}
Following this line the resolution of unity in the antisymmetric
part of the $A$--body Hilbert space is just the antisymmetric
product of single--particle unity operators.
Be $\ket{\hat{Q}}$ the unnormalized Slater determinant of
single--particle states $\ket{q}$ and $\ket{Q}$ the normalized
Slater determinant
\begin{eqnarray}
\ket{\hat{Q}}
&=& 
\frac{1}{A!}
\sum_{\pi}
\sign(\pi)
\left(
\ket{q_{\pi(1)}}\otimes\cdots\otimes\ket{q_{\pi(A)}}
\right)
\\
\ket{Q}
&=& 
\frac{1}{\braket{\hat{Q}}{\hat{Q}}^\half}\; \ket{\hat{Q}}
\ .
\nonumber
\end{eqnarray}
Then the unity operator can be written as the projection of the
$A$--body unit operator onto the antisymmetric subspace of the
Hilbert space
\begin{eqnarray}
\label{ASEinsOp}
\EinsOp^{(A)}
&=& 
\int \mbox{d{\Large $\mu$}}(Q)\;
\ket{Q} \bra{Q}\ ,
\end{eqnarray}
with a measure 
\begin{eqnarray}
\mbox{d{\Large $\mu$}}(Q)
&=& 
\braket{\hat{Q}}{\hat{Q}}
\prod_{k=1}^A
\frac{1}{\braket{q_k}{q_k}}
\frac{\mbox{d}^3r_k\;\mbox{d}^3p_k}{(2\pi)^3}\;\sum_{m_s(k)}\;\ ,
\end{eqnarray}
that accounts for antisymmetrization by means of the ratio
between the norm of the Slater determinant
$\braket{\hat{Q}}{\hat{Q}}$ and the norm of the corresponding
product state $\prod_{k=1}^A\,\braket{q_k}{q_k}$ (see also
\cite{OhR93}).  In a sampling where the values of $\vec{r}_k$
and $\vec{p}_k$ are chosen according to Monte Carlo methods this
measure determines the probability to find the state $\ket{Q}$
belonging to this configuration
$Q=\{\vec{r}_1,\vec{p}_1;\vec{r}_2,\vec{p}_2;\dots\}$ in Hilbert
space. If for example two fermions with the same spin
are close in $\vec{r}$ and $\vec{p}$ then this measure is small
because the norm
$\braket{\hat{Q}}{\hat{Q}}=\mbox{det}\{\prodkl\}$ will be small.

Coherent states are Gaussian wave--packets with fixed width,
but the single--particle states of FMD 
\begin{eqnarray}
\braket{\vec{x}}{q}
&\propto&
\exp\left\{ 
   -\frac{(\vec{x}-\vec{r})^2}{2\, a} + i\, \vec{p}\cdot\vec{x}
    \right\}
\otimes\ket{\chi,\phi}\otimes\ket{\xi}
\end{eqnarray}
contain more degrees of freedom, for instance the width
parameter $a$, the spin angles $\chi$ and $\phi$ and the
isospin--3 component $\xi$.  Since coherent states are already
complete the additional degrees of freedom $a_R$ and $a_I$ do
not bother. They can be integrated keeping track of the
normalization as it is shown in
\fmref{ASEinsOpBreite} and the isospin is summed over like the spin
\begin{eqnarray}
\label{ASEinsOpBreite}
\EinsOp^{(1)}
&=& 
\frac{1}{\Omega_R\; \Omega_I}\;
\int \frac{\mbox{d}^3r\;\mbox{d}^3p}{(2\pi)^3}\;\sum_{m_s,\xi}\;
\int_{\Omega_R} \mbox{d}a_R\;
\int_{\Omega_I} \mbox{d}a_I\;
\frac{\ket{q}\bra{q}}
{\braket{q}{q}}\ ,
\end{eqnarray}
where $\Omega_R$ and $\Omega_I$ denote the intervals the width
$a=a_R+ia_I$ is integrated over.
\begin{eqnarray}
\int_{\Omega_R} \mbox{d}a_R\; = \Omega_R\ ,\quad
\int_{\Omega_I} \mbox{d}a_I\; = \Omega_I\ .
\nonumber
\end{eqnarray}
The measure then changes to
\begin{eqnarray}
\mbox{d{\Large $\mu$}}(Q)
&=& 
\braket{\hat{Q}}{\hat{Q}}
\prod_{k=1}^A
\frac{1}{\braket{q_k}{q_k}}
\frac{\mbox{d}^3r_k\;\mbox{d}^3p_k}{(2\pi)^3}\
\frac{\mbox{d}a_R}{\Omega_R}\
\frac{\mbox{d}a_I}{\Omega_I}\;\sum_{m_s(k),\xi(k)}\;\ .
\end{eqnarray}

\subsubsection{The partition function}

Once the resolution of unity is given in terms of model states
the partition function can be evaluated.
Eq.~\fmref{ASEinsOp} is very useful in calculating traces by
means of Monte Carlo sampling \cite{OhR93}. For instance the
canonical partition function is given by
\begin{eqnarray}
\label{PartFunc}
Z(T)
&=&
\Tr\left(\exp\left\{-\Operator{H}/T \right\}\right)
\\
&=&
\int \mbox{d{\Large $\mu$}}(Q)\;
\bra{Q} \exp\left\{-\Operator{H}/T \right\} \ket{Q}
\ .
\nonumber
\end{eqnarray}

\subsubsection{Example}

In the following the above considerations are illustrated with
the example of $A$ identical fermions in a common
one--dimensional harmonic oscillator potential \cite{Sch96}.
Starting from the Hamilton operator
\begin{eqnarray}
\Operator{H}
&=&
\sum_{n=1}^A
\OphHO(n)\ ,
\quad
\OphHO(n) = \frac{\Operator{\vec{k}}^2(n)}{2 m}
          + \half m \omega^2 \Operator{\vec{x}}^2(n)
\end{eqnarray}
the mean energy of the $A$--fermion system can be derived from
the partition function $Z(T)$ \fmref{PartFunc} as the
derivative with respect to $T$
\begin{eqnarray}
\EnsembleMean{\Operator{H}}
&=&
T^2 \frac{\partial}{\partial T} \ln(Z(T))
\\
&=&
\frac{
\int \mbox{d{\Large $\mu$}}(Q)\;
{\mathcal W}(T)\;
\summn {\mathcal O}_{nm}(T) 
\left[T^2 \frac{\partial}{\partial T} 
\bram\exp\Big\{-\OphHO/T\Big\}\ketn\right] 
}{
\int \mbox{d{\Large $\mu$}}(Q)\;
{\mathcal W}(T)
}\ ,
\nonumber
\end{eqnarray}
where two the abbreviations ${\mathcal W}(T)$ and 
${\mathcal O}^{-1}(T)$  are introduced as
\begin{eqnarray}
{\mathcal W}(T)
&=&
\frac{\mbox{det}\Big(\brak \exp\{-\OphHO/T\}\ketl\Big)}
     {\mbox{det}\Big(\prodkl\Big)}\ ,
\\[3mm]
\left({\mathcal O}^{-1}(T)\right)_{kl}
&=&
\brak \exp\{-\OphHO/T\}\ketl \ .
\nonumber
\end{eqnarray}
\begin{figure}[hhht]
\unitlength1mm
\begin{picture}(120,65)
\put(30,0){\epsfig{file=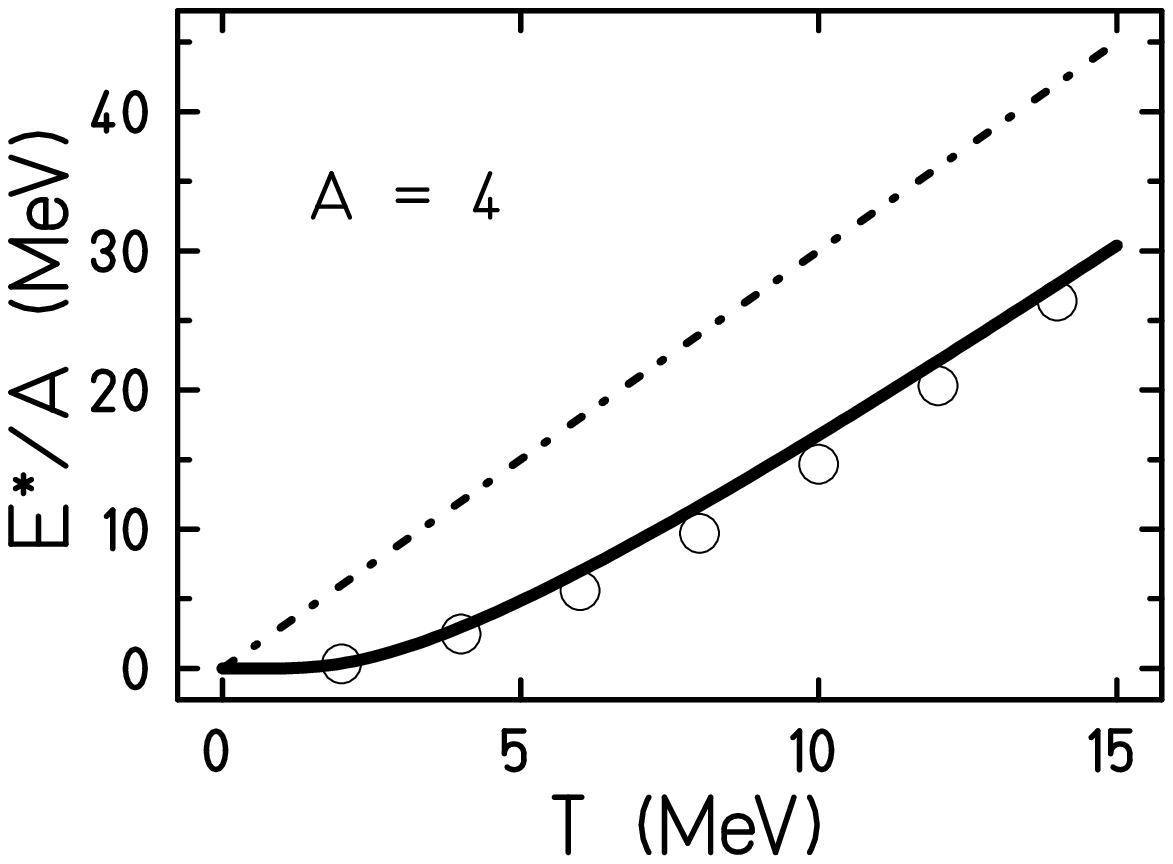,height=60mm}}
\end{picture}
\mycaption{Excitation energy as a function of temperature,
calculated with FMD model states (circles) or with
eigenfunctions of the harmonic oscillator (solid line).
The dashed--dotted line shows the classical result.
}{P-5.1-1}
\end{figure} 
The matrix elements can be evaluated in closed form \cite{Sch96}.
Figure \xref{P-5.1-1} shows the result of a Monte--Carlo
simulation for a system of four identical fermions in a harmonic
oscillator with $\hbar\omega=8~\MeV$ by open circles.
The solid line shows the same result, but calculating the
partition function with eigenstates of the Hamilton operator,
which obviously gives the same relation. For comparison the
classical dependence is shown as a dashed--dotted line.

\subsection{Canonical versus ergodic ensemble}
\label{ErgodicE}

Fermionic Molecular Dynamics is a deterministic microscopic
transport theory.  Given the Hamilton operator and a state
$\ket{Q(t_0)}$ at a certain time $t_0$ the state $\ket{Q(t)}$ is
known for all times.

Expectation values are well--defined in FMD so that one can
easily calculate quantities like the excitation energy of a
nucleus or the probability of finding the system in a given
reference state.  But it is not obvious how intensive
thermodynamical quantities, such as the temperature, might be
extracted from deterministic molecular dynamics with wave
packets.  In classical mechanics with momentum--independent
interactions the partition function
\begin{eqnarray}
&&
Z_{\mbox{\scriptsize classical}}(T)
=
\int
\prod_{k=1}^A
\frac{\mbox{d}^3r_k\;\mbox{d}^3p_k}{(2\pi)^3}\;
\exp\left\{-\frac{1}{T}\,{\cal H}_{\mbox{\scriptsize classical}}(\vec{r}_1,\vec{p}_1,\cdots)\right\}
\\
&=&
\int
\prod_{k=1}^A
\mbox{d}^3 p_k\;
\exp\left\{-\frac{1}{T}\,\sum_{i=1}^{A} \frac{\vec{p}_i^{\,2}}{2 m_i}\right\}
\times
\int
\prod_{l=1}^A
\frac{\mbox{d}^3r_l}{(2\pi)^3}\;
\exp\left\{-\frac{1}{T}\,{\cal V}(\vec{r}_1,\vec{r}_2,\cdots)\right\}
\nonumber
\end{eqnarray}
is a product of a term with the kinetic energy and a term
containing the interactions. Therefore, the momentum distribution
can be used to determine the temperature $T$. In the
quantum case eq.~\fmref{PartFunc} has to be employed which does
not show this factorization. A simple example for this behaviour
is the ground state of the free Fermi gas where finite momenta
are occupied, nevertheless the temperature is zero. Another
example is the ground state of a nucleus for which the momentum
distribution has a smeared out Fermi edge due to the finite size
and not because of temperature.

In this section time averaging is compared with a statistical
ensemble.  If the system is ergodic both are equivalent and
statistical properties of molecular dynamics can be evaluated by
means of time averaging.

For this the ergodic ensemble is defined by the statistical
operator $\Rerg$ as
\begin{eqnarray}
\label{EER}
\Rerg
&\; := \;&
\lim_{t_2\rightarrow\infty}\;
\frac{1}{(t_2 - t_1)}\;
\int_{t_1}^{t_2} \mbox{d}t\;
\ket{Q(t)}\bra{Q(t)}\ .
\end{eqnarray}
The ergodic mean of an operator $\Operator{B}$ is given by
\begin{eqnarray}
\label{EEM}
\hspace{-8mm}\ErgodicMean{\Operator{B}}
:= \;
\Tr\left(\Rerg\;\Operator{B}\right)
=
\lim_{t_2\rightarrow\infty}\;
\frac{1}{(t_2 - t_1)}\;
\int_{t_1}^{t_2} \mbox{d}t\;
\bra{Q(t)}\Operator{B}\ket{Q(t)}\ .
\end{eqnarray}
In general the statistical operator $\Rerg$ is a functional
of the initial state $\ket{Q(t_1)}$,
the Hamilton operator $\Operator{H}$ and
the equations of motion.
If the ergodic assumption is fulfilled,
the statistical operator should only depend 
on $\erw{\Operator{H}}$,
which is actually a constant of motion.
Thus the average in the ergodic ensemble is always performed
at the same expectation value of the Hamilton operator.
In our notation this is denoted by the
condition "$\erw{\Operator{H}}$" in eq.~\fmref{EEM}.

\subsubsection{Canonical ensemble of fermions in a harmonic oscillator}

With the statistical operator of the canonical ensemble for $A$
identical fermions (all spin up) in a one--dimensional common
harmonic oscillator potential $\HHO$ given by
\begin{eqnarray}
\label{RHOexakt}
\Operator{R}(T)
&=&
\frac{1}{Z(T)}
\exp\left\{-\frac{\HHO}{T}\right\}
\\
\HHO
&=&
\sum_{l=1}^A
\Operator{h}(l)
\ ,\qquad
\Operator{h}(l)
=
\omega \; \sum_{n=0}^\infty 
\Big(n + \half\Big)\; \Operator{c}_n^+\Operator{c}_n\ ,
\nonumber
\end{eqnarray}
the statistical mean of an operator $\Operator{B}$ 
is calculated as
\begin{eqnarray}
\label{HOMean}
&&\hspace{-1mm}
\EnsembleMean{\Operator{B}} 
:=
\Tr\left(\Operator{R}(T)\;\Operator{B}\right)
\\
&&\hspace{-1mm}=
\frac{1}{Z(T)}
\int
\frac{\mbox{d}r_1\;\mbox{d}p_1}{2\pi}
\cdots
\frac{\mbox{d}r_A\;\mbox{d}p_A}{2\pi}\;
\bra{\hat{Q}}
\Operator{B}\;\exp\left\{-\frac{\HHO}{T}\right\}
\ket{\hat{Q}}
\nonumber\\
&&\hspace{-1mm}=
\frac{1}{Z(T)}
\sum_{n_1 < \cdots <n_A}
\bra{n_1,\cdots,n_A}
\Operator{B}
\ket{n_1,\cdots,n_A}
\;
\exp\left\{-\frac{E(n_1,\cdots,n_A)}{T}  \right\}\ .
\nonumber
\end{eqnarray}
As already mentioned the FMD states are a representation of the
unit operator and hence can be used to calculate
traces.  For numerical convenience, however, the mathematically
identical third line in eq. \fmref{HOMean} is used, where
$\ket{n_1,\cdots,n_A}$ denotes the Slater determinant composed
of single--particle oscillator eigenstates $\ket{n_1} ,\cdots ,
\ket{n_A}$ and
\begin{eqnarray}
E(n_1,\cdots,n_A)
=
\omega\; \sum_{i=1}^A \; \left( n_i + \half  \right)
\end{eqnarray}
are the eigenenergies of $\HHO$.
In eq. \fmref{HOMean} the subscript $T$ indicates that the
average is taken at a constant temperature $T$.

In the following a system of four fermions in a
common one--dimensional harmonic oscillator is investigated.
The frequency of the oscillator is chosen to be 
$\omega=0.04~\fm^{-1}$ in order to get a spacing of 8~MeV
between the single--particle eigenstates.
For the canonical ensemble 
\figref{HOFermionen} shows the dependence of the excitation
energy on the temperature (l.h.s.)
and displays how the lowest eigenstates are occupied
in the four--fermion system for five different temperatures (r.h.s.).

\begin{figure}[hhht]
\unitlength1mm
\begin{picture}(120,55)
\put( 0,0){\epsfig{file=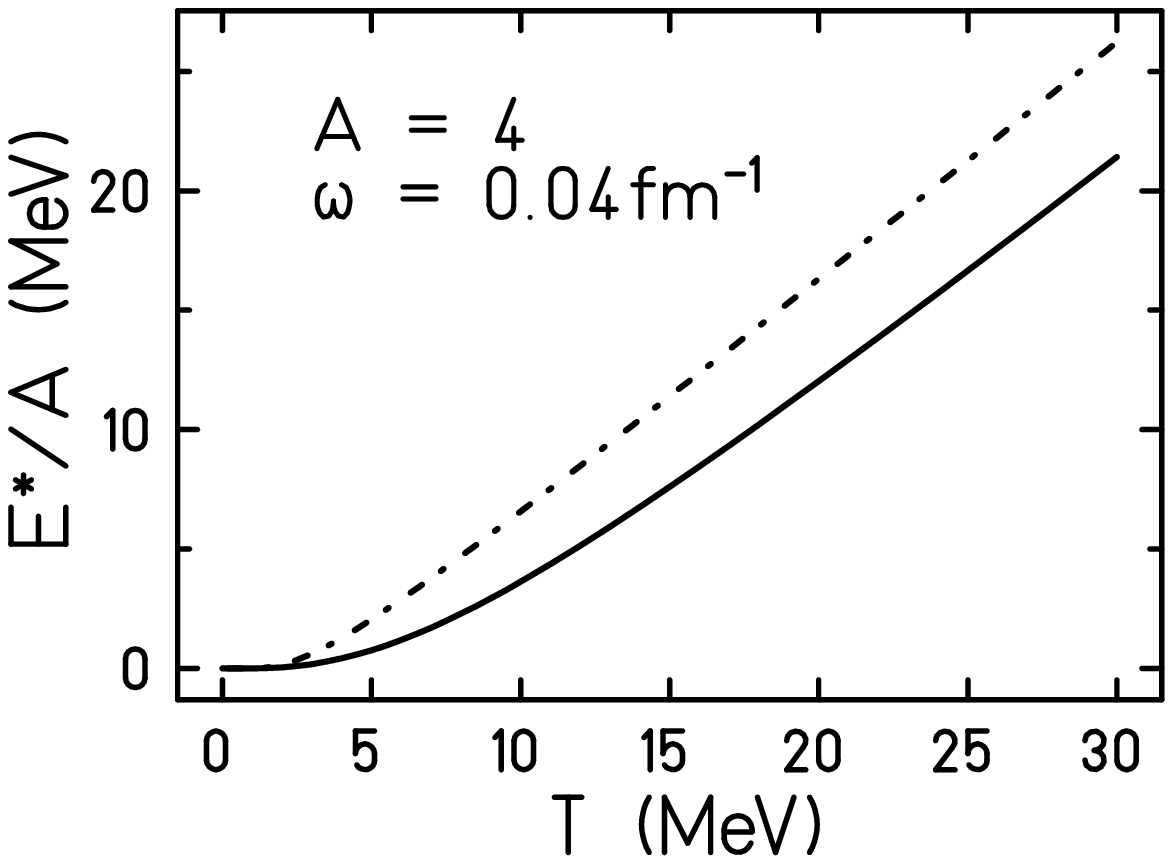,height=50mm}}
\put(80,0){\epsfig{file=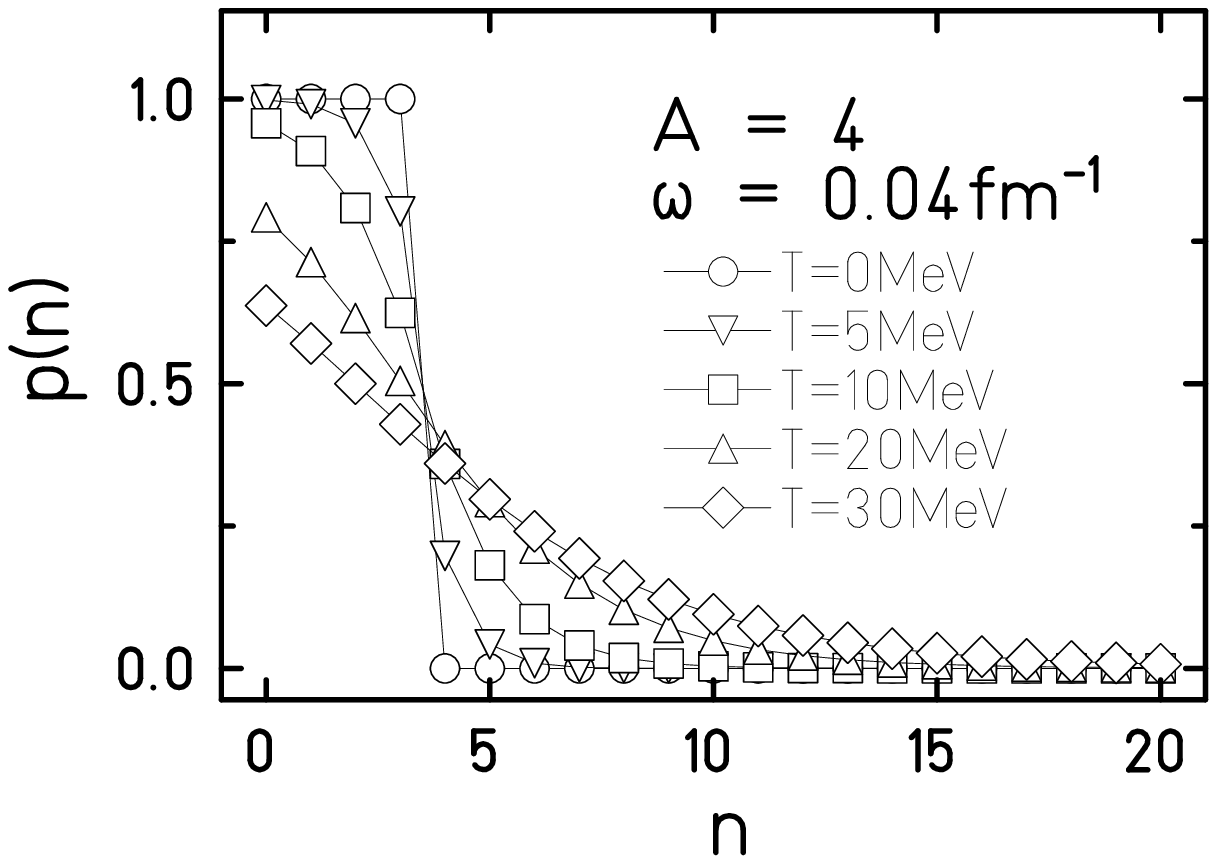,height=50mm}}
\end{picture}
\mycaption{A system of four fermions in a common oscillator
described by the canonical ensemble.
L.h.s.: Excitation energy as a function of 
temperature (solid line). The dashed--dotted line
shows the result for a product state (Boltzmann statistics).
R.h.s.: Occupation numbers $p(n)$ of the oscillator
eigenstates for five temperatures 
(eq. \fmref{ExaktHOBesetz}).
The lines are drawn as a guide for the eye.}{HOFermionen}
\end{figure} 

The mean occupation probabilities are given by
\begin{eqnarray}
\label{ExaktHOBesetz}
p(n) = \EnsembleMean{\Operator{c}_n^+\Operator{c}_n}\ ,
\end{eqnarray}
where $\Operator{c}_n^+$ denotes the creation operator
of a fermion in the oscillator eigenstate $\ket{n}$.

\subsubsection{Ergodic ensemble of fermions in a harmonic oscillator}

In this section the averages of the occupation numbers in the
ergodic ensemble are evaluated and compared with those of the
canonical ensemble discussed in the previous section.  As
pointed out already, in Fermionic Molecular Dynamics the time
evolution of Gaussian wave packets in a common oscillator is
exact, and thus the occupation probabilities of the eigenstates
of the Hamilton operator do not change in time.  In order to
equilibrate the system a repulsive short--range interaction
$\Operator{V}_{I}$ is introduced.
The strength of the interaction is chosen such that
the resulting matrix ele\-ments of $\VI$
are small compared to the level spacing $\omega$
and the excitation energy $E^*$.
The contribution of $\erw{\VI}$
to the total energy is of the order of $0.1\dots 1.0$~MeV.

The initial state is prepared in the following way.
Three wave packets with a width of 
$a = 1/m\omega$ are put close to the origin
at $x = (-d, 0, d)$
--- with $d=0.5/\sqrt{m\omega}$ ---
whereas the fourth packet with the same width is pulled away from 
the centre in order to obtain the desired energy.
As the mean momenta are all zero, the excitation
is initially only in potential energy which 
has to be converted into thermal energy by means of the
small interaction $\VI$.
 
The initial system, which is far from equilibrium,
is evolved over about 2000 periods
of the harmonic oscillator ($2\pi/\omega=157$~fm/c).
The equilibration time is rather large
as we are using a very weak interaction in order
not to introduce correlations which would destroy
the ideal gas picture implied in the canonical ensemble \fmref{RHOexakt}
of non--interacting particles.
The time averaging of the occupation numbers 
\fmref{EEHOBesetz} starts at time $t_1=10000$~fm/c 
in order to allow a first equilibration.
\begin{eqnarray}
\label{EEHOBesetz}
\ErgodicMean{\Operator{c}_n^+\Operator{c}_n}
&=&
\lim_{t_2\rightarrow\infty}\;
\frac{1}{(t_2 - t_1)}\;
\int_{t_1}^{t_2} \mbox{d}t\;
\bra{Q(t)} \Operator{c}_n^+\Operator{c}_n \ket{Q(t)}
\end{eqnarray}

\begin{figure}[tttt]
\unitlength1mm
\begin{picture}(120,55)
\put( 0,0){\epsfig{file=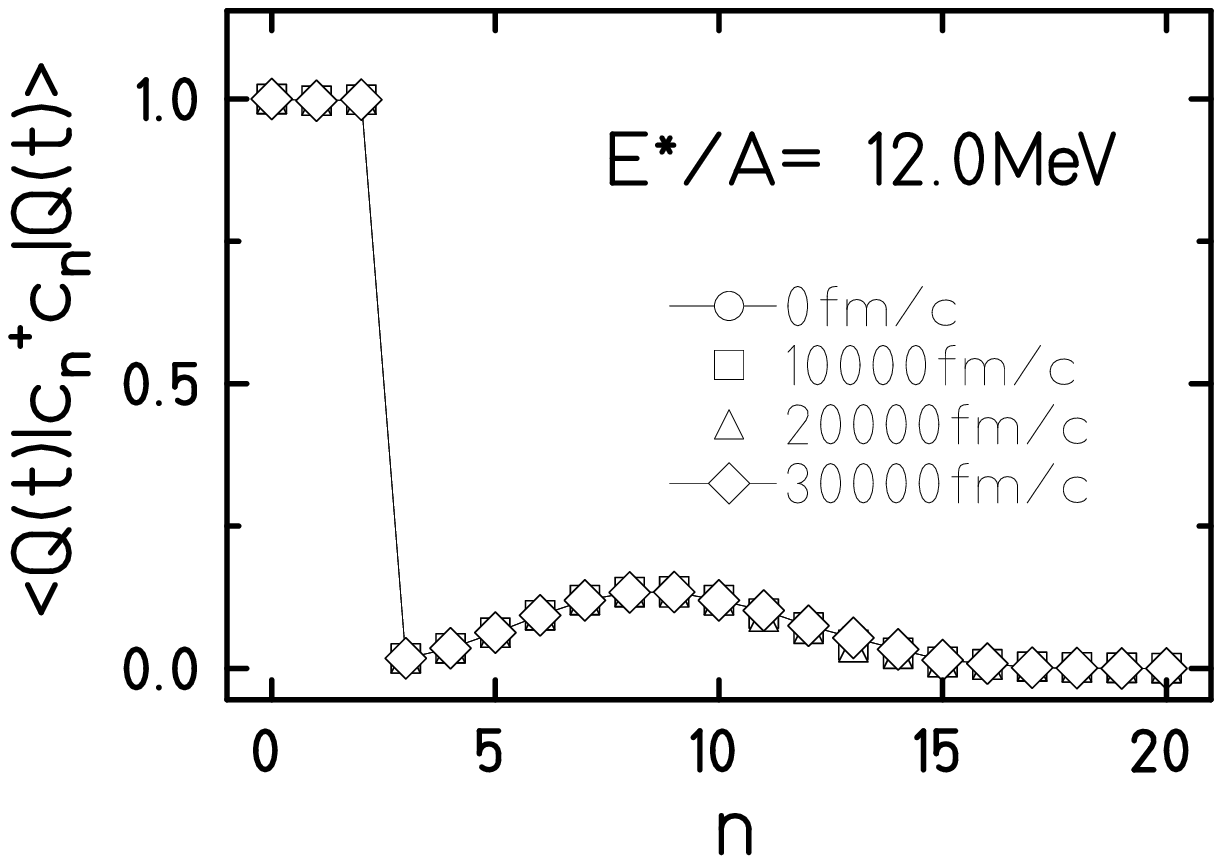,height=50mm}}
\put(80,0){\epsfig{file=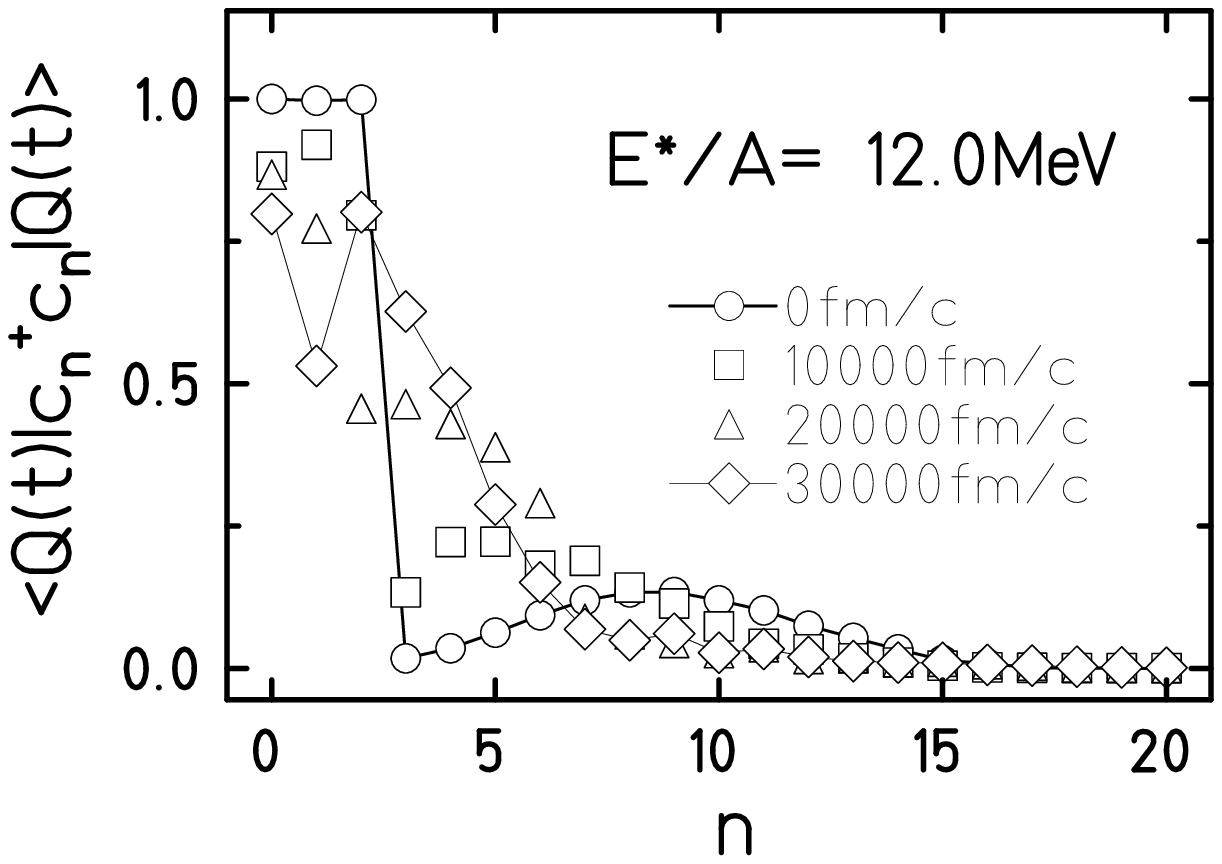,height=50mm}}
\end{picture}
\mycaption{Time evolution of the occupation probabilities 
for four fermions in a common harmonic oscillator potential
without (l.h.s.) and with two--body interaction (r.h.s.).  The
distributions at $t=0$ and $t=30000$fm/c are connected by a
solid line.}{HOUnitaer}
\end{figure} 
Figure \xref{HOUnitaer} gives an impression of
how the occupation numbers evolve in time.
The part to the left shows the time evolution without interaction
which is just a unitary transformation in the one--body space.
Thus the occupation numbers do not change in time
although the wave packets are swinging. 
This has been expected since the $\Operator{c}_n^+$ 
create eigenstates of the Hamiltonian $\HHO$.
It also serves as an accuracy test of the integrating routine.
The part to the right displays the evolution with interaction
at three later times.
The occupation probabilities are reshuffled due to the interaction
and they fluctuate in time.
In \figref{OccNumTime} (l.h.s.) the chaotic time dependence of 
$\bra{Q(t)} \Operator{c}_n^+\Operator{c}_n \ket{Q(t)}$
for $n=0, 3\; \mbox{and}\; 6$ is depicted.

\begin{figure}[hhhh]
\unitlength1mm
\begin{picture}(150,55)
\put( 0,0){\epsfig{file=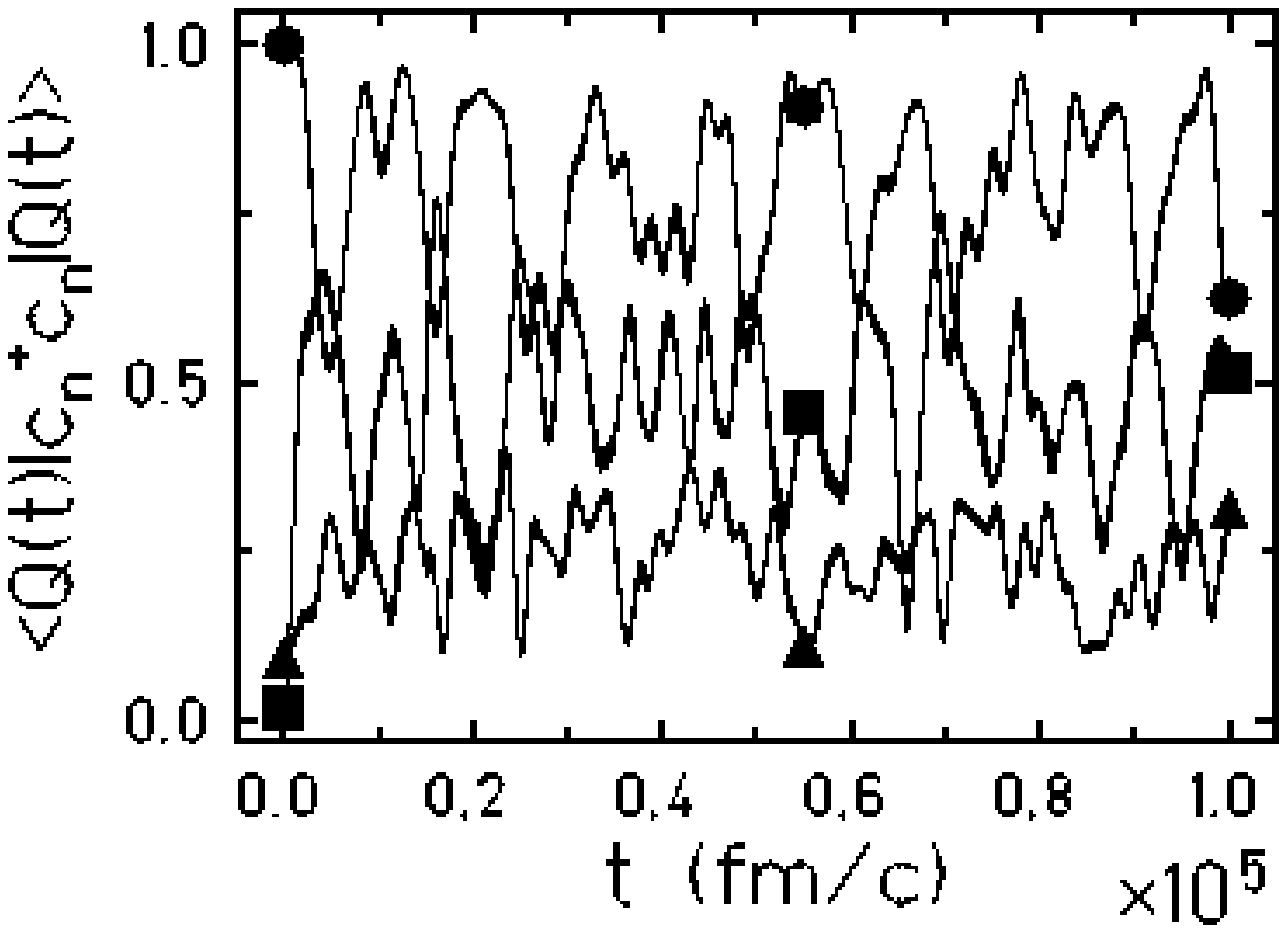,height=50mm}}
\put(80,0){\epsfig{file=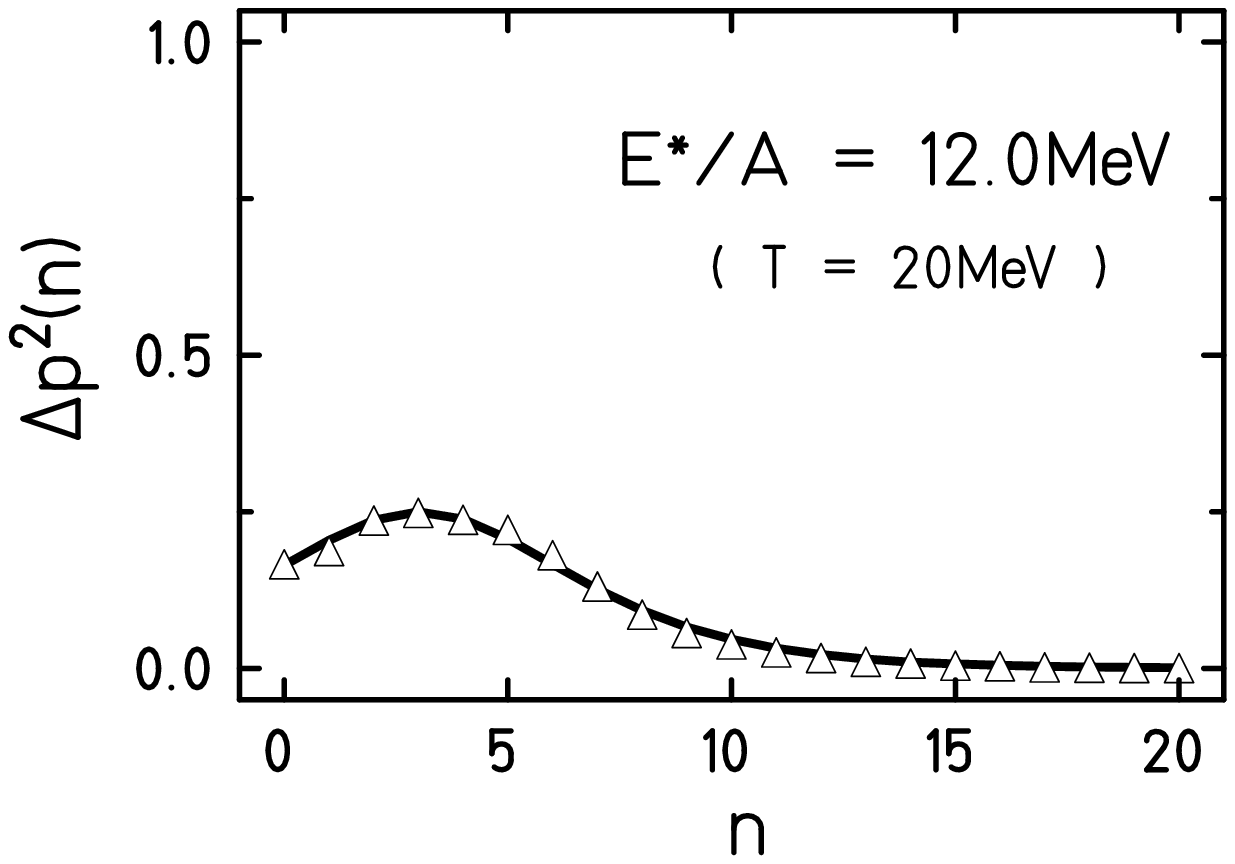,height=50mm}}
\end{picture}
\mycaption{L.h.s.: Occupation probabilities 
versus time ---
$n=0$: circles, 
$n=3$: squares,
$n=6$: triangles.
R.h.s.: Variance of the fluctuations 
$\Delta p^2(n)$
calculated in the canonical ensemble (solid line)
and in the ergodic ensemble (triangles).}{OccNumTime}
\end{figure} 

The result of time averaging is seen in \figref{HOBesetz}
(symbols) for four different initial displacements which
correspond to four different excitation energies of the fermion
system.  To each case we assign a canonical ensemble which has
the same mean energy.  The solid lines in \figref{HOBesetz} show
the corresponding distributions of occupation probabilities for
these canonical ensembles.  Their temperatures $T$ are also
quoted in the figure.  It is surprising to see that there is
almost no difference between the ergodic and the canonical
ensemble:
\begin{eqnarray}
\label{CCIdentity}
\ErgodicMean{\Operator{c}_n^+\Operator{c}_n}
&\approx&
\;\;\EnsembleMean{\Operator{c}_n^+\Operator{c}_n}
\qquad \forall \; n
\ ,
\end{eqnarray}
provided both have the same excitation energy
\begin{eqnarray}
E^* = \ErgodicMean{\HHO - E_0}
&=&
\;\;\EnsembleMean{\HHO - E_0}\ ,
\quad E_0 = 8\; \omega\ .
\end{eqnarray}
The relation between $E^*$ and $T$ is given by eq. \fmref{HOMean}
and displayed in \figref{HOFermionen}.

\begin{figure}[tttt]
\unitlength1mm
\begin{picture}(150,110)
\put(0,0){\epsfig{file=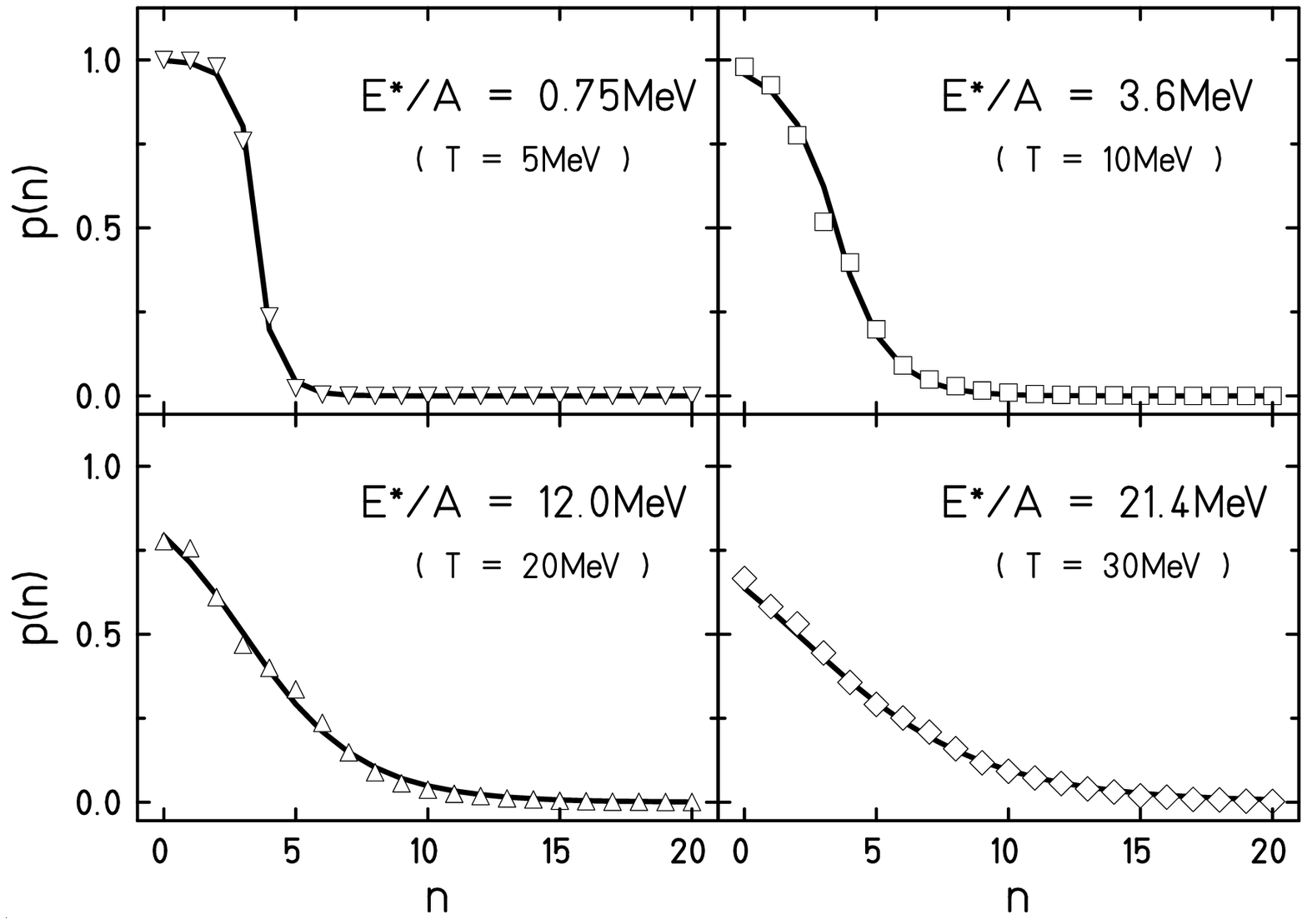,width=150mm}}
\end{picture}
\mycaption{Occupation numbers calculated in the ergodic ensemble
(symbols, eq. \fmref{EEHOBesetz}) compared with the 
canonical ensemble (solid line, eq. \fmref{ExaktHOBesetz}).
}{HOBesetz}
\end{figure} 

This result is not trivial because, firstly,
the system is very small, consisting of only four particles,
and secondly, the equations of motion are approximated by FMD.
The one to one correspondence between the occupation probabilities
of the ergodic ensemble and the ones of the canonical ensemble,
which has the same mean energy $\erw{\Operator{H}}$ 
as the pure state, is an impressive demonstration
that the system is ergodic and that
the FMD many--body trajectory covers the phase space
according to Fermi--Dirac statistics.

Not only the one--body distributions
of the two ensembles coincide,
but also the variances of the fluctuations $\Delta p^2(n)$,
\begin{eqnarray}
\Delta p^2(n) 
&:=&
\EnsembleMean{\big(\Operator{c}_n^+\Operator{c}_n\big)^2}
-
\EnsembleMean{\Operator{c}_n^+\Operator{c}_n}^{\hspace{-2.0ex}2}
\;\ ,
\end{eqnarray}
as is demonstrated in \figref{OccNumTime} (r.h.s.).
The ergodic mean converges to the result 
of the canonical ensemble which is $\Delta p^2(n) = p(n) (1-p(n) )$.

\subsubsection{Canonical and ergodic ensemble for distinguishable particles}

In this section it is shown that time averaging results in 
quantum Boltzmann statistics if the fermions are replaced by 
distinguishable particles.
For this end the antisymmetrized many--body state
is replaced by a product state of Gaussian wave packets.
The resulting equations of motion differ from the FMD case in 
the skew--symmetric matrix ${\mathcal A}_{\mu\nu}(Q)$ 
(given in eq. \fmref{Bewgl})
which does not couple the generalized velocities of different
particles any longer.

For product states the ergodic ensemble is again 
investigated at different energies and compared with
the canonical ensembles with the same mean energies.
The appropriate relation between temperatures and excitation 
energies in the canonical ensemble for distinguishable particles
\begin{eqnarray}
\label{EMeanDP}
E^* = \EnsembleMean{\HHO - E_0}
=
4\; \frac{\omega}{2}\;
\Big[
\mbox{coth}\left( \frac{\omega}{2\, T}  \right) -1
\Big]
\ ,
\quad E_0 = 2\; \omega
\end{eqnarray}
is shown by the dashed--dotted line in \figref{HOFermionen}.

Since distinguishable particles are not affected by 
the Pauli principle, the occupation numbers for the
many--body ground state look quite different. 
For instance for zero temperature
all particles occupy the eigenstate $\ket{0}$ of
the harmonic oscillator (\figref{HOBoltzmann}, l.h.s.). 

\begin{figure}[h]
\unitlength1mm
\begin{picture}(150,55)
\put( 0,0){\epsfig{file=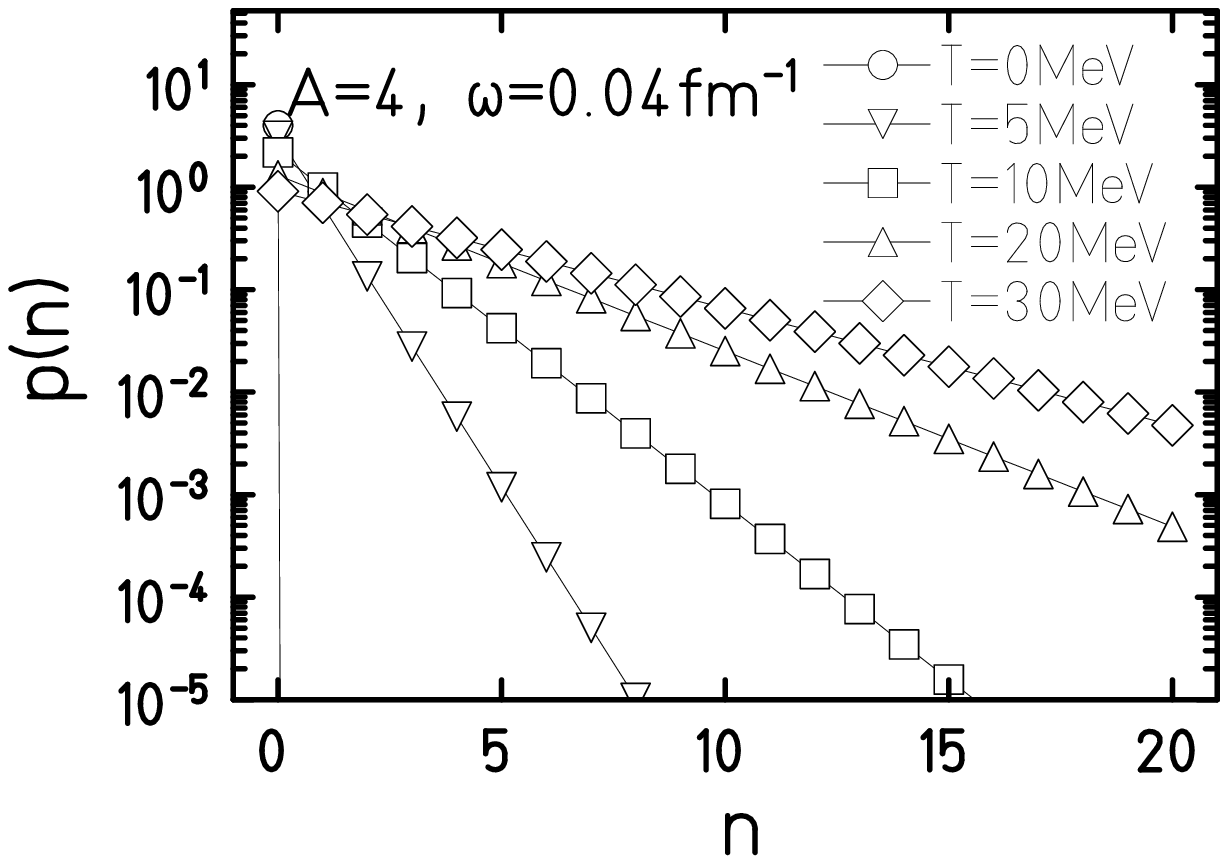,height=50mm}}
\put(80,0){\epsfig{file=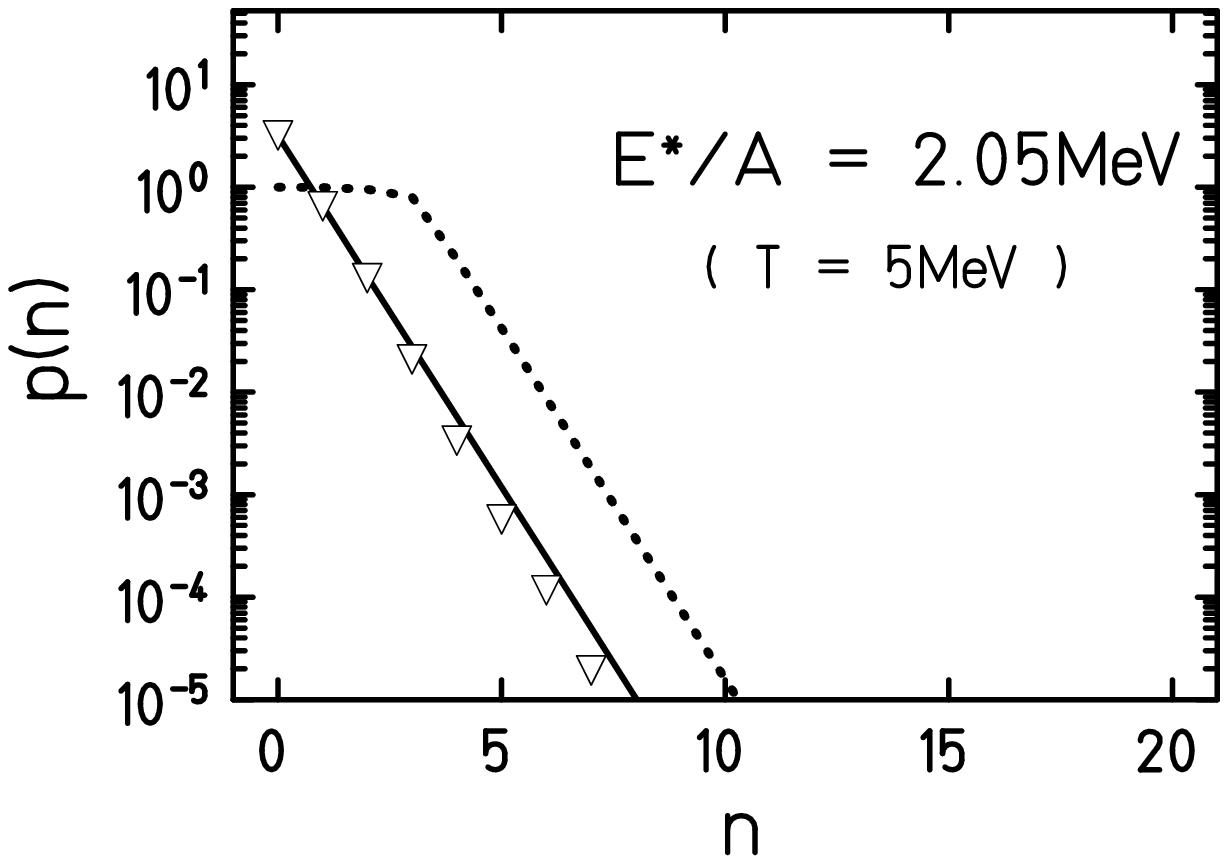,height=50mm}}
\end{picture}
\mycaption{Occupation probabilities for a product state (Boltzmann statistics).
L.h.s.: Occupation probabilities $p(n)$ of the oscillator
eigenstates for five temperatures for
the canonical ensemble.
R.h.s: Occupation probabilities calculated in the ergodic ensemble
(symbols) compared with those calculated in the 
canonical ensemble (solid line) for an excitation energy
of $E^*=2.05$~AMeV which corresponds to a temperature
$T=5$~MeV in the canonical ensemble.
The dotted line shows the result for Fermi--Dirac statistics
at the same temperature.}{HOBoltzmann}
\end{figure} 

The initial single--particle states of the interacting system
are chosen analogue to the fermion case.
Again the time evolution of the system exhibits ergodic behaviour
for all excitation energies.
As an example \figref{HOBoltzmann} (r.h.s.) is showing
the case of $E^*/A = 2.05$~MeV ($T = 5$~MeV) 
after a time averaging of about 2000 periods.
The ergodic ensemble (triangles) and the Boltzmann canonical ensemble
(solid line) are the same within the size of the symbols.
The result for Fermi--Dirac statistics 
with the same temperature is included
to demonstrate the difference (dotted line).

This result shows that equations of motion which are not
influenced by the Pauli principle lead to the quantum Boltzmann
distribution.  The only difference to "true classical" equations
is the presence of the width parameters as dynamical variables.
Only if they are removed from the equations of motion the
statistical behaviour of the ergodic ensemble is that of
classical statistics.

\subsection{Caloric curve for finite nuclei}
\label{SubSec-5.3}

As an outlook of this chapter a method will be presented that
allows to investigate the caloric curve of finite charged
self--bound Fermi systems like nuclei.

The concept of determining the temperature is to bring a
reference system, for which thermodynamic relations between
temperature and measurable quantities are known, into thermal
equilibrium with the investigated system.  The weakly
interacting ideal gas, where the temperature is given by the
mean kinetic energy of the particles, may serve as an
example. The reference system is called a heat bath if its heat
capacity is much larger than that of the system and it is called
a thermometer if its heat capacity is much less.

\begin{figure}[hhht]
\unitlength0.1cm
\begin{picture}(120,65)
\put(30,0){\epsfig{file=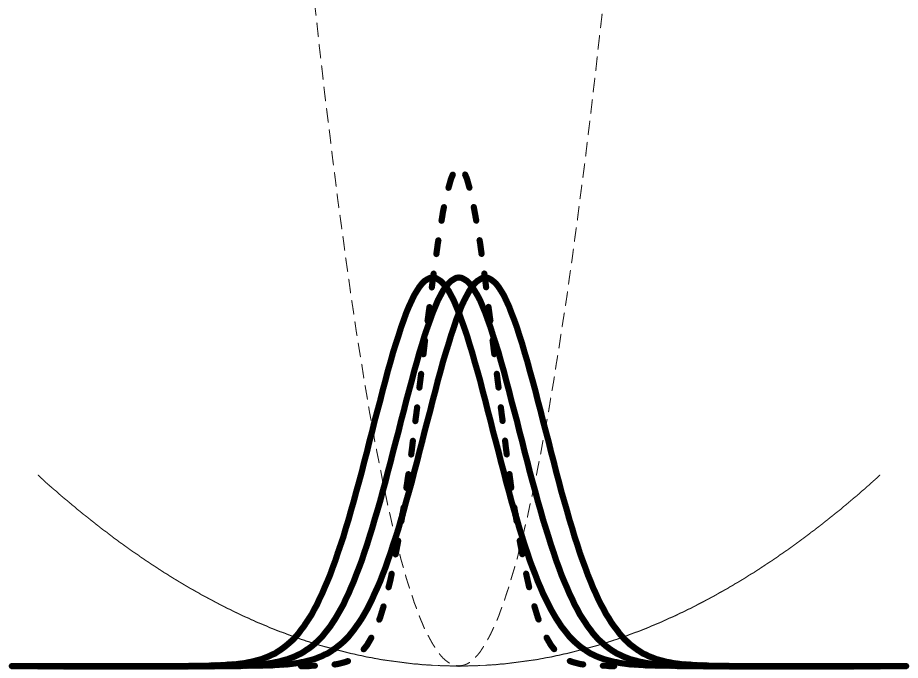,height=60mm}}
\end{picture}
\mycaption{Sketch of the setup: the self--bound excited nuclear
system is represented by Gaussian wave packets (solid lines)
which are enclosed in a broad container potential. For the
thermometer only one wave packet is shown (dashed line), it is
contained in a different oscillator.}{P-5.3-1}
\end{figure} 
As the nuclear system is quantal and strongly interacting its
temperature cannot be deduced from the momentum distribution or
the mean kinetic energy of the nucleons. Therefore, the concept
of an external thermometer which is coupled to the nuclear
system is used in the present investigation. The thermometer
consists of a quantum system of distinguishable particles moving in a
common harmonic oscillator potential different from the
container potential as shown in \figref{P-5.3-1}.

The time evolution of the whole system is described by the FMD
equations of motion. For this purpose the many--body trial
state is extended and contains now both, the nucleonic degrees
of freedom and the thermometer degrees of freedom
\begin{eqnarray}
\ket{Q}
=
\ket{Q_n} \otimes \ket{Q_{th}}
\ .
\end{eqnarray}
The total Hamilton operator including the thermometer is given
by
\begin{eqnarray}
\Operator{H}
=
  \Operator{H}_n(\omega)
+ \Operator{H}_{th}
+ \Operator{H}_{n-th}
\  ,
\end{eqnarray}
where $\Operator{H}_n(\omega)$ denotes the nuclear Hamiltonian
with an additional external field which serves as a container.
$\Operator{H}_{th}$ is the Hamilton operator of the thermometer
system, i.e. the Hamiltonian of a harmonic
oscillator. The week repulsive interaction between thermometer
wave packets and nucleons is given by $\Operator{H}_{n-th}$.

The determination of the caloric curve is done in the following
way. The nucleus is excited by displacing all wave packets from
their ground--state positions randomly. Both, centre of mass
momentum and angular momentum are kept fixed at zero. To allow a
first equilibration between the wave packets of the nucleus and
those of the thermometer the system is evolved over a long time
(14000~\fm/c). After that a time averaging of the energy of the
nucleonic system as well as of the thermometer is performed over
2000 steps covering a time interval of 2000~\fm/c. During this
time interval the mean of the nucleonic excitation energy
\begin{eqnarray}
E^*
&=&
\frac{1}{N_{steps}} \sum_{i=1}^{N_{steps}}
\bra{Q_n(t_i)} \Operator{H}_n \ket{Q_n(t_i)}
- E_0(N,Z)
\end{eqnarray}
is evaluated.  $E_0(N,Z)$ denotes the FMD ground--state energy
of the isotope under consideration.  The time--averaged energy
of the thermometer $E_{th}$ which is calculated during the same
time interval determines the temperature $T$ through the
relation for an ideal gas
\begin{eqnarray}
T
&=&
\omega_{th}
\left[
\ln\left(
\frac{E_{th}/N_{th}+\frac{3}{2}\omega_{th}}
     {E_{th}/N_{th}-\frac{3}{2}\omega_{th}}
\right)
\right]^{-1}
\ .
\end{eqnarray}
The system is then cooled and after another 2000~\fm/c, in which
the system equilibrates, the averaging is done again. Repeating
this procedure one follows the caloric curve from high
excitations to low excitations.

The relation between the excitation energy and the
temperature is evaluated for the three nuclei 
\element{16}{O}, \element{24}{Mg} and \element{40}{Ca} using the
same container potential with $\hbar\omega=1~\MeV$. 

\begin{figure}[h]
\unitlength0.1cm
\begin{picture}(120,65)
\put(  5,0){\epsfig{file=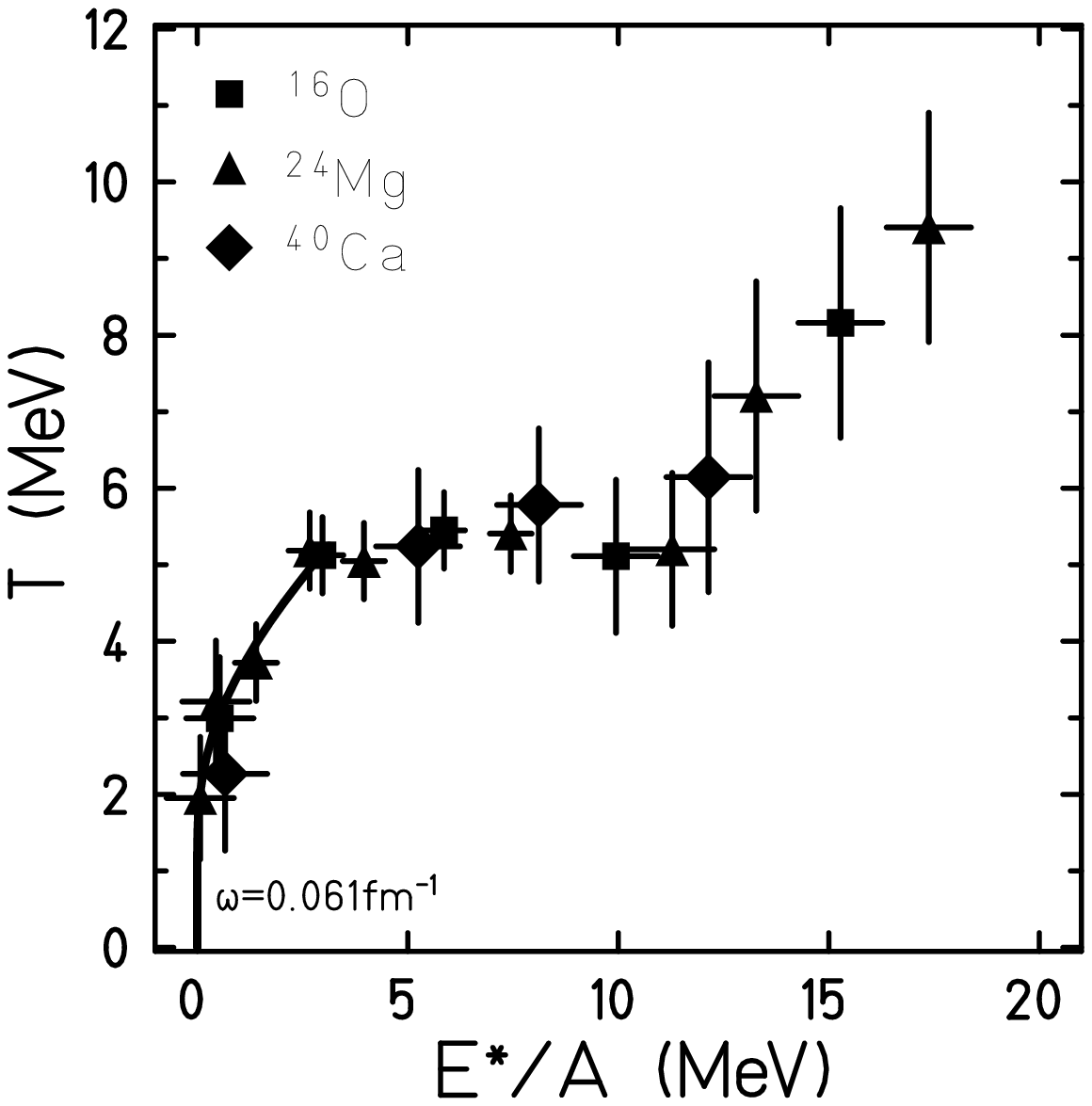,height=60mm}}
\put( 80,0){\epsfig{file=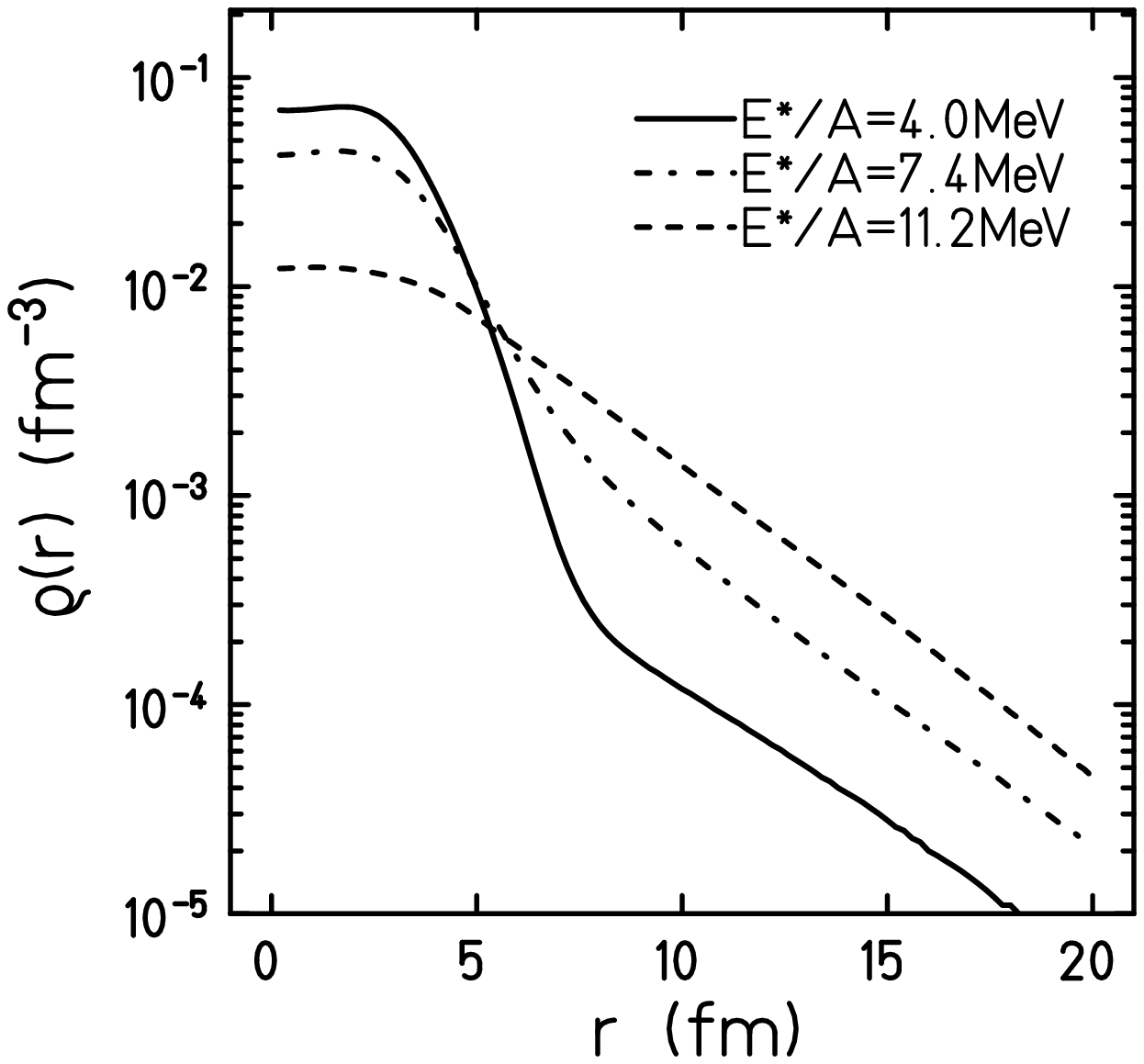,height=60mm}}
\end{picture}
\mycaption{L.h.s.: caloric curve of \element{16}{O}, 
\element{24}{Mg} and \element{40}{Ca} at $\hbar\omega=1~\MeV$,
r.h.s.: time averaged radial density distribution of \element{24}{Mg}
at various excitation energies in the coexistence region.
}{P-5.3-2}
\end{figure} 

The caloric curves shown in the graph on the left hand side of
\figref{P-5.3-2} clearly exhibit three different
parts. Beginning at small excitation energies the temperature
rises steeply with increasing energy as expected for the shell
model.  The nucleons remain bound in the excited nucleus which
behaves like a liquid drop of fermions.  At an excitation energy
of $3~\MeV$ per nucleon the curve flattens and stays almost
constant up to about $11~\MeV$.  This plateau at
$T\approx5~\MeV$ indicates the coexistence of liquid and
vapour phases, the latter consisting of evaporated nucleons
which are in equilibrium with the residual liquid drop due to the
containment.  Around $E^*/A\approx11~\MeV$ all nucleons are
unbound and the system has reached the vapour phase. This is
indicated by the steep rise of the caloric curve beyond this
point.  One has to keep in mind that the plateau is not the
result of a Maxwell construction as in nuclear matter
calculations.  In the excitation energy range between 3 and
11~\MeV\ per particle an increasing number of nucleons is found
in the vapour phase outside the liquid phase.  This can be seen
in the density plot on the right hand side of \figref{P-5.3-2},
where the radial dependence of the time--averaged density for
\element{24}{Mg} at three excitation energies in the coexistence
region is shown. For small excitations ($E^*/A=4.0~\MeV$) the
nucleus, which due to recoil is bouncing around, is surrounded
by very low density vapour (solid line).  The dashed line
($E^*/A=7.4~\MeV$) and the dashed--dotted line
($E^*/A=11.2~\MeV$) show that with increasing energy the vapour
contribution is growing and the amount of liquid decreasing.
However, in the high energy part of the plateau the averaged
one--body density shown here does not represent the physical
situation adequately. The time--dependent many--body state shows
the formation and disintegration of several small drops, which
due to time--averaging cannot be seen in \figref{P-5.3-2}. Above
$E^*/A\approx13~\MeV$ only vapour is observed.
\begin{figure}[hhhh]
\unitlength0.1cm
\begin{picture}(120,75)
\put( 40,-5){\epsfig{file=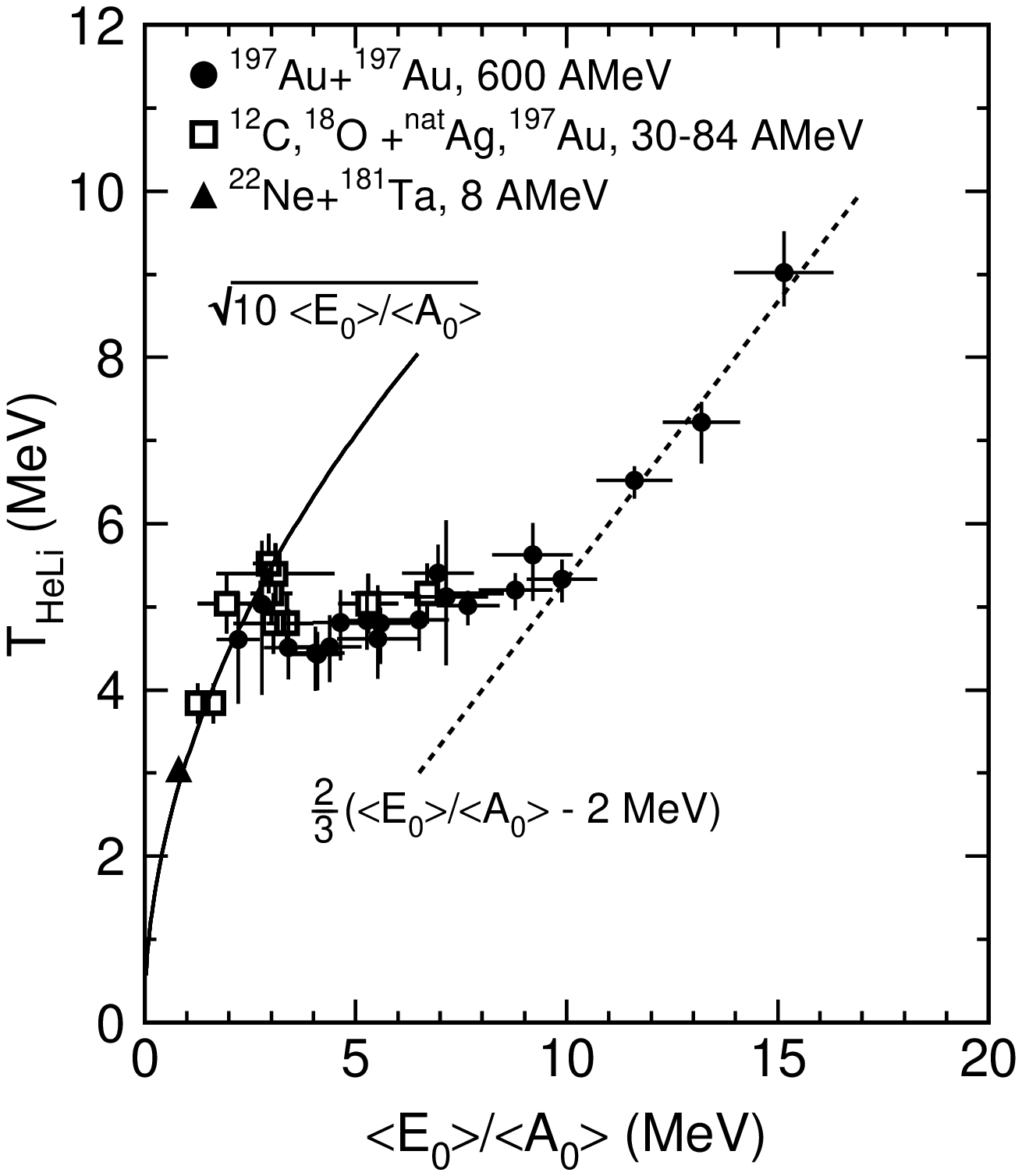,height=80mm}}
\end{picture}
\mycaption{Caloric curve extracted from spectator fragmentation 
by the ALADIN collaboration (picture taken from ref. \cite{Poc95}).}{P-5.3-3}
\end{figure} 

In how far the experimental result, \figref{P-5.3-3}, where the
temperature is determined from isotope ratios \cite{Poc95}, can
be compared to our time--averaged temperature and excitation
energy is still under investigation.

The results on the caloric curve of nuclei present another
example for the wide applicability of Fermionic Molecular
Dynamics.  A correct equilibrium behaviour is of course a
necessary condition for further investigations of
non--equilibrium situations as in nucleus--nucleus collisions.
These investigations are followed up presently.

\\[5mm]
\begin{center}
{\large\bf Acknowledgment}
\end{center}

We would like to thank M.~Petrovici who kindly provided us with
the data on \element{19}{F}--\element{27}{Al} collisions and
J.~Pochodzalla who supplied us with the caloric curve of nuclei.
We also thank Ph.~Chomaz and P.~Danielewicz for discussions and
T.~Neff and R.~Roth for carefully reading the manuscript.

\appendix
\newpage
\section{Interaction matrix elements}
\subsection{Approximation of the matrix elements}

A large fraction of the numerical effort of FMD is caused by
the evaluation of the interaction matrix elements
\begin{eqnarray}
{\mathcal V}(Q) 
\equiv
\erw{\Operator{V}}
= 
\Tr\left(\Operator{v}\, \Operator{\rho}^{(2)}\right)
=
\frac{1}{2}
\sumklmn
\brakl \Operator{v} \ketmn
\OO
\ .
\end{eqnarray}
This effort grows with $A^4$ because Gaussian wave packets are
not orthogonal. In order to reduce the computation time 
two types of approximations were tested \cite{Sch93,Sch96}.
Both approximation schemes use that antisymmetrization
effectively reduces the strength of the interaction.

The first ansatz \cite{Sch93} 
\begin{eqnarray}
\label{VApprox-1}
{\mathcal V}(Q) 
&\approx&
\sum_{k<l}
\frac{\brakl \Operator{V}
\ketkl_a}{\prodklkl}
\exp\left\{
-\frac{1}{4} c_{kl}
\right\}
\nonumber\\
&&\\
&&
c_{kl} 
= 
\sum_{m\ne k,l}^A \Bigg(
\frac{\prodkmmk}{\prodkmkm}
+ \frac{\prodlmml}{\prodlmlm}
\Bigg)
\ .
\nonumber
\end{eqnarray}
takes all matrix elements $\brakl\Operator{V}\ketkl$ and
$\brakl\Operator{V}\ketlk$ into account which are scaled with
an overall factor counting the overlapping wave packets.

The second ansatz \cite{Sch96} reduces this effort even further since it
needs only the direct matrix elements
\begin{eqnarray}
\label{VApprox-2}
{\mathcal V}(Q) 
&\approx&
\sum_{k<l}
\frac{\brakl \Operator{V} \ketkl_a}{\prodklkl}
\left( (1 - \gamma_{kl}) + \gamma_{kl} 
\exp\Bigg\{ - \frac{\prodkllk}{\prodklkl}
\Bigg\} \right)
\nonumber \\
&&\\
\gamma_{kl} &=& \erf\left\{
\half + 
\sum_{m\ne k,l}^A \Bigg(
\frac{\prodkmmk}{\prodkmkm}
+ \frac{\prodlmml}{\prodlmlm}
\Bigg) \right\}
\nonumber \ .
\end{eqnarray}
Both schemes \fmref{VApprox-1} and \fmref{VApprox-2} provide a
reasonable approximation for the expectation value of the
two--body interaction. In order to reproduce the
nuclear binding energies it can be necessary to readjust the
strength of the original interaction.

\subsection{Coulomb interaction}

This section briefly shows how the second approximation
\fmref{VApprox-2} works for the Coulomb interaction \fmref{VCoulomb}
\begin{eqnarray} 
\hspace*{-8mm}
\bra{\xvec_i,\xvec_j}\Vsim_{c}(i,j)\ket{\xvec_k,\xvec_l}
&=&
\frac{1.44\MeV\fm}{|\xvec_i-\xvec_j|}\;\Psim^p \otimes \Psim^p
\delta^3(\xvec_i-\xvec_k) \delta^3(\xvec_j-\xvec_l)
\end{eqnarray}
where $\Psim^p$ denotes the projection operator on the protons.
The diagonal matrix elements are
\begin{eqnarray}
\frac{\bra{q_k q_l}\vsim\ket{q_k q_l}}{\prodklkl}
&=& 
\frac{1.44\MeV\fm}{r_{kl}}
\quad \erf\Bigg\{
\sqrt{\frac{\lambda_{klkl}}{2}} \; r_{kl}
\Bigg\}
P_{kk}^p\; P_{ll}^p
\,
\end{eqnarray}
with
\begin{eqnarray}
r_{kl}
&=&
|\vecrk - \vecrl |
\\
\lambda_{klkl}
&=&
\frac{2\, a_{kR} \, a_{lR}}
{a_{kR} |a_l|^2 + a_{lR} |a_k|^2}
\nonumber\\
\erf(x)
&:=&
\frac{2}{\sqrt{\pi}} \int_{0}^{x} du \exp\{-u^2\}
\nonumber\\
P_{kl}^p 
&=&
\bra{\xi_k} \Operator{P}^p \ket{\xi_l}
=
\half (1 + \xi_l) \braket{\xi_k}{\xi_l}
\end{eqnarray}
The isospin variable $\xi_k$ takes values $\xi_k=1$ for protons
and $\xi_k=-1$ for neutrons, respectively.
\begin{figure}[hhht]
\unitlength1mm
\begin{picture}(120,60)
\put( 0,5){\epsfig{file=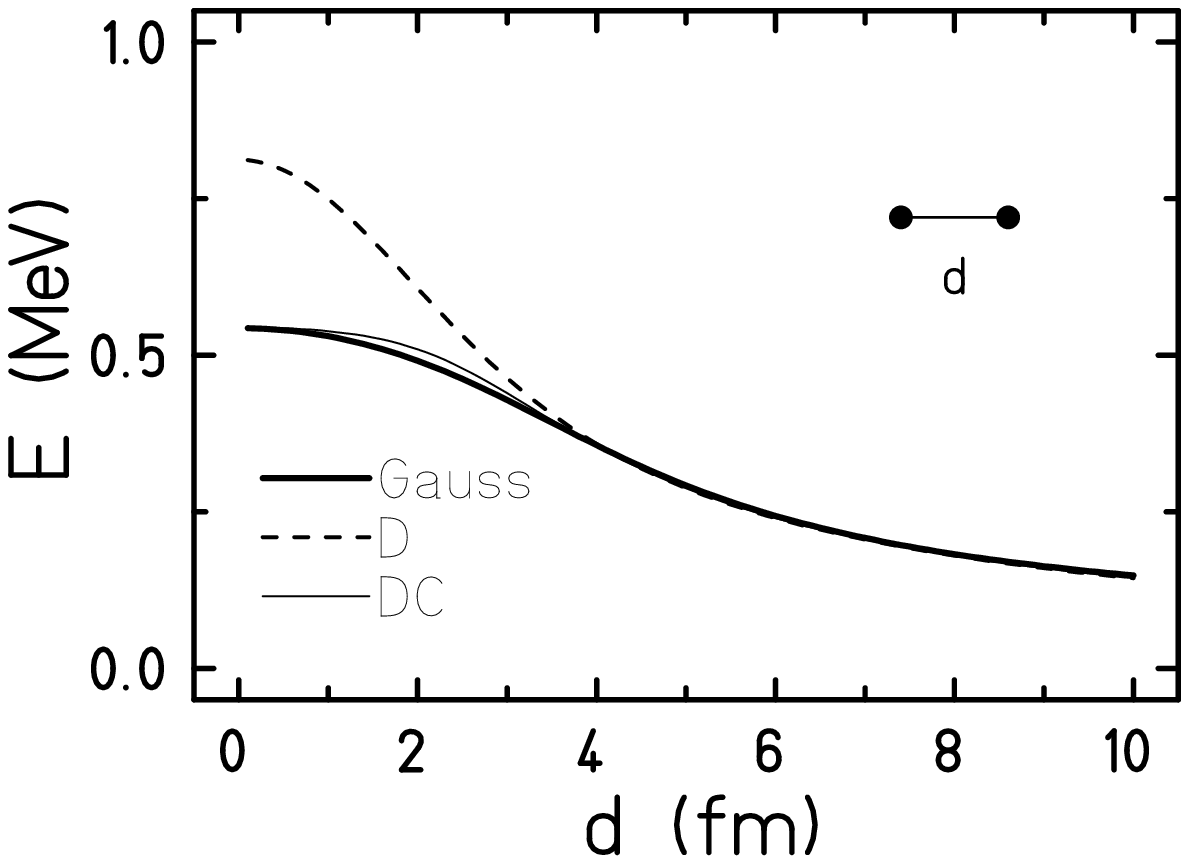,height=50mm}}
\put(80,5){\epsfig{file=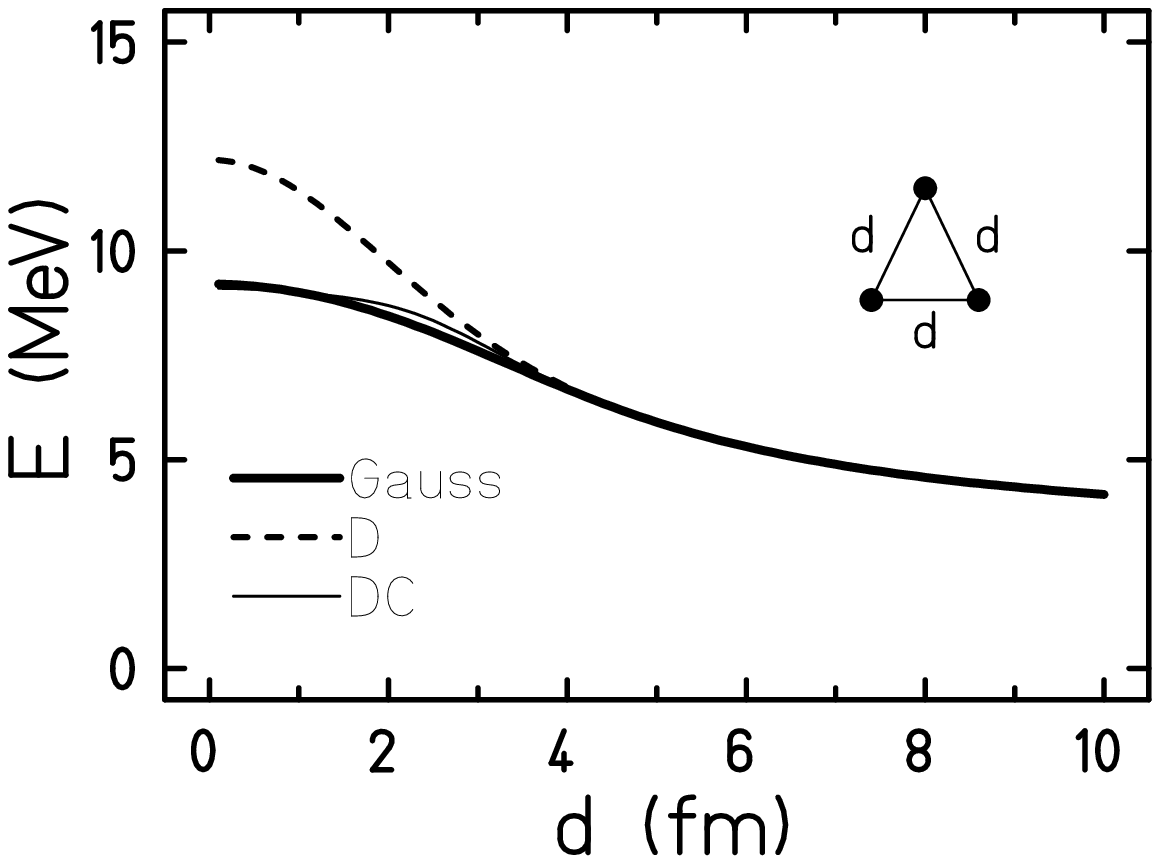,height=50mm}}
\end{picture}
\mycaption{Approximation of the Coulomb energy. The l.h.s shows
the Coulomb energy of two protons with the same spin component
as a function of the relative distance, the r.h.s. shows the
Coulomb energy of a \element{12}{C} nucleus as a function of the
relative distance between the three alpha clusters. The thick
solid line represents the exact solution, the dashed line the
direct term only and the thin line the result of the
approximation.}{P-A.1-1}
\end{figure} 

Figure \xref{P-A.1-1} shows the result of the approximation for
two systems, two identical protons and a \element{12}{C}
nucleus. The approximation (thin line) compares nicely to the
exact result (thick line).

\end{document}